\documentclass{iopart}

\usepackage{graphicx}
\usepackage{epsfig}
\usepackage{bm}
\usepackage{amssymb}
\usepackage{mathrsfs}
\usepackage{dcolumn}
\usepackage{iopams}
\usepackage{setstack}
\usepackage{pifont}
\usepackage[normalem]{ulem}
\usepackage{type1cm}
\usepackage{tabularx}

\newcommand{\beq}{\begin{equation}}
\newcommand{\eeq}{\end{equation}}
\newcommand{\bea}{\begin{eqnarray}}
\newcommand{\eea}{\end{eqnarray}}

\newcommand{\calC}{\mathcal{C}}

\begin{document}

%
\topical{Lattice methods for strongly interacting many-body systems}

\author{Joaqu\'{\i}n E. Drut$^{1,2}$, Amy N. Nicholson$^3$}

\address{$^1$ Theoretical Division, Los Alamos National Laboratory, Los Alamos, NM 87545-0001, USA}
\address{$^2$ Department of Physics and Astronomy, University of North Carolina, Chapel Hill, NC 27599-3255, USA}
\address{$^3$ Maryland Center for Fundamental Physics, Department of Physics, University of Maryland, College Park, MD 20742-4111, USA}
\eads{\mailto{drut@email.unc.edu}, \mailto{amynn@umd.edu}}
\date{\today}

\begin{abstract}\\
Lattice field theory methods, usually associated with non-perturbative studies of quantum 
chromodynamics, are becoming increasingly common in the calculation of ground-state and
thermal properties of strongly interacting non-relativistic few- and many-body systems, blurring
the interfaces between condensed matter, atomic and low-energy nuclear physics. While some 
of these techniques have been in use in the area of condensed matter physics for a long time, 
others, such as hybrid Monte Carlo and improved effective actions, have only recently found their 
way across areas. 

With this topical review, we aim to provide a modest overview and a status update on a few 
notable recent developments. For the sake of brevity we focus on zero-temperature, non-relativistic 
problems. After a short introduction, we lay out some general considerations and proceed to 
discuss sampling algorithms, observables, and systematic effects. We show selected results on ground- 
and excited-state properties of fermions in the limit of unitarity. The appendix contains details on group theory on the lattice.
\end{abstract}
\tableofcontents
\title[Lattice methods for many-body systems]
\maketitle

\section{Introduction \label{Sec:Introduction}}

The term `lattice methods' is commonly used to refer to a set of computational physics tools whose unifying idea 
consists in describing the spacetime continuum as a (generally hypercubic) mesh, of finite volume $V$ and 
finite spacing $b$. Quantum fields are then defined on such a spacetime grid and the many-body problem 
thus becomes a problem with a finite number of degrees of freedom. As a matter of convenience, 
the corresponding partition function is typically represented in terms of Euclidean path integrals.

The fundamental reason for employing lattice methods to treat a given problem is that in many cases of interest 
the resulting path integral can be evaluated stochastically using computers, in a fashion that is both efficient 
and in principle exact. Indeed, assuming the problem is such that a proper probability measure can be defined, 
the stochastic calculation of the lattice path integral will suffer only from statistical uncertainties, which can be 
precisely quantified. Furthermore, errors associated with the use of finite volumes $V$ and finite lattice 
spacings $b$, are also controlled and can be systematically removed by using larger $V$, smaller $b$, and/or improved actions. Thus, lattice methods, 
when applicable, surpass perturbative as well as mean-field approaches in the quest for the reliable quantitative 
characterization of quantum many-body systems. This is true in particular if those systems are close to a phase 
transition, or if they are in any way strongly correlated, by which we mean that the low-energy excitations (and thus 
the long-distance properties) of the system are dramatically different from those of a system with the same 
degrees of freedom but no interactions.

On the other hand, lattice methods are not the only available numerical approaches for strongly coupled 
many-body physics. Other potentially exact techniques such as Green's function Monte Carlo 
(GFMC)~\cite{GFMC0,GFMC01,KalosWhitlockBook}, Coupled Cluster (CC)~\cite{CCQC, CCNP} and No-Core Shell Model (NCSM)~\cite{NCSM0} are in 
the mainstream, especially in modern studies of nuclear structure (see 
e.g.~\cite{GFMC1, GFMC2, GFMC3, GFMC4, GFMC5, GFMC6, GFMC7, CC1, CC2, CC3, CC4, NCSM1, NCSM2, NCSM3}).
Still, it should be pointed out that GFMC involves the uncontrolled fixed-node approximation, and while this has not prevented practitioners from obtaining 
remarkably accurate results, it is regarded by many
as intellectually somewhat dissatisfying. Uncontrolled approximations aside, none of these methods scales with the size of the 
problem as favorably as current lattice approaches. In addition, lattice methods have robust technical foundations 
and benefit from a continuous stream of attention from the Lattice QCD community.

Apart from explaining the motivations for using and developing lattice methods, and setting the broad context, 
the purpose of this Introduction is to establish the scope for the rest of the article. We have decided to focus our 
attention on methods for non-relativistic systems, more specifically for ground-state calculations. 
There are several excellent books and reviews on relativistic systems, particularly on those with
gauge degrees of freedom (see e.g. Refs.~\cite{RotheBook, DeGrandDeTarBook, MILC_RMP}).
As expected, there are differences in the treatment of spatial and temporal lattice directions between
relativistic and non-relativistic systems, some of which are worth mentioning at this point. As a result of Lorentz invariance, 
relativistic actions naturally involve first-order derivatives for both space and time, while Galilei-invariant non-relativistic actions have second
derivatives in the spatial directions. The resulting spectral properties are such that temporal lattice
spacings need to be much smaller than the spatial ones for non-relativistic systems, a point we shall return to below. 
The reduced symmetry of non-relativistic systems also leads to different critical properties 
of the lattice theory, which affect their approach to the continuum limit as well as the numerical complications
that arise in the vicinity of critical points.
In addition, there are less obvious and rather serious challenges for relativistic systems, such as the fermion doubling problem
of chiral theories, which are generally absent (or are altogether irrelevant) in non-relativistic systems.

The topic of finite temperature merits a separate article by itself (see e.g. Refs.~\cite{RRdosSantos, BDM_4, BFPReview, Kooninetal}).
We shall also avoid the all-important issues of time-dependent response functions and their analytic continuation to real time, 
which are topics that deserve their own review articles (see e.g.~\cite{JaynesBook, JarrellGubernatis, LINPRO}).

The motivation for the investigation of lattice methods specifically for non-relativistic systems is at least twofold. 
On one hand, the behavior of the vast majority of atomic and condensed-matter systems is governed by electronic 
Hamiltonians in which relativistic effects enter only as a small correction. The symmetry group is thus not given by 
the Poincar\'e set of Lorentz boosts and spacetime translations, but rather by rotations and Galilei transformations, 
or their discrete counterparts when a physical lattice is present.
It should be stressed, however, that the richness of low-energy condensed-matter physics allows for an 
immense array of Hamiltonians whose characterization defies the above simple-minded binary approach.
At this point in time perhaps the paradigm of this situation is given by graphene, whose low-energy dynamics 
is controlled by a (relativistic-like) linear dispersion relation plus a (certainly non-relativistic) Coulomb interaction.
While a few groups are currently using lattice field theory methods to elucidate the effects of strong coupling in 
graphene (see e.g. Refs.~\cite{DL1, DL2, DL3, Handsetal1, Handsetal2, Handsetal3, Broweretal}), we shall not discuss 
this system beyond this point.

The other main motivation comes from low-energy nuclear physics, where the description is predominantly given
by non-relativistic actions, as the mass of the nucleons ($\sim 1$ GeV) is much larger than the mass of the
low-energy excitations ($m^{}_\pi \simeq 0.1$ GeV), which are pseudo-Goldstone bosons associated 
with the spontaneous breaking of an approximate chiral symmetry (see Ref.~\cite{MTNF_Meissner}). Lattice calculations of nuclei are
still in their infancy, but are currently being aggressively pursued by the Bonn-Raleigh 
group~\cite{EpelbaumLO, EpelbaumNLO, EpelbaumNNLO}, 
having performed calculations of the lightest nuclei up to $^{12}$C~\cite{Epelbaum34612_1, Epelbaum34612_2}, 
including a first study of the Hoyle state~\cite{EpelbaumHoyleState, MeissnerFewBody}.
Although they are beyond the scope of this review, the efforts of the NPLQCD collaboration~\cite{NPLQCD1, NPLQCD2}, 
dedicated to ab initio nuclear physics starting from Lattice QCD, should not be overlooked in this context.

Finally, it should be pointed out that there are dynamic synergistic activities among the various areas mentioned above. 
Indeed, with the advent of ultracold atom experiments in the last few years 
(see Ref.~\cite{AtomsReview1, AtomsReview2, ReviewExperiments} for reviews of the current experimental and theoretical situation), 
the nuclear- and atomic-physics communities are working in close collaboration with each other.
This applies in particular to the study of ultracold Fermi gases close to a Feshbach resonance. The latter are effectively 
spin-$1/2$ systems whose interaction can be tuned to vanishing range $r^{}_0$ and large s-wave scattering length $a_S^{}$. This 
seemingly ``unnatural" situation is actually very much relevant for nuclear structure in general and neutron 
matter in particular, where the ranges are short ($r^{}_0 \simeq 2.7$ fm) and the scattering lengths are large 
($a^{}_S \simeq -18$ fm). The stark difference between the nuclear scales ($\sim$ fm) and the atomic ones 
($\sim$\AA) testifies for the property of {\it universality} of the underlying 
dynamics~\cite{Universality1, UniversalityBook, BraatenHammerReview}. 

Universality is particularly evident in the case of the so-called unitary Fermi gas, which is defined as the limit 
$0 \leftarrow k^{}_F r^{}_0 \ll 1 \ll k^{}_F a^{}_S \rightarrow \infty$, where $k^{}_F = (3 \pi^2 n)^{1/3}$ 
is the Fermi wave number and $n$ is the density. By definition, the unitary limit has no intrinsic scales other than its 
density $n$ and temperature $T$. As a consequence, every dimensionful thermodynamic quantity must come as a power 
of $k^{}_F$, or equivalently the Fermi energy $\varepsilon^{}_F = \hbar ^2 k^2_F/(2M)$ (where $M$ is the mass of the fermions) 
times a universal function of $T/\varepsilon^{}_F$ (or equivalently $\beta \mu$, where $\beta^{-1} = k_B^{} T$ and $\mu$ is 
the chemical potential). In particular, one of the most sought-after universal functions corresponds to the energy per particle
(in units of $\varepsilon^{}_F$), which is typically denoted by $\xi$ and often referred to as the Bertsch parameter.

For the same reasons, the unitary Fermi gas displays a number of dynamical symmetries and exact relationships. Indeed, 
as shown in Refs.~\cite{MSW,NishidaSonConformal}, the unitary limit respects a non-relativistic conformal symmetry 
(Sch\"odinger algebra~\cite{Hagen, Niederer:1972zz}). This symmetry is directly related to an exact relation between 
the scaling dimensions of the primary operators of the non-relativistic conformal field theory (which is defined 
in free space) and the spectrum of the system in a harmonic trap~\cite{Mehen,Castin1,Castin2,noteSon}.

Aside from those relations, valid in the limit of infinite scattering length, there are a set of exact universal relations, now 
widely known as the Tan relations~\cite{ShinaContact1,ShinaContact2,ShinaContact3, ZhangLeggett,Werner,BraatenPlatter1,
BraatenPlatter2,SonThompson,TaylorRanderia} (see Ref.~\cite{BraatenReview} for a recent review), that encode the physics 
at short distances and which are due to the short-range nature of the interaction. The latter property allows one to write down 
explicit relations in terms of the probability of finding two particles of unequal spin at short distances from each other; this 
probability is commonly referred to as the ``contact''. At unitarity, the contact constitutes another universal property of quantum
many-body physics. The Tan relations are valid at all temperatures and couplings, and have also been generalized to spatial 
dimensions other than 3, as well as to bosonic and fermionic systems. The precise calculation of the contact in most of these cases 
is still an open problem as of this writing.

In this review we shall be only indirectly concerned with the physics of the unitary limit, mostly as a test case
for the various methods. While this limit is easy to define and implement in few- and many-body calculations, it has 
remained challenging to arrive at precise numerical answers for various quantities such as the Bertsch parameter and the contact.
This is true not only of analytic methods, most of which break down in one way or another when approaching unitarity, but also 
for purely numerical methods. For these reasons, this problem is currently one of the most important benchmarks for few- and 
many-body methods in atomic and nuclear physics.

The rest of the article is organized as follows: in Sec.~\ref{Sec:Generalities} we discuss the general process of formulating the 
many-body problem on a lattice, and give a first glimpse into improvement procedures. In Sec.~\ref{Sec:Algorithms} we present a 
sample of the current algorithms on the market for generating field configurations to be used in the stochastic estimation of the path 
integral. Sec.~\ref{Sec:Observables} demonstrates some of the various observables which may be calculated on a given set of configurations, 
focusing on ground-state observables. Sec.~\ref{Sec:Analysis} then discusses difficulties associated with the extraction of the ground-state, 
and provides some techniques for dealing with these problems. In Sec.~\ref{Sec:Systematics} we consider the issue of systematic errors 
induced by the use of a finite set of lattice points. Finally, we present a small selection of results obtained using the methods outlined in this 
article in Sec.~\ref{Sec:Results}, followed by our Conclusions (Sec.~\ref{Sec:Summary}). An appendix reviewing the concept of group theory 
on the lattice is also provided (App.~\ref{Sec:GroupTheory}).


\section{General considerations \label{Sec:Generalities}}

The main objective of this section is to develop the connection between textbook many-body formalism and
the expressions commonly used in actual numerical calculations. Ultimately, the goal is to develop expressions that 
allow us to evaluate expectation values of quantum operators $\hat O$ in a stochastic fashion, such that, up to 
systematic effects (finite volume, finite lattice spacing, etc.), the exact result is given (schematically) by
\beq
\langle \hat O \rangle = \frac{1}{\mathcal Z} \sum_{n} \mathcal P^{}_n O^{}_n
\eeq
where $\mathcal Z = \sum_{n} \mathcal P^{}_n$ and $\mathcal P^{}_n$ is a positive semi-definite probability measure
whose form depends on the problem under consideration, though generally not on the observable $\hat O$.
Here, $n$ parametrizes the values of a set of auxiliary variables (typically fields that live in spacetime), and it is the sum over
their (usually unfathomably large) domain that is performed using Monte Carlo methods. If one succeeds in doing so,
the result is then given by
\beq
\langle \hat O \rangle \simeq \frac{1}{N_{s}} \sum_{n=1}^{N_{s}} O^{}_n,
\eeq
where $N_{s}$ is the number of samples of the auxiliary variables, distributed according to $\mathcal P^{}_n$, and
the uncertainty in the approximation is of order $1/\sqrt{N_{s}}$ if the samples are uncorrelated and assuming the central limit theorem holds. The quantity $O^{}_n$ is the expectation value of $\hat O$ in the $n$-th auxiliary configuration.

Naturally, limitations arise in the kinds of problems that can be treated in this fashion, as it is not always possible
(or at least not easy) to define a positive semi-definite $\mathcal P^{}_n$. This is of course the well-known sign-problem that
appears in many problems of interest. We shall return to these and related issues below, and restrict ourselves to cases where the sign problem is absent.

Rather than focusing on specific observables, it is often more useful to attempt to re-write the partition function $\mathcal Z$ 
as a sum of positive terms, as mentioned above. As we shall see, in many cases this sets a natural framework for the calculation
of any observable.

\subsection{Starting point: the partition function \& the transfer matrix}

As mentioned in the Introduction, we shall focus on ground-state approaches for non-relativistic systems. 
However, it is convenient to set up the formalism by considering the general case of non-zero temperature $T$.
In the grand-canonical ensemble, the thermodynamic properties of a quantum many-body system are fully specified by
the partition function
\beq
\mathcal Z = e^{-\beta \Omega} = \Tr e^{- \beta (\hat H - \mu \hat N)},
\eeq
where $\Omega$ is the grand thermodynamic potential, the trace is over all multi-particle states, i.e. over the Fock space, 
and $\beta = ({k^{}_B T})^{-1}$. In second-quantization language the Hamiltonian is given, in general, by
\beq
\label{generalH}
\hat H = h^{(1)}_{\alpha \beta} \hat a^{\dagger}_\alpha \hat a^{}_{\beta} 
+ h^{(2)}_{\alpha \beta \delta \gamma} \hat a^{\dagger}_\alpha \hat a^{\dagger}_{\beta} \hat a^{}_\gamma \hat a^{}_{\delta} 
+ h^{(3)}_{\alpha \beta \gamma \nu \mu \delta} 
	\hat a^{\dagger}_\alpha \hat a^{\dagger}_{\beta} \hat a^{\dagger}_\gamma 
	\hat a^{}_{\delta} \hat a^{}_{\mu} \hat a^{}_{\nu} 
+ \cdots , 
\eeq
and the particle number is simply
\beq
\hat N = \hat a^{\dagger}_\alpha \hat a^{}_{\alpha},
\eeq
where sums over repeated indices are implied, and $\hat a^{\dagger}_{\alpha}$ ($\hat a^{}_{\alpha}$) is the 
creation (annihilation) operator of particles in state $\alpha$, for a given choice of single-particle 
states $|\alpha \rangle$.

In Eq.~(\ref{generalH}) we have written a rather general Hamiltonian including not only the conventional one- 
and two-body terms, but also three- and higher-body interactions. We take the point of view that the Hamiltonian 
in question is {\it effective}, in the sense that it describes the physics of the problem below a certain scale $\Lambda$, 
and all the processes above that scale are encoded in the so-called low-energy constants $h^{(j)}(\Lambda)$, and
higher-body interactions~\cite{Lepage:1989hf, Lepage:1997cs}. This is expected to be a practicable approach when 
there is a clear separation of scales in the system. While this is the philosophy of modern effective field theory (EFT), 
with few exceptions most Monte Carlo 
calculations include only two-body interactions. Such a simplified theory can be an accurate description of dilute atomic 
systems or neutron matter in the crust of neutron stars (where the range of the interaction is small compared to the 
inter-particle distance), but fails at higher densities as well as for nuclear structure calculations. Concerning the latter, 
we expect vigorous research in this direction in the coming years, as the application of lattice methods to nuclear physics (with 
nucleons and pions as the elementary degrees of freedom) is currently gathering significant momentum, as mentioned 
in the Introduction. In this review we shall focus on two-body interactions only, but we shall comment on the difficulties of 
implementing higher-body interactions below.

Following the usual route (see e.g. Refs.~\cite{NegeleOrlandBook,ZinnJustinBook}), one may introduce a coherent-state 
representation and go on to the language of fields (with Grassmann fields for fermions and complex fields for bosons), such 
that the partition function of the problem at hand becomes
\beq
\mathcal Z = \int \mathcal D\psi^\dagger \mathcal D\psi \ e^{- S[\psi^\dagger, \psi]}.
\eeq
From this point on we shall focus on the fermionic case, such that $\psi$ denotes in general a multicomponent Grassmann 
field $\psi_\alpha^{}$, where $\alpha = 1, \dots, N^{}_f$, and $N^{}_f$ is the number of fermion flavors present in the problem. 
The Euclidean action is given by
\beq
S[\psi^\dagger, \psi] = \int d\tau d^3x
\left( 
\psi^\dagger_\alpha (\partial_\tau - \mu) \psi_\alpha^{} + \mathcal H[\psi^\dagger, \psi]
\right),
\eeq
and the Hamiltonian density reads
\beq
\mathcal H = \psi^\dagger_\alpha \hat H^{}_0 \psi_\alpha^{} + \dots , 
\eeq
where we have implicitly summed over the flavors $\alpha$, and the interaction terms are in the ellipsis.
Here, $\hat H^{}_0$ is a differential operator corresponding to the coordinate space representation of $h^{(1)}_{\alpha \beta}$ in
Eq.~(\ref{generalH}). The full Hamiltonian is of course given by
\beq
H =  \int d^3x \; \mathcal H[\psi^\dagger, \psi].
\eeq

The fields are defined in all of spacetime, where it should be kept in mind that at finite temperature the time direction is compact,
extending from $\tau = 0$ to $\tau = \beta$. Fermionic (bosonic) fields obey anti-periodic (periodic) boundary conditions at the endpoints 
of the temporal direction. This is the starting point for the formulation of the problem in terms of lattice field theory.

We have thus started from a physical perspective by considering the grand-canonical partition function, and moved on to the language
of fields, which provides the connection with modern methods often seen in Lattice QCD. 
In spite of the (sometimes considerable) technical differences, all of the methods we shall discuss here, and the vast majority of the 
ground-state methods currently in use, actually share a common essential aspect. Indeed, they all rely on the so-called ``power" method 
to extract from a given matrix (here the transfer matrix, defined below) the largest eigenvalue and eigenvector (here the ground state) 
(see e.g.~Ref.~\cite{MatrixComputations}). The idea of starting with the finite-temperature formalism and taking the 
zero-temperature $\beta \to \infty$ limit is indeed nothing other than taking powers of the transfer matrix
\beq
{\mathcal T}^{}_t =  e^{- t \hat H},
\eeq
as a filtering process to systematically project out the excited states, assuming that one starts with a trial state that has a non-zero 
(and ideally very large) projection on the true ground state. In the language of fields this amounts to the determination of the mass gap
(or in general, spectral properties) of the fermion propagator by selecting an optimal source.

We shall return to this idea below. In order to appreciate the connection between the various methods, which typically come from different 
areas, it is important to be able to jump between the ``fields'' and ``transfer-matrix'' languages. We anticipate that for many readers this
connection will seem trivial. However, it is our experience that this is not the norm.

\subsection{Putting the problem on the lattice}

The next step is to discretize spacetime into $N^3_{x} \times N^{}_\tau$ points, and define fields on such a lattice. Throughout this review we shall consider a hypercubic lattice with lattice spacings $b_s, b_{\tau}\equiv \frac{b_s^2}{M}$ corresponding to the spatial and temporal directions, respectively, as that is by far the most common 
choice. We denote with $\psi^{}_{{\bf n},\tau}$ the 
Grassmann field defined on all the space-time points ${\bf n}, \tau$, and $\psi^{\dagger}_{{\bf n},\tau}$ denotes its hermitian conjugate.
The lattice Hamiltonian is then
\beq
H = \sum_{{\bf n},{\bf m}} \psi^\dagger_{{\bf n}} \left [ H^{}_0 \right]_{{\bf n},{\bf m}} \psi^{}_{{\bf m}} + \dots,
\eeq
where ${\bf n},{\bf m}$ are points on the spatial lattice, and $\left [ H_0 \right]_{{\bf n},{\bf m}}$ is a particular 
choice of a discrete representation for the continuum Hamiltonian $\hat H_0^{}$. 
We have again omitted the interaction terms, which we turn to below.

In the absence of interactions, the dispersion relation for a non-relativistic system is of course $E = p^2/(2M)$, where $M$ is the 
fermion mass. The choice of a discretized derivative, to represent the momentum operator in coordinate space, determines the form 
of the Hamiltonian of the non-interacting system on the lattice. The various choices differ in their approach to the continuum limit, i.e. 
in their behavior at high energies. For the Laplacian operator, the simplest choice is given by the second difference formula
\beq
\label{DD1}
\Delta f_{\bf j} \equiv \sum_{k=1,2,3} \frac{1}{b_s^2} \left [ f^{}_{{\bf j} + {\bf e}_{k}} + f^{}_{{\bf j} - {\bf e}_{k}} - 2 f^{}_{\bf j} \right ],
\eeq
where ${\bf e}_{k}$ is the unit vector in the $k$-th direction. This yields the correct behavior at low energies, as it gives
the following kinetic-energy spectrum
\beq
E({\bf p}) = \frac{2\hbar}{M b_s^2} \sum_{k=1,2,3} \sin^2 \left (\frac{q^{}_k}{2}\right),
\eeq
where ${\bf p} = \hbar {\bf q}/b^{}_s$, and $q^{}_k = \frac{2\pi n^{}_k}{N^{}_x}$ takes on discrete values as $n^{}_k = -N^{}_x/2, \dots, N^{}_x/2$.
Clearly, the above reproduces $E({\bf p}) \simeq \frac{{\bf p}^2}{2M}$ for ${\bf q}^2 \ll 1$.

One may reproduce the correct behavior at {\it all} energies by using a 
definition in momentum space:
\beq
\label{DD2}
\Delta f^{}_{\bf j} \equiv \sum_{\bf n} \frac{e^{i  2\pi {\bf j}\cdot{\bf n} / N_x^{} }}{N_x^{3/2}}
\left(\frac{2\pi {\bf n}}{b_s^{}N_x^{}}\right)^2
{\tilde f}_{\bf n}^{},
\eeq
where
\beq
{\tilde f}_{\bf n}^{} \equiv 
\sum_{{\bf k}}
\frac{e^{-i 2\pi {\bf n}\cdot{\bf k} / N_x^{}}}{ N_x^{3/2} }
f_{\bf k}^{}
\eeq
is the discrete Fourier transform of $f$. This yields
\beq
E({\bf p}) = \frac{{\bf p}^2}{2M},
\eeq
by definition, at all energies.

The form of Eq.~(\ref{DD1}) has the advantage of being as local as possible, in the sense that the calculation of
the Laplacian at point $\mathbf j$ requires input only from its nearest-neighbors ${\mathbf j} \pm {\mathbf e}_{k}$. This means that 
it can be performed very quickly on a computer. It should be kept in mind, however, that the memory layout in
current computers is linear, such that nearest-neighbors in different directions can be physically very far from each other. This fact can affect
performance considerably.

The form of Eq.~(\ref{DD2}), on the other hand, is as non-local as possible, but has the benefit of being closer to the continuum 
limit than any other implementation. References~\cite{Privitera1,Privitera2} have analyzed the properties of these choices in the 
presence of interactions using a dynamical mean-field approach (see also Ref.~\cite{WernerCastinArXiv})

This is a first glimpse at what is commonly referred to as ``improvement''. The action resulting from Eq.~(\ref{DD2}) is an 
improved version of that of Eq.~(\ref{DD1}), in the sense that it is closer to the continuum limit. In fact, Eq.~(\ref{DD2}) would 
lead to what we call a ``perfect'' action for the non-interacting system. We shall return to the issue of improved actions 
in Sec.~\ref{Improvements}. Other forms of improved dispersion relations, which in some sense interpolate between
cases of Eq.~(\ref{DD1}) and Eq.~(\ref{DD2}), have been described in Ref.~\cite{Lee2}.

The issue of choosing a proper discretization for the temporal derivative $\partial_\tau$ requires some care.
Indeed, as noted in Ref.~\cite{BSS}, one should be careful when transferring fermion
operators to the lattice. In particular, the generator of time translations, at finite temporal lattice spacing $b^{}_\tau$, will not be the 
discretized version of the Hamiltonian shown above; rather, we have
\beq
\psi^{}_\tau - \psi^{}_{\tau - 1} =  - (1 - B_\tau^{}) \psi^{}_{\tau - 1},
\eeq
where 
\beq
B_\tau^{} \equiv \exp \left( - b_\tau H(\tau) \right),
\eeq
for a general time-dependent Hamiltonian $H$. Therefore, the generator of translations is an effective Hamiltonian given by
\beq
b^{}_\tau H^{\mathrm{eff}}(\tau) = 1 - e^{-b^{}_\tau H(\tau)}.
\eeq

Taking this into account, the fermionic part of the Euclidean-time lattice action, in the absence of interactions, will be
\beq
S =\sum_{a,b} \psi^{\dagger}_{a} \left [ K_0 \right]_{ab} \psi^{}_{b}
\eeq
(we use collective indices $a,b$ to denote points on the spacetime lattice), where
\bea
\label{hugematrK0}
K_0 = 
\left( \begin{array}{ccccccc}
1 & 0 & 0 & 0 & \dots & B_{0}^{} \\
-B_{0}^{} & 1 & 0 & 0 & \dots & 0 \\
0 & -B_{0}^{} & 1 & 0 &  \dots & 0 \\
\vdots & \vdots & \vdots & \ddots & \vdots & \vdots \\
0 & 0 & \dots & -B_0^{} & 1 & 0 \\
0 & 0 & \dots & 0 & \hspace{-.4cm} -B_0^{} & 1
\end{array} \right), \quad 
\eea
and where the entries are blocks of size $N^3_{x} \times N^3_{x}$, with $1$ being the unit matrix and $B_{0}^{} = \exp({-b^{}_\tau (H_0 - \mu)})$.
Notice that this discretization amounts to a one-sided temporal derivative, which should be contrasted with the relativistic case,
where a symmetric discretization is used.

Thus far, aside from the fact that we have only considered fermionic degrees of freedom in the problem, the formulation we have 
described is potentially general enough to cover a large number of problems in condensed-matter and nuclear physics.
Unfortunately, once interactions are taken into account many of those problems are not easily tractable with current methods. Special techniques are needed to even 
begin to consider certain cases (e.g. systems with repulsive interactions) due to the appearance of the sign or 
signal-to-noise problem. We shall therefore restrict our attention even further whenever necessary, but we will return to 
signal-to-noise issues below, as it is an active topic of research in all areas.

In order to proceed, we shall focus on two-body interactions, such that in general,
\bea
\label{Eq:FullHamiltonian}
H = \sum_{{\bf n},{\bf m}} \psi^\dagger_{{\bf n}} \left [ H^{}_0 \right]_{{\bf n},{\bf m}} \psi^{}_{{\bf m}} + 
\sum_{{\bf n},{\bf m},{\bf j},{\bf k}} \psi^\dagger_{{\bf n}} \psi^\dagger_{{\bf m}} 
\left [ V^{}_2 \right]_{{\bf n}{\bf m}{\bf j}{\bf k}} \psi^{}_{{\bf j}} \psi^{}_{{\bf k}}
\eea
%
%
%
In fact, this problem is already challenging. Indeed, while taking the exponential of a one-body operator is relatively easy,
it is much harder to do so for a two-body operator. Similarly, the only field integral we know how to handle is the Gaussian case. It is for these reasons that auxiliary fields are introduced in order to deal with the interaction.
Unfortunately, even for the simplest interactions there are cases where a sign problem arises, such as the case of
the point-like repulsive potential. For a purely attractive potential, on the other hand, we may ``decouple" the two-body
interaction by using the Hubbard-Stratonovich ``trick", which we now turn to.

\subsection{Hubbard-Stratonovich transformations}

Almost every lattice calculation is based on the fact that two-body interactions (from the Hubbard model to QCD) can be represented by mediation through bosonic fields. In QED this field 
is the physical photon field, in QCD it is the gluon field, and in theories with contact interactions (Hubbard model, Nambu-Jona Lasinio, 
Gross-Neveu, etc.) one uses a non-propagating ``auxiliary field''. This auxiliary field is introduced using the so-called Hubbard-Stratonovich (HS) transformation 
(in any of its many incarnations) in which the four-fermion operator of the interaction is effectively replaced by a path integral
over a field $\phi$ that couples to a two-fermion operator~\cite{HS1, HS2, Hirsch}.  Although the HS transformation is an extremely useful step toward 
putting the problem on a computer, it should be pointed out that there is an alternative that is particularly powerful for the 
case of contact interactions, namely the worm algorithm (see e.g. Refs.~\cite{Worm0,Worm1,Worm2}).

To fix ideas we shall consider the case of spin-$1/2$ fermions with a contact interaction, such that the second
term in Eq.~(\ref{Eq:FullHamiltonian}) is given by
\beq
V = g \sum_{{\bf n}} \psi^\dagger_{{\bf n},\uparrow} \psi^{}_{{\bf n},\uparrow} \psi^\dagger_{{\bf n},\downarrow} \psi^{}_{{\bf n},\downarrow} \; .
\eeq
As mentioned above, we will focus on time-independent potentials, such that we may evaluate the above expression at any time slice $\tau$.
The HS transformation is based on the fact that, at each point $a = {\bf n}, \tau$ in spacetime, we may write
\beq
\!\!\!\!\!\!\!\!\!\! e^{b^{}_\tau g \psi^\dagger_{a,\uparrow} \psi^{}_{a,\uparrow} \psi^\dagger_{a,\downarrow} \psi^{}_{a,\downarrow} } 
= \frac{1}{\sqrt{2 \pi}} \int_{-\infty}^{\infty}\!\!\!\!\! d\phi \;
e^{- \frac{\phi^2}{2} -\phi \sqrt{b^{}_\tau g}(\psi^\dagger_{a,\uparrow} \psi^{}_{a,\uparrow} + \psi^\dagger_{a,\downarrow} \psi^{}_{a,\downarrow})}.
\eeq
A discrete version is also available:
\beq
\!\!\!\!\!\!\!\!\!\! e^{b^{}_\tau g \psi^\dagger_{a,\uparrow} \psi^{}_{a,\uparrow} \psi^\dagger_{a,\downarrow} \psi^{}_{a,\downarrow} } 
= \frac{1}{2} \sum_{\phi = \pm 1} 
e^{-\phi \sqrt{b^{}_\tau g} (\psi^\dagger_{a,\uparrow} \psi^{}_{a,\uparrow} + \psi^\dagger_{a,\downarrow} \psi^{}_{a,\downarrow})},
\eeq
and a continuous but compact form:
\bea
\!\!\!\!\!\!\!\!\!\! e^{b^{}_\tau g \psi^\dagger_{a,\uparrow} \psi^{}_{a,\uparrow} \psi^\dagger_{a,\downarrow} \psi^{}_{a,\downarrow} } 
 = 
\frac{1}{2\pi}  \int_{-\pi}^{\pi}\!\!\!\!\! d\phi \;
e^{-\sin (\phi) \sqrt{b^{}_\tau g} (\psi^\dagger_{a,\uparrow} \psi^{}_{a,\uparrow} +  \psi^\dagger_{a,\downarrow} \psi^{}_{a,\downarrow})}.
\eea

Implementing an HS transformation, the partition function becomes
\beq
\mathcal Z = \int \mathcal D \phi \mathcal D \psi^\dagger \mathcal D \psi \; \rho[\phi] e^{S^{}_{} [\phi,\psi^{\dagger},\psi]},
\eeq
where the function $\rho[\phi]$ represents a contribution whose form depends on the specific choice of HS transformation. For these three 
possibilities (by no means an exhaustive list), we have
\[
\!\!\!\!\!\!\!\!\!\!\!\!\!\!\!\!\!\!\!\!\!\!\!\!\!\!\!\!\!\!\!\!\!\!\!\!\!   \rho[\phi] = \left\{ 
  \begin{array}{l l}
    \prod_{\bf n} e^{-\frac{1}{2}\phi^{2}_{\bf n} } & 
\mbox{Continuous, unbounded (Refs.~\cite{HS1, HS2})}\\ \\
    \prod_{\bf n} (\theta(-\pi + \phi^{}_{\bf n})\theta(\pi - \phi^{}_{\bf n}))& 
\mbox{Continuous, bounded (Refs.~\cite{Lee1,Lee2})}\\ \\
    \prod_{\bf n} (\delta^{}_{\phi^{}_{\bf n},1} + \delta^{}_{\phi^{}_{\bf n},-1}) & 
\mbox{Discrete (Ref.~\cite{Hirsch}).}\\
  \end{array} \right.
\]
The fermionic part of the Euclidean-time lattice action will be
\beq
S =\sum_{a,b} \psi^{\dagger}_{a} \left [ K[\phi] \right]_{ab} \psi^{}_{b}, 
\eeq
and the dynamics of the fermions is fully specified by the 
matrix $K$, which has a general structure given by
\bea
\label{hugematrK}
K[\phi] = 
\left( \begin{array}{ccccccc}
1 & 0 & 0 & 0 & \dots & B_{N_\tau^{}}^{}[\phi]\\
-B_{1}^{}[\phi] & 1 & 0 & 0 & \dots & 0 \\
0 & -B_{2}^{}[\phi] & 1 & 0 &  \dots & 0 \\
\vdots & \vdots & \vdots & \ddots & \vdots & \vdots \\
0 & 0 & \dots & -B_{N_\tau^{}\!-2}^{}[\phi] & 1 & 0 \\
0 & 0 & \dots & 0 & \hspace{-.4cm} -B_{N_\tau^{}\!-1}^{}[\phi] & 1
\end{array} \right), \quad 
\eea
where the entries are blocks of size $N^3_{x} \times N^3_{x}$. Notice that the structure of this matrix is such that each 
block column (and row) corresponds to a particular temporal slice. It should be clear from the above that the derivative
with respect to imaginary time that appears in the action is represented here as a one-sided difference. 
Also clear in this equation is the anti-periodic boundary condition along the time direction.
In fact, for ground-state studies, the upper-right block $B_{N_\tau^{}}^{}$ is in principle at our disposal, as it merely 
implements the boundary condition in the time direction, which is infinitely long at zero temperature. 
For instance, Refs.~\cite{EKLN3,LEKN1,NEKL1} use open boundary conditions, i.e. $B_{N_\tau^{}}^{} \equiv 0$; we shall
elaborate on this choice in the Algorithms section. 

The form of the $B_{k}^{}$ matrices, which in general are functions of $\phi$, depends on the form of the interaction 
as well as on the choice of HS transformation. We shall see a specific example below.


Since the resulting action is quadratic in the fermion fields, we are now in a position to perform the path integration over the fermion fields,
which leads to
\beq
\mathcal Z = \int \mathcal D \phi \; \mathcal P[\phi],
\eeq
where 
\beq
\mathcal P[\phi] \equiv \rho[\phi] \det K[\phi]
\eeq
is the probability of the specific configuration $\phi$. Here and in the following we shall neglect overall multiplicative
constants in partition functions whenever possible (e.g. when integrating out fermions or performing gaussian integrals of any kind).
(Extensive explanations on gaussian integration over various fields and dimensions appear in Ref.~\cite{ZinnJustinBook})

This formulation, which results from the grand-canonical approach to quantum statistical mechanics, is the 
most natural one in connection with finite-temperature studies (where the length of the time direction is 
proportional to the inverse temperature) as well as with its relativistic counterpart.

\subsection{The canonical approach}

As particle number is conserved in non-relativistic quantum mechanics, it is often natural to start from the 
canonical ensemble, which leads to a formulation based on a Slater determinant trial wavefunction for $N$ 
particles and leads to a ``partition function" (technically this is not a partition function as we are not truly summing 
over all possible quantum states, but it is a kind of probability sum) of the form
\beq
\label{eq:Zcanonical}
\mathcal Z^{}_{N} = \int \mathcal D \phi \; \mathcal P^{}_N[\phi],
\eeq
where
\beq
\mathcal P^{}_N[\phi] \equiv \rho[\phi] \det K^{}_{N}[\phi],
\eeq
\beq
[K^{}_{N}]^{}_{\alpha\beta} = \langle \alpha |B_{N_\tau^{}}^{} \cdots B_{3}^{} B_{2}^{} B_{1}^{} | \beta \rangle,
\eeq
and $| \alpha \rangle$, $\alpha = 1,...,N$ are the single-particle states in the Slater determinant. One may
define at most $N_x^{3}$ linearly independent single-particle states, such that $N = N_x^{3}$ is the maximum
number of particles we may place on the lattice, as befits fermions. In general, however, one takes 
$N \ll N_x^{3}$ to avoid lattice spacing effects (we shall return to systematic effects in general in 
Sec.~\ref{Sec:Systematics}). 

Ground-state properties are obtained by extrapolation from finite $\beta$
(i.e. finite $N_\tau^{}$) to the $\beta \to \infty$ limit by fitting to analytic formulas. The performance of calculations 
and subsequent extrapolations in this approach depends strongly on the choice of the single-particle basis and 
trial wavefunctions. For example, Ref.~\cite{EKLN4} uses functions that include some degree of pairing correlations, 
as does Ref.~\cite{PhysRevA.84.061602}, while Ref.~\cite{JED} uses simple Slater determinants of free-particle states.

These two formulations (the canonical and the grand canonical of the previous sections) are of course related 
in more than one way, starting from the fact that the matrices $B_{k}^{}$ that enter the imaginary time propagation 
have exactly the same form given the choice
of dispersion relation and HS transformation. In the Algorithms section we shall elaborate on the differences 
and similarities between these two approaches.

As a particular example, if we choose the discrete form of the HS transformation, then
\beq
\label{BkExample}
[B_{k}^{}]^{}_{{\bf n},{\bf n'}} = [\mathcal K \mathcal V^{}_k]^{}_{{\bf n},{\bf n'}},
\eeq
where the kinetic-energy factor is diagonal in momentum space and given by
\beq
\label{Kop}
[\mathcal K]^{}_{{\bf p},{\bf p}'} = e^{-b^{}_\tau (p^2/(2M) - \mu)}\delta^{}_{{\bf p},{\bf p}'}
\eeq
and the potential-energy factor is diagonal in coordinate space and reads
\beq
\label{Vop}
[\mathcal V^{}_k]^{}_{{\bf n},{\bf n'}} = (1 + \sqrt{C} \phi(k,{\bf n}))\delta^{}_{{\bf n}, {\bf n'}}.
\eeq
where $C = e^{b^{}_\tau g} - 1$. (Notice that if $g$ is negative, which corresponds to a repulsive interaction, then $C$ 
is negative as well, and its square root is complex. This brings about the sign problem, as mentioned before.)

As the path integral is to be performed over $\phi(k,{\bf n})$, the actual form of the operator $\mathcal V^{}_k$ will 
obviously vary wildly throughout a calculation. The operator $\mathcal T$, on the other hand, is constant.
Therefore, in practice, $\mathcal T$ is computed only once at the beginning of a calculation and stored in memory.
Furthermore, it is best to compute it in momentum space where it is diagonal (it is a fully dense matrix in coordinate 
space) and apply it as needed by using fast Fourier transforms (FFT). 

The strategy of using FFT to apply operators that are non-local in coordinate space but diagonal in momentum space,
is commonly referred to as ``Fourier acceleration" (see e.g.~\cite{Batrouni,Daviesetal1,Daviesetal2,Katzetal1}). 
To the uninitiated this may appear to be a ridiculously convoluted 
way to proceed. Why not do everything in coordinate space instead of switching back and forth? The reason is that applying 
a dense matrix of dimension $M\times M$ to a vector is an operation whose computational cost scales as $M^2$, whereas 
using FFT for this purpose changes the scaling to $M \log M$. Moreover, modern FFT libraries, such as FFTW~\cite{FFTW05} are
so advanced that their performance is very difficult to surpass, even in cases where the matrix in question is somewhat sparse in its 
coordinate-space representation (e.g. when using a nearest-neighbor finite-difference formula for the momentum operator).


\subsection{A comment on many-body forces\label{ManyBodyForces}}

As mentioned above, in this article we restrict ourselves to two-body forces. Higher-body forces, however,
play a crucial role in nuclear physics and should therefore be somehow accounted for. In Ref.~\cite{Epelbaum34612_1} 
a strategy is used to include interactions beyond two-body in a perturbative fashion.

If the perturbative approach is justified, i.e. if the couplings $g_{n}$ (assume $n > 2$) to the interaction operators
$\hat H_n$ are indeed much smaller for the higher-body forces in question than for the two-body sector, then one may 
expand the action around vanishing $g_{n}$, such that the partition function becomes
\beq
\!\!\!\!\!\!\!\!\!\!\!\!\!\!\!\!\!\!\!\!\!\!\!\!\!
\mathcal Z = \left . \mathcal Z\right |_{g_n = 0, n>2}  + \sum_{n>2} g_{n} \left. \frac{\partial \mathcal Z}{\partial g_n} \right |_{g_n = 0} 
+ \frac{1}{2}\sum_{n,m>2} g_{n} g_{m} \left. \frac{\partial^2 \mathcal Z}{\partial g_n \partial g_m} \right |_{g_n,g_m = 0} 
+ \dots.
\eeq
In practice this is performed by introducing sources coupled to one-fermion operators.
The evaluation of observables will therefore include propagators reflecting the insertion of the interaction.
A different way to proceed is to attempt to perform an HS transformation on the many-body interaction in question.
Unfortunately this introduces a sign problem in almost any case of interest.


\subsection{Improved Actions \& Transfer Matrices\label{Improvements}}

At this point we have presented the various steps
necessary to connect the formalism of finite-temperature many-body quantum mechanics with the expressions
amenable to numerical calculation. Before moving on to specific algorithms to sample the probability measure, 
and then to the problem of computing observables, it is worth pointing out a recent development which modifies 
the zero-range Hamiltonian presented above in order to allow a faster approach the continuum limit. This represents 
a kind of improved action. Improved observables can also be defined using the same methods, as we shall see 
in a later section.

The starting point is the transfer matrix. By definition,
\beq
\mathcal T = \exp(-b^{}_\tau \hat H) \simeq \int d\phi(k) \hat{\mathcal K} \hat {\mathcal V}^{}_k[\phi] + O(b^{2}_\tau),
\eeq
where $\mathcal K$ and $\mathcal V^{}_k$ are as in Eqs.~(\ref{Kop}) and (\ref{Vop}), respectively, and
we have set $\mu = 0$.
In order to reduce lattice-spacing effects in non-relativistic systems Refs.~\cite{EKLN1,EKLN3} employed the following 
strategy. One may promote the constant $C$ in Eq.~(\ref{Vop}) to an operator $C(\hat p)$ that is diagonal in momentum space, such that
a more general HS transformation is defined. In order to fix the form of the function $C(p)$, one 
parametrizes its form using a convenient set of basis functions $\mathcal{O}_{2n}({\bf p})$, 
\beq
C(\mathbf{p}) = \frac{4\pi}{M} \sum_{n=0}^{N_{\mathcal{O}}-1}C_{2n} \mathcal{O}_{2n} (\mathbf{p}) \, ,
\eeq
and then diagonalizes the transfer matrix in the two-particle subspace of the Fock space,
which is given in the momentum representation by
\bea
\label{2particleTmatrix} 
\!\!\!\!\!\!\!\!\!\!\!\!\!\!\!\!\!\!\!\!\!\!\!\!\!\!\!\!\!\!\!\!\!\! \!\!\!\!\!\!\!\!
\mathcal T_2^{}({\bm p}_\uparrow^{} {\bm p}_\downarrow^{};{\bm q}_\uparrow^{}{\bm q}_\downarrow^{}) &=& \\ \nonumber
&&  \!\!\!\!\!\!\!\! \!\!\!\!\!\!\!\! e^{-\frac{b^{}_\tau T(p)}{2}}\! \left [
\delta_{{\bm p}_\uparrow^{} {\bm q}_\uparrow^{}}\delta_{{\bm p}_\downarrow^{} {\bm q}_\downarrow^{}}
+ 
\frac{\sqrt{C({\bm p}_\uparrow^{}\!-\! {\bm q}_\uparrow^{})} 
\sqrt{C({\bm p}_\downarrow^{}\!-\! {\bm q}_\downarrow^{})}}
{2V}
\delta_{{\bm p}_\uparrow^{} + {\bm p}_\downarrow^{}, {\bm q}_\uparrow^{} + {\bm q}_\downarrow^{}}\right]
e^{-\frac{b^{}_\tau T(q)}{2}},
\eea
where $V = N^3_x$ is the lattice volume and \mbox{$T(k) = ({\bm k}_\uparrow^{2} + {\bm k}_\downarrow^{2})/(2M)$}.

The exact eigenvalues of the transfer matrix corresponding to a given choice of scattering parameters is given by
L\"uscher's formula~\cite{Luescher1,Luescher2}:
\beq
\label{Luscherformula}
p \cot \delta^{}_0(p) = -\frac{1}{a^{}_S}  + \frac{1}{2} r^{}_0 p^2 + O(p^4) = \frac{1}{\pi L}\mathcal S^{}(\eta)
\eeq
where $a$ is the scattering length, $r$ is the effective range, $\eta \equiv \frac{pL}{2\pi}$, and 
\beq
\label{Seta}
\mathcal S^{}(\eta) \equiv \lim_{\Lambda \to \infty}\sum_{\bf n}^{} 
\frac{\Theta(\Lambda^2 - {\bf n}^2)}{{\bf n}^2 - \eta^2} - 4 \pi \Lambda,
\eeq
where the sum is over all 3D integer vectors.

\begin{figure}[t]
\includegraphics[bb=0 0 1280 960,width=\columnwidth]{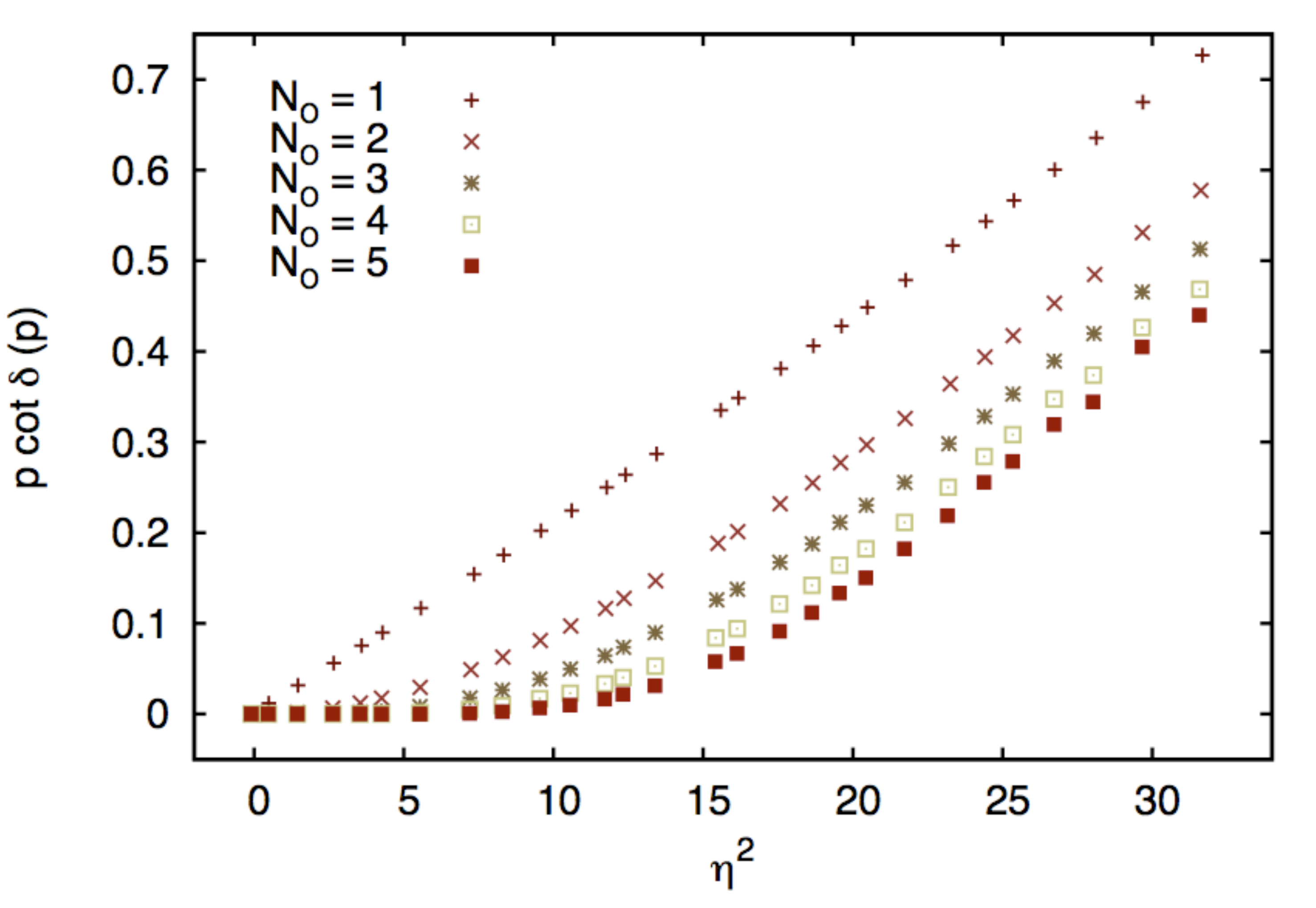}
\caption{\label{Fig:pcotdGI}(Color online) Plot of $p\cot \delta (p)$ as a function of 
$\eta^2 = {E^{}_2/p_0^2}$ (where $p_0^{} \equiv 2\pi/L$), for $N_x = 20$, $b^{}_\tau = 0.05$, 
and levels of improvement $N^{}_\mathcal{O} = 1 - 5$, for a Galilean invariant definition of the transfer matrix.}
\end{figure}

In the case of the unitary limit, where by definition 
\beq
p \cot \delta^{}_0(p) \equiv 0, 
\eeq
the above procedure yields
considerable improvement in the approach to the continuum. Indeed, with a single coefficient $C^{}_0$ 
one may tune the scattering length $a$, but the 
effective range $r$ remains finite due to lattice-spacing artifacts (systematic effects will be discussed in more detail in 
Sec.~\ref{Sec:Systematics}). Naturally, by introducing more parameters one may tune the effective-range expansion in 
Eq.~(\ref{Luscherformula}) to higher accuracy (see Fig.~\ref{Fig:pcotdGI}), thus systematically reducing finite lattice-spacing
effects. Extrapolations to the dilute limit, however, are still needed, as we shall see in the Results section.

Thus, tuning to the unitary point requires knowledge of the roots of $S(\eta)$. The first few roots are quoted in Table 3.
The result of the fit is shown in Table~\ref{Table:Cn}, for various improvement levels.
\begin{table}[t]
\begin{center}
\caption{\label{Table:Cn}
Results of fitting the coefficients $C^{}_{n}$
to the low-energy spectrum of the two-body problem at resonance, in a box of side $N^{}_x = 16$,
for an imaginary time step $b^{}_\tau = 0.05$, in lattice units, from Ref.~\cite{JED}.
}
\begin{tabularx}{\columnwidth}{@{\extracolsep{\fill}}c c c c c c}
       \hline
       $N^{}_\mathcal{O}$ & $C^{}_0$ & $C^{}_1$ & $C^{}_2$ & $C^{}_3$ & $C^{}_4$\\
       \hline
       \hline
1   &  0.68419    &  --   &  --    &  --   &  --  \\
2   &  0.53153    & 0.07896     & --    & --    & --     \\
3   &  0.49278    & 0.04366    & 0.01807    & --    & --     \\
4   &  0.47217     & 0.03711   & 0.00784    & 0.00467     & --     \\
5   &  0.45853     & 0.03331   &  0.00718     & 0.00132    & 0.00129    \\
\hline
\end{tabularx}
\end{center}
\end{table}


\section{Algorithms \label{Sec:Algorithms}}

In the previous section we outlined the general strategy that is typically followed to write down a 
two-body Hamiltonian in a form that can be treated with Monte Carlo methods. Our main development 
was expressing the partition function as a sum over auxiliary-field configurations:
\beq
\mathcal Z = \int \mathcal D \phi \; \mathcal P[\phi],
\eeq
where $\mathcal P[\phi]$ generally involves the determinant of a large matrix whenever fermions are involved.

Once a suitable probability measure $\mathcal P[\phi]$ has been chosen, one faces the task of constructing an efficient 
method to sample uncorrelated field configurations that obey such a probability. Here, the importance of 
the word ``efficient" should not be underestimated. Indeed, the possibility of evaluating a path integral stochastically 
is a remarkable step forward given the size of the domain of integration, but stochastic integration of quantum systems
would not be viable without methods for fast sampling of $\mathcal P[\phi]$. This consideration is quite general, applying to 
relativistic as well as non-relativistic problems. Indeed, none of the endeavors of the Lattice QCD community would be possible 
without fast sampling methods. In this section we describe some of these methods, specifically for the case of 
non-relativistic systems. The problem at hand is equivalent to that of constructing an efficient random number 
generator, where the output is not just one number but a field defined in space-time. 

Among the methods available to tackle this problem, one may identify two large classes: the so-called heat-bath 
methods, which allow for the direct generation of uncorrelated samples; and Markov-chain methods, based on the 
Metropolis algorithm~\cite{Metropolis}. While the former are extremely useful when studying for instance pure gauge theories \cite{Creutz,CabibboMarinari,Bazavovetal,Johnson}, 
their use is much more limited in the case of theories that involve dynamical fermions due to the complexity (non-
linearity and non-locality) of the fermion determinant present in $\mathcal P[\phi]$. 
As we shall see, however, heat-bath approaches are still useful for non-relativistic fermion theories in 
certain formulations. Markov-chain approaches, on the other hand, are the most popular and generally viable 
when dealing with fermions (as long as there is no sign problem), and we shall therefore discuss 
them first.


\subsection{Determinantal Monte Carlo}

One the simplest methods to sample a non-trivial fermionic probability is known by the name of Determinantal 
Monte Carlo (DMC). In this algorithm, one generates a sequence of samples of the auxiliary 
field $\phi$ (the Markov chain) by performing localized changes, typically of random magnitude affecting a 
random set of locations.  The new configuration, $\phi^{}_{\mbox{new}}$, is then accepted or rejected according to the Metropolis 
algorithm, i.e. with a probability of acceptance given by
\beq
p = \mbox{min}\left(1, \frac{\mathcal P [\phi^{}_{\mbox{new}} ]}{\mathcal P [\phi_{\mbox{old}} ]} \right).
\eeq 
where $\mathcal P[\phi] = \det K[\phi]$.

As successive configurations are strongly correlated with each other, snapshots of the field should 
be taken at sufficiently long intervals along the chain. Such a process of decorrelation can be time consuming, 
depending on the proximity to critical points \cite{LandauBinder}. The minimal requirements for this algorithm to function and 
yield the correct probability measure are ergodicity and detailed balance (though technically a weaker condition 
is enough, see e.g. Ref.~\cite{Toussaint} for a nice overview). Ergodicity implies that long enough chains will eventually 
touch every point in the space of possible configurations. Detailed balance, i.e. ensuring that a given update has the same 
probability as the reverse process, implies that the Markov chain has $\mathcal P[\phi]$ as a fixed point.

When moving from one configuration to the next, it is possible in some cases to update clusters of points 
at a time, as explored many years ago in the context of the Ising model \cite{SwendsenWang, BarbuZhu, Wolff}. 
However, such attempts to take long random steps in configuration space in our case will in general, though 
not always, result in a low probability of acceptance. One is therefore forced to take relatively small steps. On 
the other hand, taking very small steps should be avoided as well, as they yield high acceptance but 
slow decorrelation and are therefore very inefficient. The rule of thumb is then to compromise by choosing 
updates such that the acceptance rate is in the range 40-60\%.

The DMC method is not only among the simplest, but is also one of the oldest and remains in wide use
(see e.g. \cite{BDM_1, BFPReview} for a recent use of DMC); it is also very effective, but unfortunately not particularly efficient. 
The main drawback of the ``plain" version of DMC presented here is that localized updates of the field represent a huge cost from the 
computational point of view. Indeed, strongly correlated systems are generally not amenable to global 
modifications of the field, and the (integrated) size of the randomly updated regions generally does not scale up 
with the volume of the system. As a consequence, the cost of a full update (a sweep), which is a measure of
the cost of obtaining a useful configuration, scales as $N N^{2}_{\tau} V^3$, where $N$ is the number 
of particles, $N^{}_{\tau}$ is the number of lattice points in the imaginary time direction, and $V = N^{3}_{x}$ is 
the spatial volume. This estimate results from the evolution of $N$ particles in imaginary time over $N^{}_{\tau}$ 
steps, with a cost per step equal to that of dense matrix-vector multiplication, which on a basis of size $V$ scales 
as $V^2$. The cost of local updates contributes an extra factor of $N^{}_{\tau} V$, because the auxiliary field lives 
in all of spacetime, which yields the final result. This estimate does not account for the possibility of Fourier acceleration, 
in which case the cost of a time step is proportional to $V \log V$, resulting in a cost per sweep of $N N^{2}_{\tau} V^2 \log V$.


\subsection{Hybrid Monte Carlo}

As of this writing, the use of Hybrid Monte Carlo (HMC) in non-relativistic physics is surprisingly scarce, both
at zero as well as finite temperature. While this powerful algorithm was developed many years ago in the field of 
Lattice QCD~\cite{HMC1,HMC2}, its application to strongly correlated systems in other fields, in particular condensed matter and low-energy nuclear physics is still in its infancy (with some exceptions, see Ref.~\cite{Lee2,DL1}).

The HMC method, as DMC, is based on the generation of a Markov sequence of configurations that are accepted 
or rejected as dictated by the Metropolis algorithm. The fundamental difference between DMC and HMC (in its 
various forms) is that the latter enables {\it global} updates of the field, which reduces the cost per sweep to 
$N N^{}_{\tau} V^2$, or $N N^{}_{\tau} V \log V$ when Fourier acceleration is possible. This dramatic improvement is accomplished by introducing molecular dynamics (MD) into the updating procedure in the following way.

Given the probability measure $\mathcal P[\phi]$ and corresponding ``effective'' action 
\beq
S^{}_\mathrm{eff}[\phi] \equiv -\log \mathcal P[\phi], 
\eeq
one introduces a fictitious gaussian-distributed momentum field $\pi$, which modifies the partition function 
only by a multiplicative constant, i.e. in a way that is immaterial to the dynamics of the system:
\beq
\label{Eq:Zphipi}
\mathcal Z = \int \mathcal D \phi \mathcal P[\phi] \to \int \mathcal D \phi \mathcal D \pi \mathcal P[\phi,\pi],
\eeq
where
\beq
\mathcal P[\phi,\pi] = \exp{\left(-\sum_{{\bf n},\tau} \frac{\pi^2_{{\bf n},\tau}}{2} \right)} \mathcal P[\phi] = 
\exp{\left (-\mathcal H_{\mbox{MD}} \right)},
\eeq
and
\beq
\mathcal H^{}_{\mbox{MD}} \equiv \sum_{{\bf n},\tau} \frac{\pi^2_{{\bf n},\tau}}{2} + S^{}_\mathrm{eff}[\phi].
\eeq
Since we have not modified the true dynamics of the problem, this new probability $\mathcal P[\phi,\pi] $ 
is physically equivalent to the original one. The reason it is advantageous to work with this enlarged problem is
that one may use MD as the updating strategy to explore configuration space. 
Taking a gaussian distributed $\pi$ and any choice of $\phi$ as the initial conditions, one solves the 
classical evolution equations dictated by $\mathcal H^{}_{\mbox{MD}}$:
\bea
\label{MDEoM}
\dot \phi^{}_{{\bf n},\tau} &=& \pi^{}_{{\bf n},\tau}, \nonumber \\
\dot \pi^{}_{{\bf n},\tau}  &=& F^{}_{{\bf n},\tau}[\phi] \equiv -\frac{\delta S^{}_\mathrm{eff}[\phi]}{\delta \phi^{}_{{\bf n},\tau}},
\eea
where the dots indicate a derivative with respect to the fictitious MD time $t^{}_{\mbox{MD}}$.
Notice that these equations are global, i.e. they affect each and every point in space-time, such that upon 
integration along a trajectory of length $T^{}_{\mbox{MD}} \sim 1$ one obtains a fully updated field configuration,
i.e. one has performed a sweep. If these equations were integrated 
exactly, the fictitious energy given by $\mathcal H_{\mbox{MD}}$ would be conserved, and the probability of
acceptance of any trajectory would be exactly 1. It is this combination of MD
with the Metropolis algorithm that gives the ``Hybrid" in HMC. In practice, integrators are precise but not exact, 
such that a Metropolis step, comparing the probabilities at the beginning and at the end of the trajectory, is 
needed to guarantee that the correct probability distribution is being sampled.
Resampling of the fictitious momentum $\pi$ at the beginning of each trajectory is needed to actually
perform the integral over $\pi$ that appears in Eq.~(\ref{Eq:Zphipi}). This also takes the random walk in a 
new direction and into a different MD classical orbit, which favors faster decorrelation.

In order to maintain detailed balance, as required by the Metropolis algorithm, the MD equations (\ref{MDEoM}) 
must be integrated in a reversible and area-preserving fashion (see e.g. Refs.~\cite{HMC1,HMC2,Toussaint}). 
This is accomplished by implementing a symplectic and symmetric integrator, such as the leap-frog algorithm, or 
more sophisticated approaches like the Omelyan integrators~\cite{Omelyan}.

It should be stressed that, in contrast to the DMC algorithm, the current form of HMC applies only to the case in 
which the field $\phi$ is continuous. This is a prerequisite if classical differential equations are to be used for 
the MD evolution, and it is somewhat unfortunate, as there are a variety of formulations that 
rely on discrete fields, which have the obvious advantage of requiring significantly less storage space than 
their continuous counterparts.

\subsection{Pseudofermions and connection to the finite temperature case}

One of the most common versions of the HMC algorithm, used routinely in Lattice QCD, utilizes the 
so-called pseudofermion fields $\chi$ to represent the determinant via the following relation:
\beq
\det M[\phi] \propto \int \mathcal D \chi^\dagger \mathcal D \chi \exp \left(- \chi^\dagger M^{-1}_{}[\phi] \chi \right),
\eeq
where we assume that all the indices are contracted in the exponent on the right-hand side,
and it is assumed that the dynamics encoded in $M$ is such that the gaussian integral is well defined.
In practice, this integral is evaluated stochastically, exploiting its gaussian form. 
The name ``pseudofermion" arises from the fact that these are bosonic complex-valued fields whose 
dynamics is given by $M^{-1}_{}$, which satisfies antiperiodic boundary conditions in the time direction.
To our knowledge, pseudofermions have never been used in non-relativistic ground-state calculations, 
possibly because in that case the determinants are substantially smaller than in the finite-temperature 
and/or relativistic cases, such that an exact calculation of $\det M$ is feasible and even convenient in
comparison.

In this case the MD force takes the form
\beq
F^{}_{{\bf n},\tau}[\phi] =  
\left[\chi^\dagger M^{-1}_{}[\phi]\right] 
\frac{\delta M^{}_{}[\phi]}{\delta \phi^{}_{{\bf n},\tau}} 
\left[M^{-1}_{}[\phi]\chi \right] ,
\eeq
where again all the indices are implicitly contracted. It is easy to see that a fermion sign problem will ensue
unless $M$ is positive semidefinite. The latter happens, in particular, in cases with an even number of flavors,
where (under certain circumstances, e.g. spin symmetry) we may define a matrix $K$ such that 
$\det M = \det K^\dagger K$.

A less common version of the HMC algorithm combines the advantages of both the DMC and HMC algorithms,
without using pseudofermions, resulting in the so-called Determinantal Hybrid Monte Carlo algorithm. The idea behind this approach is to make use of the fact that
\beq
\label{Eq:Mand1pU}
\det M = \det(1 + \mathcal U[\phi])^{N_f^{}},
\eeq
where
\beq
\mathcal U[\phi] = B_{N_\tau^{}}^{} \cdots B_{3}^{} B_{2}^{} B_{1}^{}
\eeq
(see Sec.~\ref{Sec:Generalities}).

Notice that the determinant on the left-hand side of Eq.~(\ref{Eq:Mand1pU}) is over a spacetime matrix,
whereas the one on the right-hand side is purely spatial. The fact that this simplification is possible is
directly related to the absence of backward temporal propagation in non-relativistic theories.

The zero-temperature version of this discussion is directly connected to the use of a Slater determinant 
$| \alpha_0 \rangle$ as the starting point or ``guess" for the ground-state wavefunction, which we mentioned
in Section~\ref{Sec:Generalities}. In that case,
\beq
\mathcal Z =  \int \mathcal D \phi \det(\mathcal U[\phi])^{N_f^{}},
\eeq
where the determinant is to be taken over the space generated by the single-particle (s.p.) orbitals $\varphi_j$ that make 
up $| \alpha_0 \rangle$, such that
\beq
F^{}_{{\bf n},\tau}[\phi] =  N_f^{} \tr \left[
{\mathcal U[\phi]}^{-1}
\frac{\mathcal U[\phi]}{\delta \phi^{}_{{\bf n},\tau}}\right]
\eeq
where, as with the determinant, both the trace and the inverse are to be interpreted as restricted to the
subspace generated by the s.p. orbitals $\varphi_j$.


\subsection{Heat-bath approach}

An interesting recent development in the field of lattice fermion calculations concerns exploiting the fact that
the non-relativistic ground-state problem can be formulated in a way that the fermion determinant is identically
equal to 1, for all the field configurations~\cite{EKLN3,LEKN1,NEKL1,EKLN4,EKLN1,EKLN2}. Indeed, as mentioned 
in the Introduction, ground-state calculations 
give us the freedom to choose the form of the boundary condition in the time direction, and in particular we 
may choose an open boundary condition, such that the dynamics is governed by the triangular matrix
\bea
K[\phi] = 
\left( \begin{array}{ccccccc}
1 & 0 & 0 & 0 & \dots & 0 \\
-B_{1}^{} & 1 & 0 & 0 & \dots & 0 \\
0 & -B_{2}^{} & 1 & 0 &  \dots & 0 \\
\vdots & \vdots & \vdots & \ddots & \vdots & \vdots \\
0 & 0 & \dots & -B_{N_\tau^{}\!-2}^{} & 1 & 0 \\
0 & 0 & \dots & 0 & \hspace{-.4cm} -B_{N_\tau^{}\!-1}^{} & 1
\end{array} \right). \quad \label{hugematrlowtriangle}
\eea
Not only is $\det K = 1$, but the triangular form also gives us direct access to the inverse:
\bea
K^{-1}_{}[\phi] = 
\left( \begin{array}{ccccccc}
1 & 0 & 0 & 0 & \dots & 0 \\
B_{1}^{} & 1 & 0 & 0 & \dots & 0 \\
B_{2}^{} B_{1}^{} & B_{2}^{} & 1 & 0 &  \dots & 0 \\
\vdots & \vdots & \vdots & \ddots & \vdots & \vdots \\
\prod^{N_\tau^{}\!-1}_{k=1}B_{k}^{} & \dots & \dots & \dots & B_{N_\tau^{}\!-1}^{} & 1
\end{array} \right). \quad \label{hugematrlowtriangleinv}
\eea
This constitutes a case in which, trivially, one can sample uncorrelated field configurations directly according to 
the relevant probability measure, and it thus falls into the category of heat-bath approaches.
This is a dramatic simplification of the original problem that comes with advantages and disadvantages. 

On the positive side, a constant
probability measure means that the cost of generating configurations is essentially insensitive to the number of 
particles and the size of the lattice, depending only on the speed of the uniform random number generator
of choice.
Field generation in this approach is therefore instantaneous compared to the other methods. Furthermore, a single set of configurations and propagators may be generated and then used for any number of observables desired. 
In addition, a constant probability implies no sign problem, regardless of spin asymmetry, nature of the 
interaction (attractive/repulsive, two-body, three-body, etc.), etc. This approach therefore constitutes an intriguing
possibility for ground-state studies of a variety of systems in condensed-matter and nuclear physics, where 
the sign problem has been a major roadblock to progress. 

On the negative side, this formulation is restricted to
zero temperature, where one may legitimately ignore the (anti-periodic) boundary condition in the time direction. 
Furthermore, a constant probability implies that importance sampling is not possible: all the configurations
are in principle equally important. In practice, the sign problem reappears here in the form of a statistical ``overlap problem": the random configurations may not have much weight for the observable 
of interest. Using a very large number of configurations ($\sim 10^8$) and a sophisticated statistical analysis 
(see Sec.~\ref{Sec:Analysis}) it is still possible to perform calculations and obtain results with unprecedented precision.

Perhaps one of the main advantages of this method is not that it ``solves" the sign problem, but rather that 
it replaces it with a new problem that we have a better chance to tackle, in particular if aided by physical insight 
into the system at hand.

\subsection{Open-ended imaginary-time evolution, re-weighting and branching random walks}


Extracting ground-state properties entails performing calculations for $\beta$ as large as possible. We will
return to the issue of extrapolating to the $\beta \to \infty$ limit from finite $\beta$ data below. It is important to underline,
however, that serious technical difficulties arise at large $\beta$. As explained most recently in Ref.~\cite{EKLN1,EKLN2},
and as we review below, an interval of exponential approach to the asymptotic ground-state properties
(even for simple observables such as the energy) is typically followed by a dramatic deterioration in the signal-to-noise ratio,
which may render the whole extrapolation effort dubious, or at least quite complicated.

The algorithms explained in the previous subsections generally involve fixing the spatial and temporal extents of the
lattice, followed by sampling configurations of the auxiliary field defined on that lattice. Notice that the heat-bath algorithm,
however, does not require a temporal direction of fixed extent, as the probability measure does not depend on the
size of the lattice in any way. 

There is a different set of algorithms that have been in use since at least the mid-1990's, which 
allow for an open-ended temporal direction, such that in practice it has been possible to reach temporal extents a factor 
of at least 2 or 3 larger than with the methods outlined above. Clearly, following this strategy is at odds with the common
practice of saving auxiliary field configurations of fixed size. These are, however, powerful algorithms that have largely 
remained unknown in various communities, and so we wish to outline them here.

The starting point is, as above, the factorization and Hubbard-Stratonovich transformation of the transfer matrix:
\beq
\label{T_HS1}
\mathcal T_t = \int \mathcal D \phi_t \ \mathcal T[ \phi_t],
\eeq
where typically we have two (or more) flavors, such that,
\beq
\label{T_HS2}
\mathcal T[\phi_t] = \mathcal T^{}_{\uparrow}[ \phi_t] \mathcal T^{}_{\downarrow}[ \phi_t].
\eeq

This is valid for each temporal slice, and we may therefore take a ``guess'' wavefunction $|\Psi^{}_0 \rangle$ and apply 
a sequence of $\mathcal T$'s for a given spacetime field configuration $\phi$. The problem remains, of course, of summing 
over all possible $\phi$ fields in order to recover the physics we are interested in. Should we choose to pick configurations 
purely randomly, we would obtain the heat-bath algorithm of the previous section. As mentioned before, this may lead to substantial 
statistical noise. 

The idea we wish to outline here, which first appeared in Refs.~\cite{PhysRevLett.74.3652, PhysRevB.55.7464},
is essentially what in certain areas is called ``re-weighting'', but with a crucial twist. Instead of using a flat probability measure, 
one may assign a weight to the auxiliary field configurations; the crucial point is in introducing this weight by defining a probability 
in which the temporal dependence factorizes completely. This property of factorization is trivially true for the heat-bath method, 
but is certainly not satisfied for the DMC, DHMC or HMC approaches, where all of spacetime is inextricably linked into the fermion 
determinant. In this fashion, one may perform an open-ended random walk in imaginary time.

The first step is to consider not a single Slater-determinant wavefunction but a sum of them, all with equal probability, making up
the initial guess for the ground state $|\Psi^{}_0 \rangle$:
\beq
|\Psi^{}_0 \rangle = \sum_{k=1}^{N^{}_w} |\varphi^{0}_k \rangle.
\eeq
%

The imaginary-time evolution from step $t$ to $t+1$ is then carried out on each Slater determinant
by applying the HS-transformed transfer matrix:
\beq
|\varphi^{t+1}_k \rangle = \int \mathcal D \phi^{}_{t+1} {\mathcal T}[\phi^{}_{t+1}] |\varphi^{t}_k \rangle.
\eeq
In principle, in the above path integral, all the field configurations at that time slice, $\phi^{}_{t+1}$, are equally important. 
To improve the efficiency of the temporal evolution we introduce the 
re-weighting strategy mentioned above.

For each evolved Slater determinant state (what is often called a ``walker'') $|\varphi^{t}_k \rangle$ we define an importance function
\beq
O^{}_T(\varphi^{t}_k) = \langle \Psi^{}_T |\varphi^{t}_k \rangle.
\eeq
With this weight we can now write the imaginary-time evolution as
\beq
|\psi^{t+1}_k \rangle = \int \mathcal D \phi^{}_{t+1} \mathcal P_k^{}[\phi^{}_{t+1}] {\mathcal T}[\phi^{}_{t+1}] |\psi^{t}_k \rangle.
\eeq
where we have defined a modified walker
\beq
|\psi^{t}_k \rangle = O^{}_T(\varphi^{t}_k) |\varphi^{t}_k \rangle
\eeq
and a probability
\beq
\mathcal P_k^{}[\phi^{}_{t+1}] = \frac{O^{}_T(\varphi^{t+1})}{O^{}_T(\varphi^{t})},
\eeq
which depends on the initial and final states $|\varphi^{(t)} \rangle$ and $|\varphi^{(t+1)} \rangle$, respectively.

At this stage, the algorithm appears to be reduced to sampling the auxiliary field $\phi^{}_{t+1}$ with the 
probability $\mathcal P_k^{}[\phi^{}_{t+1}]$, and thus construct the evolution operator for the modified walkers.
It can be shown, as in the cases we discussed before, that in the presence of an equal number of fermions for each flavor
(and for an even number of flavors) this probability is positive semidefinite, and this holds in particular for a 
particle-projected BCS state~\cite{PhysRevA.84.061602}. At the end of the problem the unmodified walkers are recovered 
simply by dividing by the relevant $O^{}_T$. 

There is however a problem in that $\mathcal P_k^{}$ is not normalized, and typically importance sampling will introduce a 
normalization automatically (which is another way of saying that a normalized probability is required). We are therefore
forced to consider the expression
\beq
|\psi^{t+1}_k \rangle = N_k^{t+1}\int \mathcal D \phi^{}_{t+1} \bar{\mathcal P}_k^{}[\phi^{}_{t+1}]
{\mathcal T}[\phi^{}_{t+1}] |\psi^{t}_k \rangle,
\eeq
where
\beq
\bar{\mathcal P}_k^{}[\phi^{}_{t}] \equiv \frac{{\mathcal P}_k^{}[\phi^{}_{t}]}{N_k^{t}} 
\eeq
and
\beq
N_k^{t} = \int \mathcal D \phi^{}_{t} \mathcal P_k^{}[\phi^{}_{t}]
\eeq
is of course the normalization of $\mathcal P_k^{}[\phi^{}_{t}]$.

The evolution in imaginary time then proceeds by obtaining $N^{}_s$ samples of $\phi^{}_{t+1}$ 
that obey the normalized probability measure $\bar{\mathcal P}_k^{}$, and using the stochastic estimate of 
the path integral as the evolution operator:
\beq
\int \mathcal D \phi^{}_{t+1} \bar{\mathcal P}_k^{}[\phi^{}_{t+1}]{\mathcal T}[\phi^{}_{t+1}]
\simeq 
\frac{1}{N^{}_s}\sum_{q = 1}^{N^{}_s} {\mathcal T}[\phi^{q}_{t+1}].
\eeq

The normalization constants $N_k^{t}$ corresponding to each walker are separately accumulated in the so-called 
``weight'' variables according to
\beq
w_k^{t+1} = N_k^{t+1} w_k^{t},
\eeq
with $w_k^{0} \equiv 1$.

The calculation of the normalization is in general not trivial. However, as explained in Refs.~\cite{PhysRevLett.74.3652, PhysRevB.55.7464},
this can be overcome by updating one lattice site at a time.

After a period of equilibration, i.e. for large enough $t$, one can expect to obtain an estimate of the ground state, which is given by
\beq
|\Psi^{}_0 \rangle \simeq \sum_k w_k^{t} |\varphi^{t}_k \rangle,
\eeq
up to an overall normalization factor.


In practice, the weights of certain walkers will quickly grow after a few time steps, while others will decrease to zero.
It is for this reason that the concept of ``branching'' is introduced. The idea is simple: every few steps in imaginary time,
consider the possibility of randomly keeping or discarding some of the walkers. The number of steps (or equivalently the
size of the time step) should be adjusted in such a way that roughly $40-60\%$ of the walkers are discarded at each instance.
Of the walkers that remain, some of them are used to generate two or more new walkers, adjusting the number with 
a given algorithm, in such a way that the total number of walkers remains approximately constant throughout the calculation.
At that point it is also important to re-orthonormalize the whole set of walkers to avoid stability issues. By allowing for the possibility
of branching, one is essentially re-starting the calculation with a many-body wavefunction that is closer to the ground state than
the original guess. In this way it is possible to evolve in imaginary time for much longer than with other algorithms.


	
\section{Observables\label{Sec:Observables}}

\subsection{Calculating the energy}

The simplest observable to compute using lattice methods is the ground-state energy of a system. One method 
for doing so is to construct the $N$-body correlation function,
\beq
\label{eq:correlator}
\calC_N(\beta) = \frac{1}{\mathcal Z}\int \mathcal{D} \phi \mathcal{D} \psi^{\dagger} \mathcal{D} \psi 
e^{-S[\psi^{\dagger}, \psi,\phi]}  
\Psi^{(A)}_{N}(\beta) \Psi^{\dagger (B)}_{N}(0) \, ,
\eeq
where $\mathcal Z$ is the partition function, and the source (sink) is given by
\beq
\Psi^{(a)}_{n}(\tau) = \int dx_1 \cdots dx_n A^{(a)}(x_1 \cdots x_n) \psi_1(x_1)\cdots \psi_n(x_n) \, ,
\eeq
where $A^{(a)}$ is an $N$-body wavefunction and the fields $\psi_i$ are chosen with the appropriate quantum 
numbers such that $\Psi^{(a)}_{n}$ has nonzero overlap with the state of interest. The simplest form for 
$\Psi^{(a)}_{n}$ is a product of non-interacting single-particle states, however, more sophisticated choices, for 
example including pairing correlations, may be incorporated into the wavefunction $A^{(a)}$ to have superior overlap 
with the ground state. This will be discussed in more detail in Sec.~\ref{Sec:Analysis}. 

Upon integrating out the $\psi$ fields we have
\begin{eqnarray}
\calC_N(\beta) &=& \frac{1}{\mathcal Z} \int \mathcal{D} \phi \mathcal{P}(\phi) \tilde{\calC}(\phi,\beta) \, ,
\end{eqnarray}
where
\begin{eqnarray}
\tilde{\calC}(\phi,\beta) &=& f(S_1(\phi,\beta), S_2(\phi,\beta), \cdots) \,
\end{eqnarray}
where $f$ is a function of single- or multi-particle propagators, depending on the form of $\Psi$. This function should be properly (anti)symmetrized for (fermionic) bosonic fields. Using the simplest example of sources and sinks composed of single-particle states $|\alpha\rangle \, , |\beta \rangle$, $f$ becomes a Slater determinant of single-particle propagators, so that $\tilde{\calC} = \det S_{\alpha, \beta}$, where
\begin{eqnarray}
\label{eq:propagator}
S_{\alpha,\beta} &=& \langle \alpha | B_{N_{\tau}} \cdots B_3 B_2 B_1 | \beta \rangle \, ,
\end{eqnarray}
and we reproduce the canonical partition function, Eq.~\ref{eq:Zcanonical}. The form of the propagators in Eq.~\ref{eq:propagator} may be recognized as a product of transfer matrices, $B_{\tau} = e^{-b_{\tau} \mathcal{H}}$, projected onto the chosen initial and final states. 

One may insert a complete set of energy eigenstates into the correlation function, Eq.~\ref{eq:correlator}, to show that for a given Euclidean time $\beta$,
\begin{eqnarray}
\label{eq:corrSpectrum}
\calC_N(\beta) &=& \sum_{m,n}\langle \Psi^{(A)}_{N} |m \rangle \langle m | \prod_{\tau=1}^{\beta}e^{-b_{\tau} \mathcal{H}}|n \rangle \langle n | \Psi^{(B)}_{N} \rangle \cr
&=& {Z}_0 e^{-\beta E_0} + {Z}_1 e^{-\beta E_1} + \cdots\, ,
\end{eqnarray}
where ${Z}_a \equiv \langle \Psi_{src} | a \rangle \langle a | \Psi_{snk}  \rangle $ is the overlap of the $a$-th energy eigenstate of the system with the chosen sources/sinks and $E_a$ is the energy associated with that state. To extract the ground state, it is often useful to construct what is commonly called the effective mass,
\beq
\label{eq:eff_mass}
m_\mathrm{eff}(\beta) =  \frac{1}{\Delta \beta} \ln \frac{\calC(\beta)}{\calC(\beta+\Delta \beta)}\underset{\beta \to \infty}{\longrightarrow} E_0 + \frac{{Z}_1}{{Z}_0} e^{- (E_1-E_0) \beta} + \cdots 
\eeq
For sufficiently large $\beta \gg E_1 - E_0$, the ground state energy will dominate as excited state contributions are exponentially suppressed. By plotting this quantity as a function of $\beta$, one looks for a plateau to indicate the elimination of excited states. The plateau region may then be fit to extract the ground state energy as well as a statistical uncertainty using standard methods.

One may also choose to use the thermodynamic definition of the energy, 
$E(\beta) \equiv -\frac{\partial \log {\mathcal Z}(\beta)}{\partial \beta}$. 
For the large-$\beta$ limit one obtains an expression similar to that derived for the correlator, 
\beq
\label{eq:Ethermo}
E(\beta) \underset{\beta \to \infty}{\longrightarrow} E_0 + b_E e^{-(E_1 - E_0) \beta},
\eeq
where
\beq
b_E^{} = \frac{{Z}_1}{{Z}_0}(E_1 - E_0).
\eeq
Taking for example the canonical partition function obtained using a Slater determinant state (Eq.~(\ref{eq:Zcanonical})), the observable calculated on each field configuration will be
\beq
\mathcal{O}_N(\beta) = E_N(\beta) = -\mbox{Tr} \left[ K_N^{-1}(\beta) \frac{\partial K_N(\beta)}{\partial \beta} \right] \, .
\eeq
In Sec.~\ref{Sec:ImprovedObservables} techniques for improving observables will be discussed, using this form of the energy as a first example.


As seen in Eqs.~(\ref{eq:eff_mass}),(\ref{eq:Ethermo}), information about excited state energies is also available from these methods. Fortunately, there exists a vast literature on sophisticated techniques developed by the lattice QCD community which one may borrow to tackle the problem of extracting excited state energies 
\cite{Luscher:1990ck,Michael198558,Blossier:2009kd,Fodor:2012gf,Edwards:2011jj,Dudek:2010wm}. Recently, the low-lying excited states of $^{12}C$ have been calculated using non-relativistic lattice methods, including a signature for the Hoyle state \cite{EpelbaumHoyleState}. 


\subsection{Scattering parameters}

\begin{figure}[h]
\includegraphics[width=0.65\columnwidth]{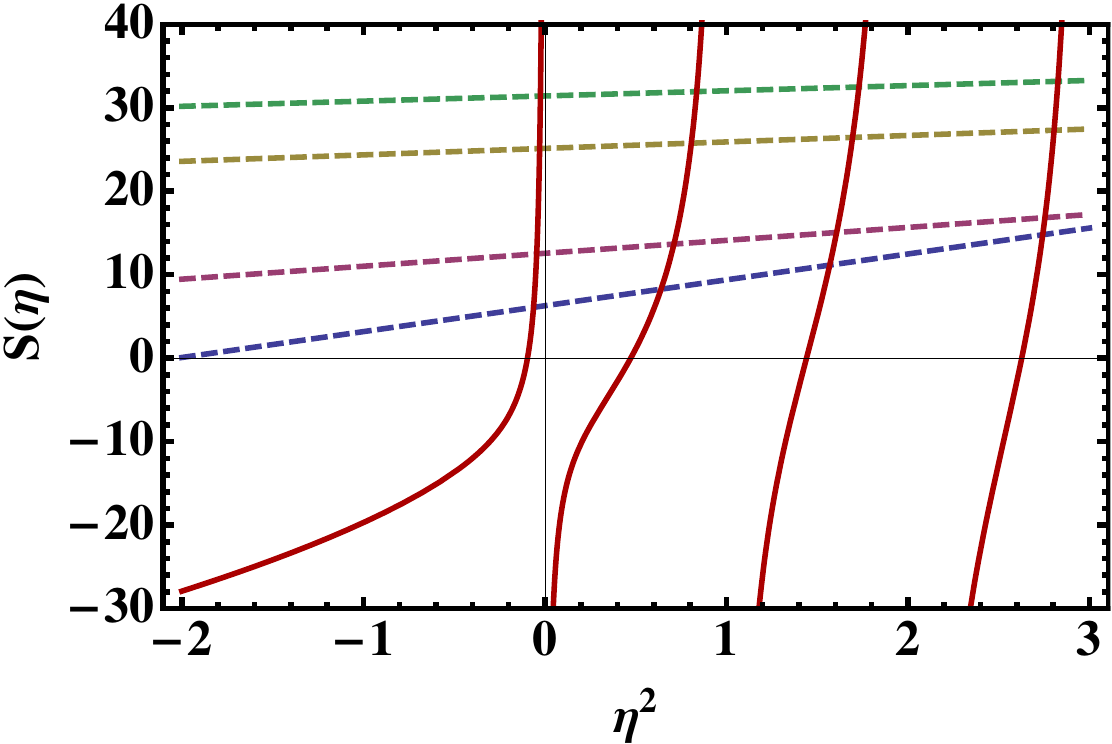}
\caption{
\label{fig:splot}%
$S(\eta)$ (solid red) and $\pi L p \cot \delta$ (dashed) vs. $\eta^2 \equiv \left(\frac{p L}{2\pi}\right)^2$. The $\pi L p \cot \delta$ correspond to the scattering parameters $r_0/a=-0.1$, with the following volumes: $L/|a|=2$ (blue), $L/|a|=4$ (magenta), $L/|a|=8$ (gold), $L/|a|=10$ (green).  The energy eigenstates for the corresponding volumes are given by the intercepts of the curves.
}
\end{figure}

Once the energies of a system have been calculated for a given set of parameters there are many quantities of interest which may then be derived. Scattering parameters form one class of quantities which can be calculated directly from the energy via the L\"uscher formula, Eq.~(\ref{Luscherformula}). According to the Maiani-Testa theorem \cite{Maiani:1990ca,Lin:2001ek}, Euclidean Green's functions cannot give information about infinite volume Minkowski scattering matrix elements except at kinematic thresholds. The L\"uscher formula gives an indirect method for calculating these parameters by relating the volume dependence of multi-particle energies to the corresponding infinite volume scattering phase shift. As shown in Fig.~\ref{fig:splot}, the energy eigenvalues for two particles in a periodic box are given by the intercepts of the corresponding $\pi L p \cot \delta$ with the function $\mathcal{S}(\eta)$, defined in Eq.~(\ref{Seta}).

In \cite{Beane:2003da} it was emphasized that the L\"uscher formula is valid for any value of the scattering length, regardless of the size of the box, so that moderately-sized lattices may be used even for scattering processes with unnaturally large scattering lengths. The authors of this work derived approximate expressions giving the explicit volume dependence of the energies of two-body, s-wave scattering states in the limits of small and large scattering length, $a/L$. In the former limit, the lowest-lying energy is given by
\beq
\label{eq:LuscherSmalla}
E_0 = \frac{4 \pi a}{ML^3}\left[ 1-c_1 \frac{a}{L} + c_2\left(\frac{a}{L}\right)^2 + \cdots \right] \, ,
\eeq
while for the latter one may use
\beq
E_0 = \frac{4\pi^2}{M L^2} [d_1 + d_2 L p \cot \delta_0 + \cdots ] \, ,
\eeq
where $c_1,c_2,d_1,d_2$ are numerical constants determined in \cite{Beane:2003da}, along with analogous expressions for bound states and first excited states. However, in addition to the universal power law dependence of these energies on the size of the box, there exist non-universal exponential corrections due to finite range effects \cite{Sato:2007ms,Bedaque:2006yi}, setting a limit on how small the lattice may be for a given scattering process at the desired accuracy. 

Generalizations of Eq.~(\ref{eq:LuscherSmalla}) to many-body systems have been accomplished using perturbation theory for systems of bosonic particles \cite{Beane:2007qr,Detmold:2008gh,Smigielski:2008pa}, and successfully applied in lattice QCD calculations of multi-meson states \cite{Beane:2007es,Detmold:2008fn,Detmold:2008yn,Detmold:2011kw,Detmold:2012wc}. The L\"uscher formula may also be generalized in many other contexts, for example, relations have been derived for asymmetric boxes \cite{Li:2003jn,Feng:2004ua}, higher partial waves \cite{Luu:2011ep,Konig:2011nz,Koenig:2011ti}, moving frames \cite{Rummukainen:1995vs,Kim:2005gf}, moving bound states \cite{Bour:2011ef,Davoudi:2011md}, and multi-channel processes \cite{Briceno:2012yi,Hansen:2012tf}.

All of these versions of L\"uscher's formula begin with the assumption that the particles exist within a periodic box, a common situation for lattice calculations. In some cases it may be advantageous to consider a different form for the IR cutoff. For example, an adaptation known as the spherical wall method can be useful for determining phase shifts at higher orbital angular momenta, as well as spin-orbit coupling and partial-wave mixing \cite{Borasoy:2007vy}. Here, a hard spherical boundary is imposed in position space, and the eigenvalue problem then becomes one which is well-known from standard nuclear theory texts: the Schr\"odinger equation is solved given the condition that the wavefunctions must vanish at this spherical boundary. The hard cutoff effectively removes the complications induced by multiple-scattering effects arising from the periodicity of the lattice, however, additional complications arise due to the lack of spherical symmetry on the lattice. This method has proven useful for setting unknown operator coefficients in lattice chiral EFT methods \cite{EpelbaumNLO,EpelbaumNNLO,Epelbaum34612_1,Epelbaum34612_2}.

Another possibility is to use the eigenvalue solution for two particles in a harmonic potential \cite{Busch:1998,Yip:2008,Suzuki:2009}. Due to progress in the development of nuclear EFT approaches using an oscillator basis \cite{Haxton:2007hx,Haxton:2002kb,Stetcu:2006ey} and for systems confined in a harmonic potential \cite{2007PhRvL..98j3202L,2007PhRvA..76f3613S,2010JPhG...37f4033S}, it has been suggested (\cite{Kolck2009,Luu:2010hw}) that the spectrum of complex nuclear systems in a harmonic potential may be useful in extracting scattering information using the relation 
\beq
\mathbf{p}^{2l+1}\cot \delta_l(\mathbf{p}) = (-1)^{l+1} (2m \omega)^{l+1/2} \frac{\Gamma \left(\frac{2l+3}{4}-\frac{\epsilon}{2}\right)}{\Gamma \left(\frac{1-2l}{4}-\frac{\epsilon}{2}\right)} \, ,
\eeq
where $\epsilon = E/\omega$.  This technique has been demonstrated for two fermions at unitarity \cite{2010AnPhy.325.1644S}, and the two nucleon system \cite{Luu:2010hw} in the continuum.  

It was suggested in \cite{NEKL1} that the use of a harmonic potential in lattice calculations may offer some advantages, such as a reduction in noise and fast convergence to the ground state due to the confinement provided by the potential. On the other hand, the potential introduces new physical scales which may induce systematic errors that must be accounted for. 
For example, because the harmonic potential is non-zero everywhere (except for the origin) there will be modifications to the short 
range behavior. These finite range effects will not be exponentially suppressed as they are using the standard L\"uscher 
relation, 
so that an extrapolation to free space ($\omega \to 0$) may be required. Systematic errors associated with harmonic potentials will be discussed in Sec.~\ref{Sec:Systematics}. 

\subsection{Improved Observables\label{Sec:ImprovedObservables}}
\subsubsection{Energy.}

The work of Refs.~\cite{EKLN1,EKLN3} tackles the problem of reducing UV lattice effects 
by defining an improved the transfer matrix, using a generalized HS transformation and a low-momentum expansion.
As we saw at the end of Sec.~\ref{Sec:Generalities}, one may then tune the coefficients $C_{n}^{}$ such that 
the two-particle energy spectrum matches the one required by L\"uscher's formula for the lowest $N^{}_{\mathcal O}$ 
eigenvalues, given the desired values of the scattering parameters.

The above procedure represents a significant step forward in mitigating lattice-spacing effects 
in MC calculations, especially considering that it requires only a small coding investment for its 
implementation in extant MC codes, and it results in minimal computational overhead.

One may use the same technique to reduce lattice-spacing effects from observables.
This is particularly useful in connection with improving finite temperature 
lattice calculations, such as those of Refs.~\cite{BDM_1, BDM_4, DLT, DLWM}. Indeed, in those calculations,
as well as in similar ground-state approaches, the transfer matrix is not the only object carrying 
lattice-spacing effects: the operators used to compute expectation values also suffer from the same problems. 

The main idea behind the method remains the same as in Ref.~\cite{EKLN1}. L\"uscher's formula provides 
us with the exact two-particle spectrum, such that the eigenvalues of the exact two-particle transfer matrix and 
its $b^{}_\tau$ derivative(s) are known:
\beq
\label{dTdtauexact}
-\frac{\partial \mathcal T^\mathrm{exact}_2}{\partial b^{}_\tau} = E_2^{} \exp\left(- b^{}_\tau E_2^{} \right)
\eeq
where $E_2^{}$ are the exact two-particle energies in a continuous box~\footnote{
We use $\mathcal T^{}_{2}$ here for didactical purposes, but one may use the symmetric expression 
$\mathcal T^{\dagger}_{2}\mathcal T^{}_{2}$ instead, with straightforward modifications, as long as this 
same expression is used in the actual Monte Carlo calculations for a double step in the imaginary-time direction.
}.

In order to match this spectrum, we take the derivative of Eq.~(\ref{2particleTmatrix}) with respect to $b^{}_\tau$, 
evaluated between eigenstates $| E_{}^{} \rangle$ of the proposed transfer matrix $\mathcal T^{}_{2}$. 
We thus obtain, using the Feynman-Hellmann theorem,
\beq
\label{dTdtauCOM}
-\frac{\partial \langle E_{}^{} | \mathcal T^{}_{2}| E_{}^{} \rangle}{\partial b^{}_\tau} = 
\langle E_{}^{} | 
e^{-\frac{b^{}_\tau p^{2}_r}{2M}}
\left [
K_2^{} + U_2^{}
\right]
e^{-\frac{b^{}_\tau q^{2}_r}{2M}} 
| E_{}^{} \rangle,
\eeq
where
\beq
K_2^{} \equiv
\left [
\frac{p^{2}_r}{2M}+
\frac{q^{2}_r}{2M}
\right]
\left [
\delta_{{\bm p}_r^{} {\bm q}_r^{}}\! + \frac{A({\bm p}_r^{})}{2V}
\right]
\eeq
and
\beq
U_2^{} \equiv -\frac{1}{2V} \frac{\partial A({\bm p}_r^{})}{\partial b^{}_\tau} = 
\frac{1}{2V} \sum_{n=0}^{N_{max}-1} D^{}_{n} {\mathcal O}_{n}^{}({\bm p}_r^{}).
\eeq

The rest of the recipe consists in taking the right-hand side of Eq.~(\ref{dTdtauCOM}) and 
fitting the coefficients $D_{n}$ such that the first 
$N^{}_\mathcal{O}$ eigenvalues of the exact expression Eq.~(\ref{dTdtauexact}) 
(which correspond to the lowest eigenvalues prescribed by L\"uscher's formula) are reproduced.
We may assume at this point that the coefficients $C^{}_{n}$ are known, such that the eigenvectors 
$| E_{}^{} \rangle$ are fixed when we set out to find the $D^{}_{n}$. In that case, the $D^{}_{n}$ are
determined by a linear system of equations of order $N^{}_\mathcal{O} \times N^{}_\mathcal{O}$:
\beq
\sum_{n=0}^{N^{}_\mathcal{O}-1} M^{}_{E n} D^{}_{n} = Y^{}_{E}
\eeq
where
\beq
M^{}_{E n} = \frac{1}{2V} \langle E_{}^{} | {\mathcal O}_{n}^{}| E_{}^{} \rangle
\eeq
and
\beq
Y^{}_{E} = E \exp\left(- b^{}_\tau E \right) -
\langle E_{}^{} | e^{-\frac{b^{}_\tau p^{2}_r}{2M}} K_2^{} e^{-\frac{b^{}_\tau q^{2}_r}{2M}} | E_{}^{} \rangle
\eeq

Once the coefficients $D_{n}$ have been determined, one can use this in a lattice calculation
simply by replacing $A \to A({\bf p})$, and taking $\partial C^{}_{n}/\partial b^{}_\tau = D_{n}$.
The results of the fits for $D_{n}$ are shown in Table~\ref{Table:Dn}.


\begin{table}[t]
\begin{center}
\caption{\label{Table:Dn}
Results of fitting the coefficients $D^{}_{n}$
to the low-energy spectrum of the two-body problem at resonance, in a box of side $N^{}_x = 16$,
for an imaginary time step $b^{}_\tau = 0.05$, in lattice units, from Ref.~\cite{JED}.
}
\begin{tabularx}{\columnwidth}{@{\extracolsep{\fill}}c c c c c c}
       \hline
       $N^{}_\mathcal{O}$ & $D^{}_0$ & $D^{}_1$ & $D^{}_2$ & $D^{}_3$ & $D^{}_4$\\
       \hline
       \hline
1   &  -14.76869    &  --   &  --   &  --   &  --  \\
2   &  -11.54894     & -1.74519     & --    & --    & --     \\
3   &  -10.74506     &  -0.96946     & -0.40164    & --    & --     \\
4   & -10.31974     & -0.82605    & -0.17494     & -0.10404   & --     \\
5   & -10.03874     & -0.74266    & -0.16064    & 0.02948    & -0.02878      \\
\hline
\end{tabularx}
\end{center}
\end{table}

\begin{figure}[t]
\includegraphics[width=0.65\columnwidth]{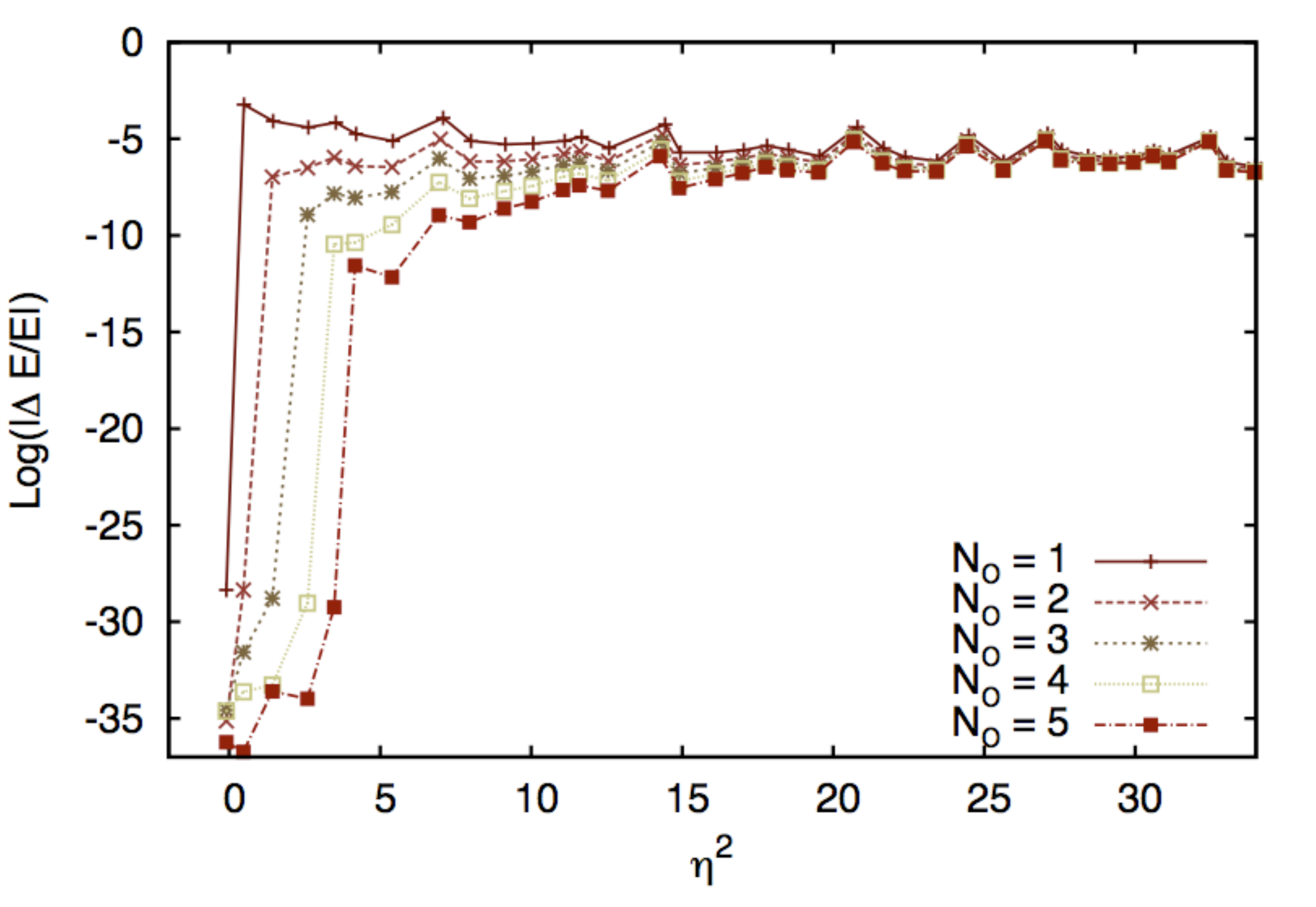}
\caption{
\label{Fig:E_Spectra}
(Color online) Logarithmic plot of the difference between the exact spectrum of the
two-body problem Eq.~(\ref{dTdtauexact}) at unitarity, and the approximate spectrum obtained with 
various levels of improvement $N^{}_\mathcal{O} = 1 - 5$, relative to the exact spectrum (see text for details),
as a function of $\eta^2 = {E^{}_2/p_0^2}$, where $p_0^{} \equiv 2\pi/L$. 
The original data are discrete; the lines are intended as a guide to the eyes.
These results correspond to a lattice of side $N_x = 20$ and temporal spacing 
$b^{}_\tau = 0.05$.}
\end{figure}

As a first illustration of the level of improvement that can be achieved for the energy, 
Fig.~\ref{Fig:E_Spectra} shows the difference between the approximate spectrum with various levels 
of improvement $E_\mathrm{approx}$ and the exact spectrum $E_\mathrm{exact}$, as a function of $\eta^2$,
through the quantity $\log_{10}(|\Delta E|)$, where $\Delta E= (E_\mathrm{approx} - E_\mathrm{exact})/E_\mathrm{exact}$.
As expected, with each new parameter a new eigenvalue is reproduced, with the concomitant reduction in 
the error.
%
\subsubsection{Contact.}

The same ideas can be applied to the calculation of Tan's ``contact" parameter. One of the many ways to define the contact $C$ (see Refs.~\cite{ShinaContact1, ShinaContact2, ShinaContact3, BraatenReview}) is through the derivative of the energy with respect to the inverse scattering length:
\beq
\frac{\partial E}{ \partial a^{-1}_{}} = -\frac{\hbar^2}{4 \pi M} C.
\eeq

The generalization of the above to finite temperature in the grand canonical ensemble is
\beq
\left(\frac{\partial \Omega}{ \partial a^{-1}_{}} \right)_{\beta,\mu} = 
-\frac{1}{\beta} \left(\frac{\partial \log \mathcal Z}{ \partial a^{-1}_{}} \right)_{\beta,\mu} 
= -\frac{\hbar^2}{4 \pi M} C(\beta),
\eeq
where $\Omega$ is the grand thermodynamic potential, and $\mu$ is the
chemical potential. In the large-$\beta$ limit,
\beq
C(\beta) \underset{\beta \to \infty}{\longrightarrow} C_0^{} + b_{C1}^{}\beta^{-1} + b_{C2}^{} e^{-\beta \delta}
\eeq
where $C_0$ is the ground-state contact,
\bea
b_{C1}^{} &=& \frac{4 \pi M}{\hbar^2} \frac{\partial \log Z_0^{}}{\partial a^{-1}_{}},
\\
b_{C2}^{} &=& -\frac{4 \pi M}{\hbar^2}\frac{Z_1^{}}{ Z_0^{}}
\left( \frac{\partial E^{}_1}{\partial a^{-1}_{}} - \frac{\partial E^{}_0}{\partial a^{-1}_{}} \right).
\eea
Since $E$ may be obtained directly from the logarithm of the partition function, we are again in a situation
where we require a derivative of the transfer matrix with respect to a parameter, in this case $a^{-1}_{}$.

In many-body lattice calculations, using these definitions involves the following expression:
\beq
\frac{\partial \mathcal T^{}_{s}[\phi]}{\partial a^{-1}_{}} = 
e^{-\frac{b^{}_\tau \hat T^{}_{s}}{2}}
\hat U_{a^{-1}_{}}^{}
e^{-\frac{b^{}_\tau \hat T^{}_{s}}{2}}
\eeq
where
\beq
\label{UU}
\hat U_{a^{-1}_{}}^{} \equiv \frac{\partial \sqrt{A}}{\partial a^{-1}_{}}\; \sum_{\bm{i}} \hat n^{}_{s,\bm{i}} \sin \phi_{\bm{i}}.
\eeq

The first step towards using these expressions in combination with the improvement procedure
is to take a formal derivative of $\mathcal T$ with respect to $a^{-1}_{}$ in the two-particle space, which we can 
treat exactly:
%
%
%
\beq
\label{dT2dainvexact}
\frac{d \mathcal T_2^{{exact}}}{d a^{-1}_{}} =
-b^{}_\tau \frac{\partial E^{}_2}{\partial a^{-1}_{}}\exp\left(- b^{}_\tau E_2^{} \right).
\eeq
In order to compute the change in the exact two-particle energy $E_2^{}$ due to a small change in the 
inverse scattering length, we use the fact that the energies are implicitly defined as solutions of 
Eq.~(\ref{Luscherformula}), which implies
\beq
\label{dE2dainv}
\frac{\partial E^{}_{2}}{\partial a^{-1}_{}} = - \frac{4\pi^3}{L}\left(\frac{d \mathcal S}{d \eta^2_{}} \right)^{-1}
\eeq
where $\eta^2 = E_2^{} L^2/(2\pi)^2$, and the derivative on the right-hand side is to be evaluated 
at the corresponding solution of Eq.~(\ref{Luscherformula}). Table~\ref{Table:etastar}
shows the first few roots of $S(\eta)$ and the corresponding values of ${d \mathcal S}/{d \eta^2_{}}$.
In the derivation of Eq.~(\ref{dE2dainv}), we have assumed that all the effective-range parameters 
other than the scattering length are kept constant.

\begin{table}[h!]
\begin{center}
\caption{\label{Table:etastar}
First few roots of $\mathcal S(\eta)$, and ${d \mathcal S}/{d \eta^2}$ evaluated at those roots.
}
\begin{tabularx}{\columnwidth}{@{\extracolsep{\fill}}c c c}
       \hline
       $k$ & ${\eta^2_k}$ & $d \mathcal S/d{\eta^2_k}$ \\
       \hline
       \hline
1   &   -0.0959007    &   123.82387   \\ 
2   &    0.4728943    &    39.75514   \\ 
3   &    1.4415913    &    82.36519   \\ 
4   &    2.6270076    &   106.24712   \\ 
5   &    3.5366199    &    84.23133   \\ 
6   &    4.2517060    &   161.88763   \\ 
7   &    5.5377008    &   212.49220   \\ 
\hline
\end{tabularx}
\end{center}
\end{table}


Having the exact target spectrum, we proceed by finding the corresponding expression in terms of 
the HS function $A({\bf p})$, which we obtain using Eq.~(\ref{2particleTmatrix}) and the 
Feynman-Hellmann theorem:
\beq
\label{dTdainvCOM}
\frac{\partial \langle E | \mathcal T^{}_{2} | E \rangle }{\partial a^{-1}_{}} = 
\langle E | e^{-\frac{b^{}_\tau p^{2}_r}{2M}}
\frac{1}{2V} \frac{\partial A({\bm p}_r^{})}{\partial a^{-1}_{}}
e^{-\frac{b^{}_\tau q^{2}_r}{2M}} | E \rangle,
\eeq
where, as before, we expand in terms of our chosen set of operators,
\beq
\frac{\partial A({\bm p}_r^{})}{\partial a^{-1}_{}}
=
\sum_{n=0}^{N^{}_\mathcal{O}} F^{}_{n} {\mathcal O}_{n}^{}({\bm p}_r^{}),
\eeq
and we determine the coefficients $F^{}_{n}$ by fitting the diagonal matrix elements in the right-hand side 
of Eq.~(\ref{dTdainvCOM}) to the exact spectrum of Eqs.~(\ref{dT2dainvexact}) and (\ref{dE2dainv}).
As with the energy, the fitting procedure can be reduced to solving a set of linear equations of order 
$N^{}_\mathcal{O} \times N^{}_\mathcal{O}$.
Illustrative results of such a fit are shown in Table~\ref{Table:Fn}.


\begin{table}[t]
\begin{center}
\caption{\label{Table:Fn}
Results of fitting the coefficients $F^{}_{n}$
to the low-energy spectrum of the two-body problem at resonance, in a box of side $N^{}_x = 16$,
for an imaginary time step $b^{}_\tau = 0.05$, in lattice units.
}
\begin{tabularx}{\columnwidth}{@{\extracolsep{\fill}}c c c c c c}
       \hline
       $N^{}_\mathcal{O}$ & $F^{}_{0}$ & $F^{}_1$ & $F^{}_2$ & $F^{}_3$ & $F^{}_4$\\
       \hline
       \hline
1   &   0.36773   &  --   &  --   &  --   &  --  \\
2   &  0.14532   & 0.07568    & --    & --    & --     \\
3   &  0.11370   & 0.02220    & 0.01957   & --    & --     \\
4   &   0.09659   & 0.01695    & 0.00415   & 0.00538   & --     \\
5   &  0.08205   & 0.01278    & 0.00406   & -0.00023   & 0.00180     \\
\hline
\end{tabularx}
\end{center}
\end{table}

\begin{figure}[h]
\includegraphics[width=0.65\columnwidth]{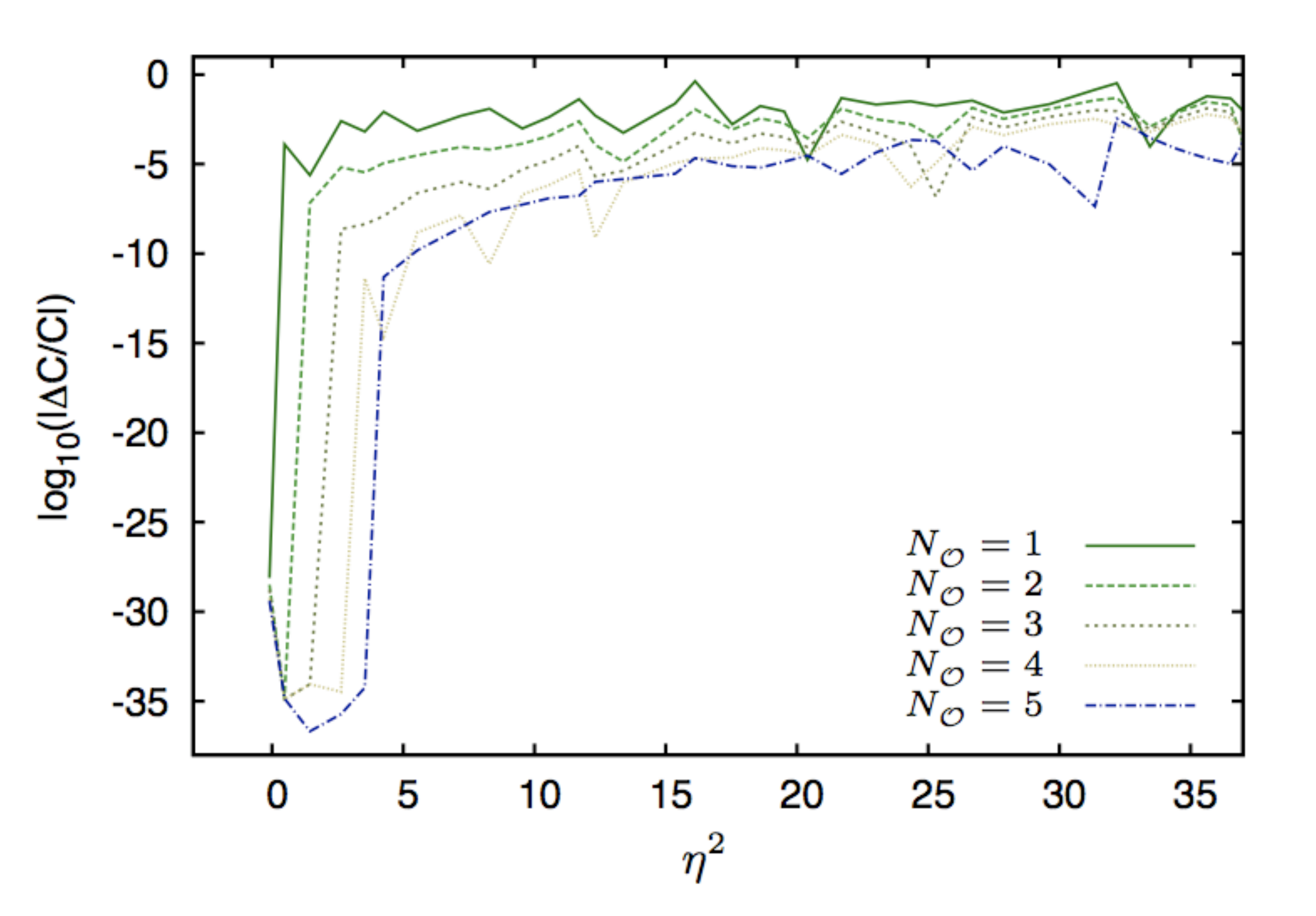}
\caption{
\label{Fig:C_Spectra}
(Color online) Logarithmic plot of the difference between the exact two-body spectrum
Eq.~(\ref{dT2dainvexact}) at unitarity, and the spectrum obtained with various levels 
of improvement $N^{}_\mathcal{O} = 1 - 5$, relative to the exact spectrum (see text for details),
as a function of $\eta^2 = {E^{}_2/p_0^2}$, where $p_0^{} \equiv 2\pi/L$. 
The original data are discrete; the lines are intended as a guide to the eyes.
These results correspond to a lattice of side $N_x = 16$ and temporal 
spacing $b^{}_\tau = 0.05$.}
\end{figure}

A first glimpse at the level of improvement that can be obtained at this point. 
This is shown in Fig.~\ref{Fig:C_Spectra}, where we display the difference $\Delta C$ 
between the spectrum with various levels of improvement and the exact spectrum, 
divided by the latter.
As in Fig.~\ref{Fig:E_Spectra}, each new parameter allows one to fit a new eigenvalue to high accuracy, 
matching the desired physics beyond the lowest momentum.

Unlike in Fig.~\ref{Fig:E_Spectra}, the improvement for eigenvalues beyond those 
explicitly fit is limited, breaking down after the eighth or ninth eigenvalue. From that point on toward 
higher energies, little improvement, if any, is observed as the order of the expansion is increased.
This behavior is only unexpected in the light of Fig.~\ref{Fig:E_Spectra}, where the situation is 
(surprisingly) much more favorable.

%
%



\section{Extracting the ground state \label{Sec:Analysis}}

\subsection{Signal-to-noise} 

Nonrelativistic systems offer an advantage over relativistic systems concerning the large Euclidean time limit: because there exist no backward propagating states, the entire time extent may be used to recover the ground state. However, there remains a serious issue when taking the large Euclidean time limit that goes beyond the simple linear algorithmic scaling with $\beta$. This issue will be referred to as the signal-to-noise problem. 

In general, one finds that signal-to-noise problems occur for theories which have multiparticle states for which the 
energy per particle is lower than for the states one is trying to study. As an example, let us consider a correlation 
function for a single particle in a two-flavor theory with an attractive interaction between different flavors, a trivial example for 
which Monte Carlo calculation is not necessary, but which nevertheless illustrates the signal-to-noise problem quite 
clearly. To determine how many configurations are necessary to achieve a desired accuracy, it is useful to study 
the signal-to-noise ratio (SNR) as a function of time. To do so, we will follow the prescription of Lepage for determining the SNR for nucleon correlators in Lattice QCD calculations \cite{Lepage:1989hd}. The signal is given by
\begin{eqnarray}
\label{eq:signalC}
\calC_1(\beta)&=&\langle K_1(\phi,\beta) \rangle \cr
&=& \frac{1}{\mathcal Z}\int \mathcal{D} \phi \mathcal{P}[\phi] K_1(\phi,\beta)  ,
\end{eqnarray}
where
\beq
K_1(\beta) = \langle 1 | B_{\beta} \cdots B_2 B_1|1 \rangle,
\eeq
the matrices $B_j$ are as in the example of Eq.~\ref{BkExample}, and $|1 \rangle$ is a single particle state of interest. 
As shown in Sec.~\ref{Sec:Observables}, the correlator is dominated by the ground state at large Euclidean time,
\begin{eqnarray}
\langle K_1(\phi,\beta) \rangle \underset{\beta \to \infty}{\longrightarrow} {Z}_0^{(1)} 
e^{- \beta E_0^{(1)}} \, ,
\end{eqnarray}
where ${Z}_0^{(n)},E_0^{(n)}$ are, respectively, the operator overlap with and energy of the 
ground-state of the $n$-particle system. For a single particle, we simply have $E_0^{(1)} = 0$, such 
that at large times the correlator will approach a constant.

Assuming that our sample size is sufficiently large that the central limit theorem holds, the noise may be estimated by calculating the standard deviation,
\begin{eqnarray}
\label{eq:Sigma2C}
\sigma^2_C(\beta)&=& \left( \langle |K_1(\phi,\beta)|^2 \rangle - | \langle K_1(\phi,\beta) \rangle |^2 \right) \cr
&=& \left( \frac{1}{\mathcal Z} \int \mathcal{D} \phi \mathcal{P}[\phi] | K_1(\phi,\beta)|^2 - |\mathcal{C}_1(\beta)|^2 \right)\, .
\end{eqnarray}
The first term in Eq.~(\ref{eq:Sigma2C}) is the ensemble average of a product of two single-particle propagators. This may be recognized as a correlator for two degenerate, distinguishable particles. At large Euclidean time, this quantity will be dominated by the ground-state of the two-particle system,
\begin{eqnarray}
\langle |K_1(\phi,\beta)|^2 \rangle \underset{\beta \to \infty}{\longrightarrow} {Z}_0^{(2)} 
e^{- \beta E_0^{(2)}} \, .
\end{eqnarray}
At zero temperature, two particles in a box with an attractive interaction will form a bound state. This means that $E_0^{(2)} < 0$, 
and the first term in Eq.~(\ref{eq:Sigma2C}) grows exponentially with time. The second term in Eq.~(\ref{eq:Sigma2C}) is simply 
the square of Eq.~(\ref{eq:signalC}), which is time-independent. Thus, in the large-$\beta$ limit it is the first term which dominates.

\begin{figure}[h]
\includegraphics[width=0.49\columnwidth]{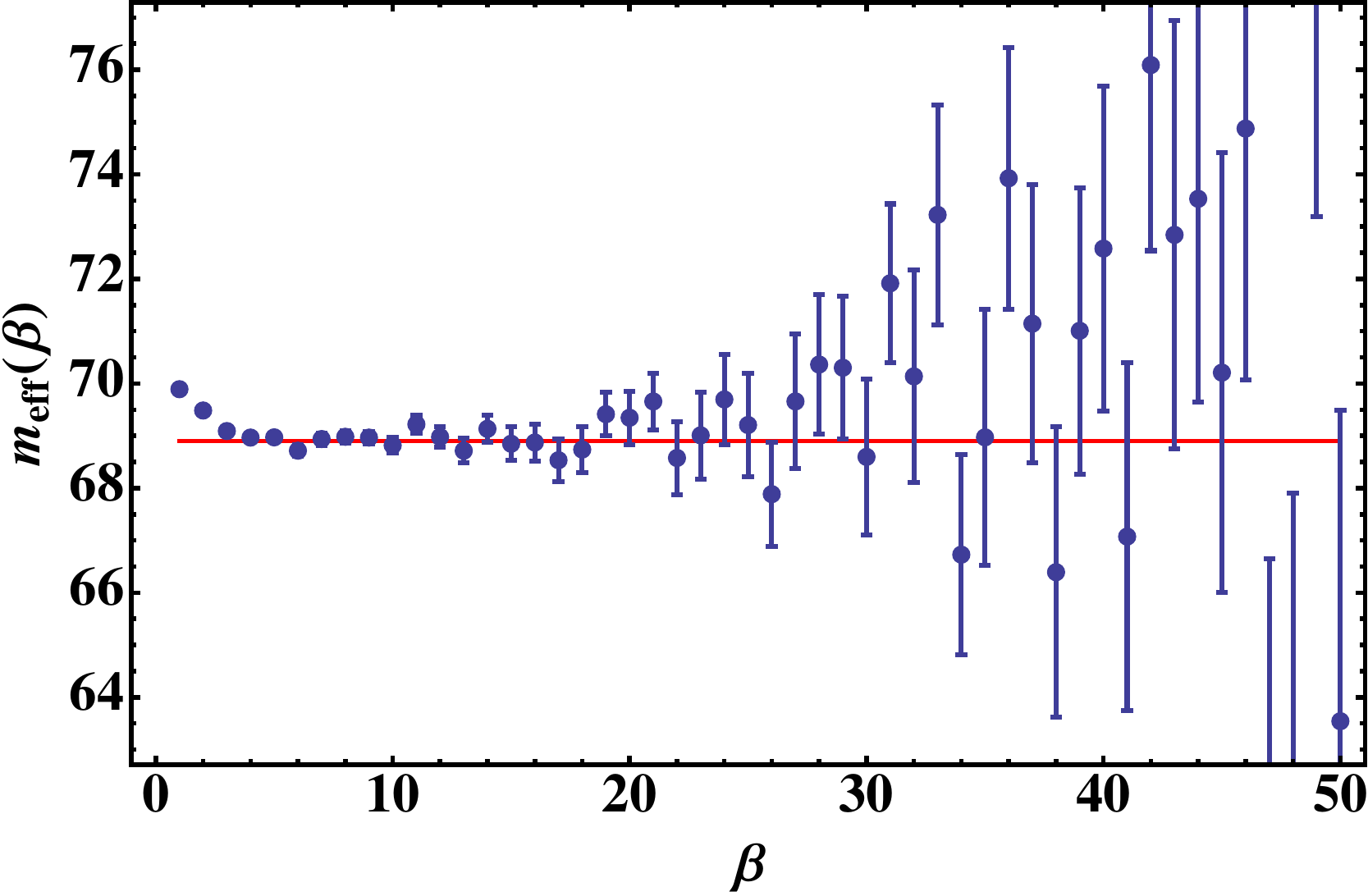}
\includegraphics[width=0.49\columnwidth]{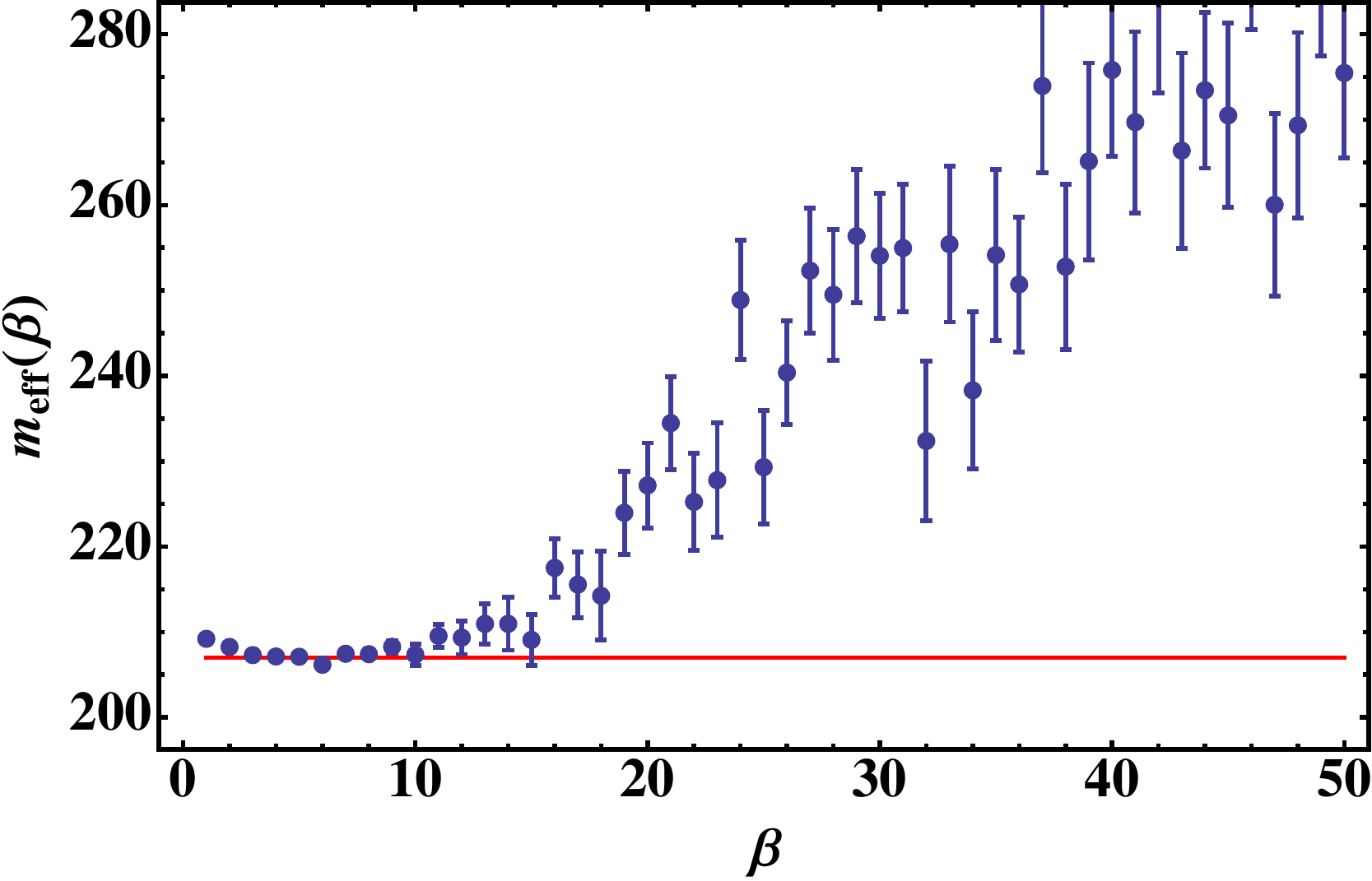}
\label{fig:statistics}
\caption{%
Effective mass plots displaying two common statistical issues: signal-to-noise problem (left), and overlap problem (right). The signal-to-noise problem is seen as an exponential growth of error bars with Euclidean time, while an overlap problem is seen as a drift with no plateau at large Euclidean time. In both cases the ground-state energy is shown as a red line.}
\end{figure}

We can now form the SNR as
\begin{eqnarray}
\label{eq:SNRR}
\mathcal{R} =  \frac{\calC_1(\beta)}{\sqrt{ \frac{1}{N_{\mbox{cfg}}} \sigma^2_C(\beta)}}\underset{\beta \to \infty}{\longrightarrow} \sqrt{N_{\mbox{cfg}}}\frac{ {Z}_0^{(1)}}{\sqrt{{Z}_0^{(2)}}} e^{\beta/2 E_0^{(2)}}\, ,
\end{eqnarray}
Because $E_0^{(2)} < 0$, the SNR is exponentially suppressed as a function of Euclidean time. Thus, an exponentially large number of configurations will be necessary to see a signal at the large times required to extract the ground state from the correlator. An example of an effective mass plot displaying a signal-to-noise problem is illustrated in Fig.~\ref{fig:statistics} (left panel).

One possibility for reducing the signal-to-noise problem lies in the time-independent prefactor in Eq.~(\ref{eq:SNRR}). The overlap factors 
${Z}^{(n)}_0$ depend on how well we are able to model the ground-state wavefunction for our source. Thus, the ability to create sources which have large overlap with the ground state of the observable of interest and poor overlap with the ground state of the system corresponding to the variance may greatly increase the number of time steps over which a signal can be extracted. This has been documented for the case of nucleon correlators in QCD and gives rise to a ``golden window" in Euclidean time in which the nucleon is in its ground state but exponential degradation of the signal has not yet set in \cite{Beane:2009kya,Beane:2009gs,Beane:2009py}. In many cases the prefactor is naturally large, for example, in the case of two-component unitary fermions the signal corresponds to a scattering state, while the noise is associated with deeply bound states \cite{EKLN5,Nicholson:2012zp}. The natural choice for a source will correspond to a scattering state and thus ${Z}_0^{(2)}$, which measures the overlap of the scattering state source with the bound states in the variance, will be much less than 1.


Note that this argument assumes that the central limit theorem holds so that we may identify the noise as coming from the standard deviation of the operator. For many-body operators, this assumption may not in practice be true. Because higher-order moments are sensitive to the energies of higher $N$-body bound states, it is possible for these moments to also grow exponentially with time. For example, the third moment of the probability distribution for the single-particle case discussed above will be controlled by the energy of the corresponding three-body state\footnote{Due to the lack of (anti-)symmetrization between the fields, the three-body state corresponding to the third moment will be composed of particles having three different flavors, with an attractive interaction between all particles. In general, the partition function may correspond to a different number of flavors from the observable; this is referred to in the lattice QCD literature as a ``partially quenched" theory.}. Thus, if $E_0^{(3)} \geq 3/2 E_0^{(2)}$, the distribution will have non-zero skew which grows exponentially with time. According to the Berry-Esseen theorem \cite{BerryEsseen1, BerryEsseen2}, the number of configurations necessary for the central limit theorem to apply at a given Euclidean time becomes, $N_\mathrm{cfg} \gg e^{\beta (2E_0^{(3)}-3E_0^{(2)})}$. 

The interpretation of such an issue is that the chosen field distribution is not peaked for configurations on which the operator is peaked. We will refer to this issue as a {\it distribution} overlap problem. This problem is particularly worrisome because the na\"ive mean that one calculates may differ significantly from the true quantity, though the error bars may be deceptively small (see Fig.~\ref{fig:statistics}, right panel). Such a problem will manifest itself in error bars which do not scale with $\sqrt{N_{\mbox{cfg}}}$, and in mean values which drift outside the error bars as the sample size is changed.

The traditional approach to solving this problem is to invoke importance sampling, in which the statistical overlap of $\mathcal{P}[\phi]$ with the desired correlation function is increased by using the correlation function itself in a reweighting procedure. For example, one may work directly with the ratio used for the effective mass, Eq.~\ref{eq:eff_mass}, and rewrite it in the following way:
\begin{eqnarray}
  \frac{\calC_N(\beta+\Delta \beta)}{\calC_N(\beta)} =  
  \frac{\int \mathcal{D} \phi \tilde{\mathcal{P}}[\phi] \mathcal{O}(\phi,\beta)}{\int \mathcal{D} \phi \tilde{\mathcal{P}}[\phi]} \, ,
\end{eqnarray}
where
\begin{eqnarray}
\mathcal{O}(\phi,\beta) &=& \frac{\tilde{\calC}_N(\phi, \beta+\Delta \beta)}{\tilde{\calC}_N(\phi, \beta)} \, , \cr
\tilde{\mathcal{P}}[\phi] &=& \mathcal{P}[\phi] \tilde{\calC}_N(\phi, \beta) \, .
\end{eqnarray}
The r.h.s. of this equation may be directly sampled using a Monte Carlo algorithm of one's choice according to the new probability measure $\tilde{\mathcal{P}}$, using the observable $\tilde{\calC}$. Note that the probability measure is operator-dependent, thus new configurations must be generated  for each new type of operator one wishes to study. Despite the added computational cost and complexity of implementing importance sampling, this technique has been used quite successfully in a variety of calculations \cite{PhysRevA.84.061602}. 


An alternative to reweighting is to use a better estimator for $\mathcal{C}(\beta)$ that is free of the distribution overlap problem. One such solution is to use the so-called cumulant expansion for the energy, which relies on the observation that the distribution of the logarithm of the correlator is nearly normally distributed \cite{EKLN2}. Such distributions have been shown to be ubiquitous in lattice calculations due to the fact that the correlator consists of many products of matrices of random numbers \cite{EKLN2,EKLN5,DeGrand:2012ik}, so that for large $N,\beta$ the distribution becomes very similar in feature to a log-normal distribution. Furthermore, the log-normal distribution has been shown to be related to systems in which an Efimov effect occurs for distinguishable particles, such as systems of unitary fermions \cite{Nicholson:2012zp}. 

One may expand the average of a generic correlator 
\begin{eqnarray}
\label{CumulantExp}
\ln \langle C(\beta,\phi) \rangle = \sum_{n=0}^{\infty} \frac{\kappa^{}_n(\beta)}{n!} \, ,
\end{eqnarray}
where $\kappa^{}_n$ is defined as the $n$-th cumulant of the distribution for $\ln C(\beta,\phi)$. This may be recognized as the cumulant generating function evaluated at unity. Note that this expansion does not place any assumptions on the particular form of the distribution other than that the expansion exists. However, in general an infinite number of cumulants need to be included for the equality to hold.

\begin{figure}[h]
\includegraphics[width=0.65\columnwidth]{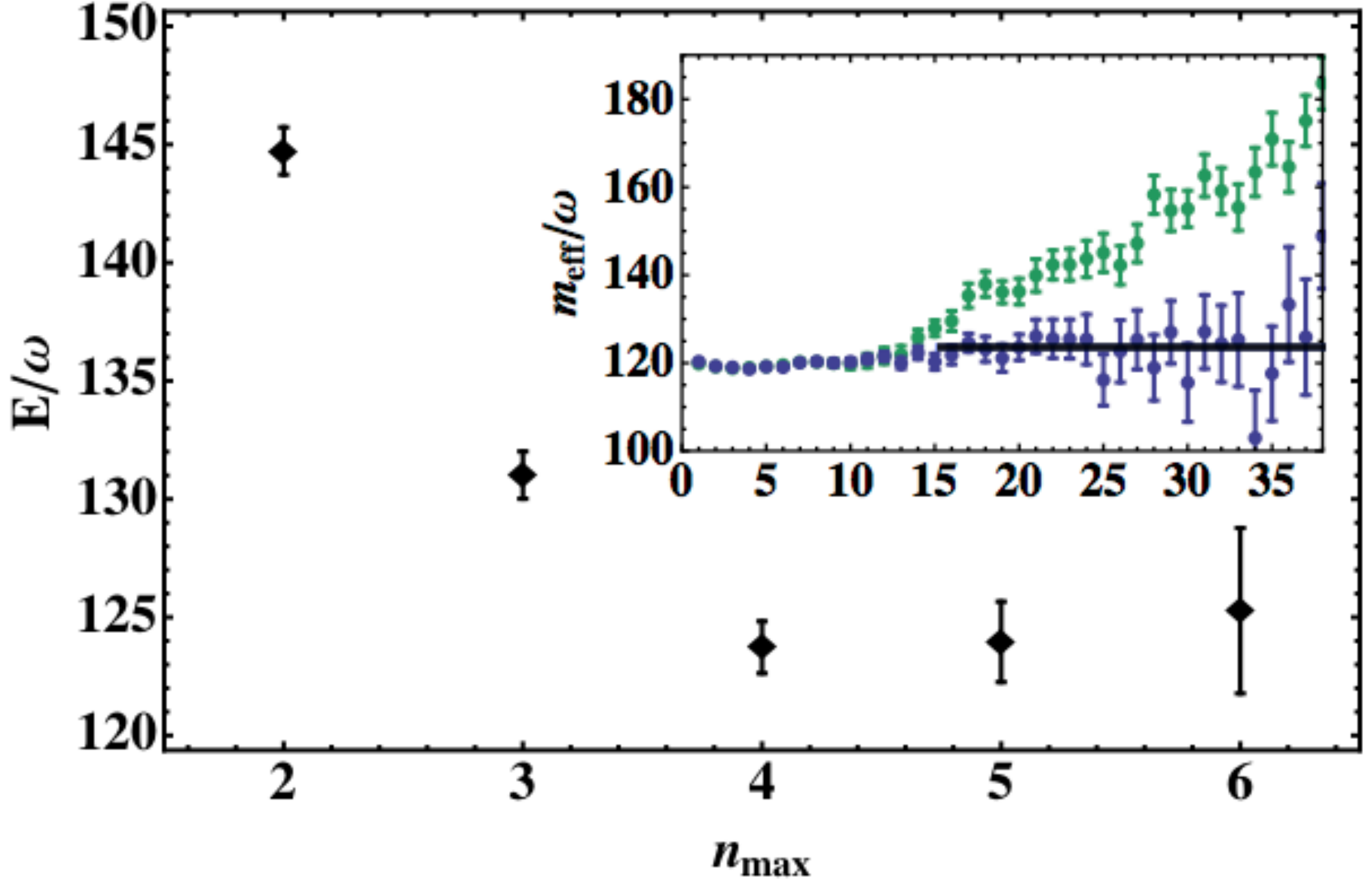}
\label{fig:cumulantExp}
\caption{%
 Energy for 50 unitary fermions in a harmonic trap of frequency $\omega$,  $10^6$ configurations; fits performed using the cumulant expansion (Eq.~(\ref{CumulantExp})) truncated at order $n^{}_{max}$.
Inset: conventional effective mass $m^{}_\mathrm{eff}(\tau) = \log \mathcal{C}(\beta)/\mathcal{C}(\beta+1)$ (green) and corresponding fitted cumulant effective mass with $n^{}_{max}=4$ (blue). Figure from \cite{EKLN5}.}
\end{figure}

On the other hand, if $\ln C(\beta,\phi)$ is nearly normally distributed, all cumulants higher than $\kappa^{}_2$ will be small and the series may be truncated at a finite order, $n^{}_{max}$. Thus the problem has been translated from directly estimating the mean of a quantity having a long-tailed distribution, which is exponentially difficult according to the Berry-Esseen theorem, to that of estimating small moments of a nearly Gaussian distribution. Care must be taken, however, in choosing the order of truncation, and any systematic errors that arise due to truncation must be estimated. An example of results obtained using the cumulant expansion, including evidence for convergence of the series, is shown in Fig.~\ref{fig:cumulantExp}. 
%

\subsection{Interpolating fields}
\label{sec:interp_fields}

The requirement for eliminating excited-state contributions may be approximated as
\begin{eqnarray}
\label{eq:betaCond}
\beta \gg \frac{\ln \left( \frac{{Z}_1}{E_0 {Z}_0}\right)}{E_1-E_0} \, .
\end{eqnarray}

For many-body systems, the spacing between energy levels may become arbitrarily small as both collective effects and single quasi-particle excitations require relatively small energy to excite. Thus, the number of time steps required to reach the ground state may become quite large for $N \gg 1$. Adding to the complication is the fact that larger systems tend to probe higher energies than small systems, so that a very fine temporal lattice spacing is required to reduce discretization errors. Finally, the exponentially growing signal-to-noise problem previously discussed may severely limit the time extents for which a signal may be obtained.

To remedy  this situation, one may construct improved interpolating fields to try to maximize the overlap with the ground state. For large systems, the ground-state wavefunction becomes increasingly complicated and difficult to parametrize, potentially leading to exponentially poor overlap with a chosen interpolating field as a function of the number of particles. To illustrate this point, we consider the simplest choice for an interpolating field, which is simply a Slater determinant of non-interacting single-particle states. Suppose the ground state of the system consists of non-interacting two-body states. If the overlap of the two-body ground state with the corresponding two-body Slater determinant is $\epsilon$, then the overlap of the $N$-body state with the $N$-body Slater determinant will be $\epsilon^N$, giving exponentially poor overlap with the true ground state. 

Two-particle correlations may be built in using an $N/2$-body Slater determinant of two-body ground states. For the unitary case, the two-body wavefunction is known and we may build two-particle correlations using this wavefunction. One possibility is to use correlators of the form
\begin{eqnarray}
C_{N_\downarrow,N_\uparrow}(\tau) = \langle \det{ S^{\downarrow\uparrow}(\tau)} \rangle\ ,
\end{eqnarray}
where
\begin{eqnarray}
S^{\downarrow\uparrow}_{i,j}(\tau) = \langle \Psi | K^{-1}(\tau,0) \otimes  K^{-1}(\tau,0) | \alpha^\downarrow_i  \alpha^\uparrow_j \rangle \ ,
\end{eqnarray}
$| \Psi \rangle$ is the two-body ground state, and 
$| \alpha^\downarrow \alpha^\uparrow \rangle = | \alpha^\downarrow \rangle \otimes | \alpha^\uparrow \rangle $ 
are the non-interacting single-particle states. This form of the Slater determinant, in which the pairing correlations 
are built into the source and not the sink, can be chosen to avoid the order-$V$ increase in computation time 
required to fully anti-symmetrize both the source and sink. It has been shown \cite{Carlson,LEKN1}
that such a form for the correlation functions greatly improves the overlap for systems of unitary fermions.
For large, fully interacting systems, 3- and higher-body correlations will also contribute to the ground-state. These 
may also be built into the interpolating field, however, each of these will increase the computational scaling with $V$.

Another way to increase the overlap of the interpolating field onto the ground-state wavefunction for systems which 
possess rotational invariance in the continuum is to project the field onto the appropriate irreducible representation 
of the cubic group. By doing so, one may eliminate contributions from low-lying excited states with 
different angular momenta. However, because rotational invariance is broken by the cubic lattice, different angular 
momentum states can mix. For example, if the ground state of interest has positive parity and zero total angular 
momentum, one may project onto the $A_1^{+}$ representation, which will give overlap with not only the $J=0$ 
state, but also $J=4, 6, \dots$. Appropriate combinations of sources projected onto different representations may 
be formed to further eliminate contributions from undesired angular momenta. For details on creating projection operators see App.~\ref{Sec:GroupTheory}.


\section{Systematic effects \label{Sec:Systematics}}

\begin{figure}[h]
\includegraphics[width=0.45\columnwidth]{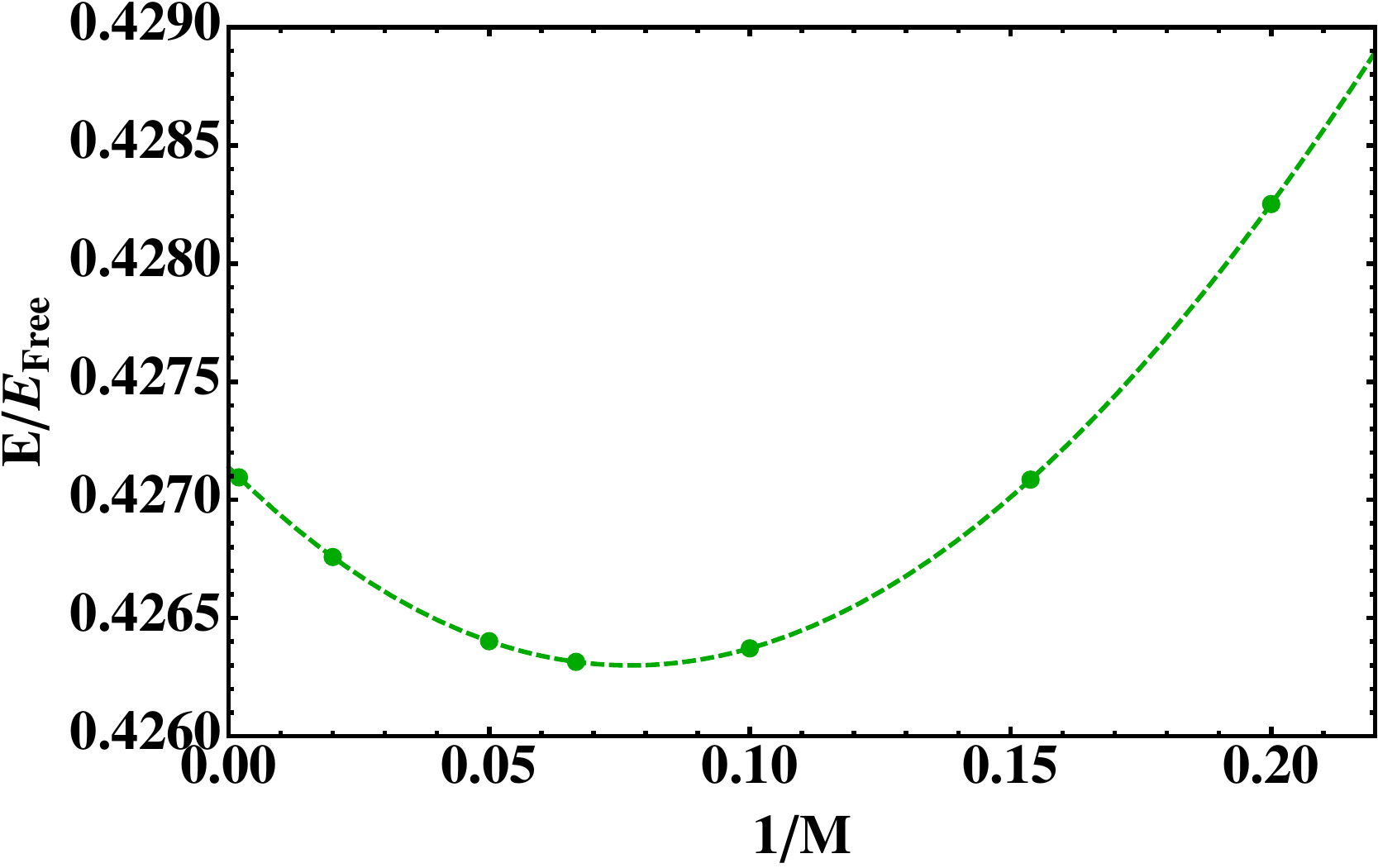}
\includegraphics[width=0.45\columnwidth]{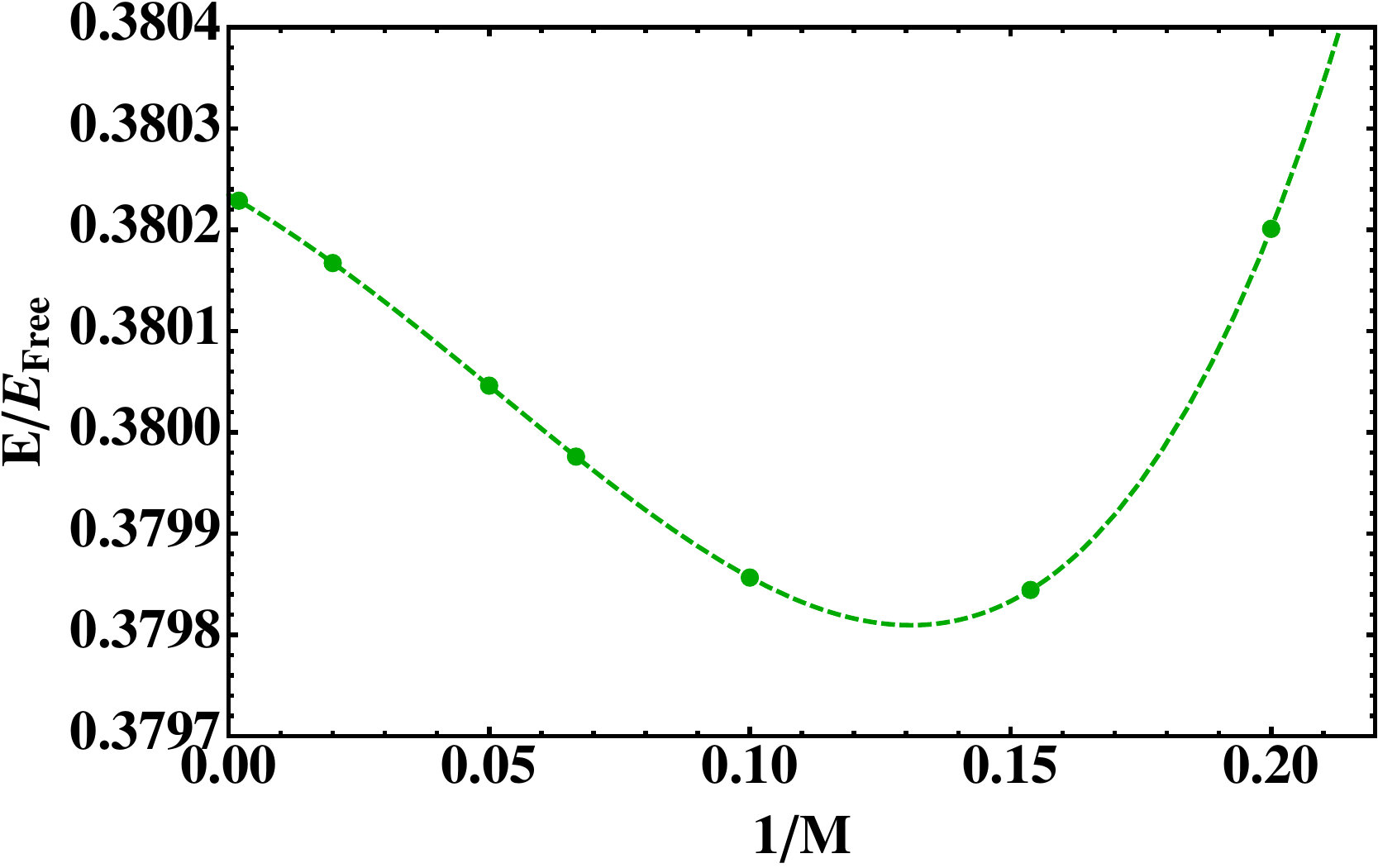}
\caption{
\label{fig:three_fermions_time_discretization}%
Ground state energy of $N=2+1$ unitary fermions in units of $E_{\mathrm{free}} = (2\pi/L)^2/M$ in the $A_1^+$ irrep as a function of $1/M$ for one (left) and four (right) tuned couplings with $L=8$.
Dashed green line corresponds to a fourth order polynomial fit in $1/M$. Figure from \cite{EKLN4}.
}
\end{figure}

In addition to the statistical errors discussed in the previous section, systematic errors arise from the discretization of space and time, as well as from the finite volume of space considered. Furthermore, there may be excited state contributions arising from the finite temporal extent. In this section, we will largely focus on errors due to spatial effects. While temporal discretization errors enter at leading order in a calculation of the energy, the temporal lattice spacing is naturally much smaller than the spatial lattice spacing, $b_{\tau} = b_s^2/M$, and may be further decreased by increasing the anisotropy parameter $M$. In addition, as discussed below we may improve the transfer matrix to systematically remove discretization effects. Temporal discretization effects are also reduced by this procedure as the transfer matrix becomes increasingly ``perfect" (see Fig.~\ref{fig:three_fermions_time_discretization}). Finally, regarding the issue of finite temporal extent, many improvements to the fitting process have been introduced in the lattice QCD literature, such as multi-exponential fits and removal of excited state contributions through the use of multiple sets of sources and sinks \cite{Luscher:1990ck,Michael198558,Blossier:2009kd,Beane:2009kya,Beane:2009gs,Beane:2009py}. In general, so long as Eq.~(\ref{eq:betaCond}) holds, we may be confident that these effects are exponentially suppressed. 

%

\subsection{Unitary fermions}

To begin our discussion of systematic effects due to spatial discretization and finite volume, we will consider the system of two-component unitary fermions. The lack of physical scales in this system greatly simplifies the analysis. For a review of the introduction of new scales via lattice effective field theory see Ref.~\cite{Lee2}. 

Recall that the relevant scale for this case is the Fermi momentum, $k_F = (3\pi^2 n)^{1/3}$, where $n = N/L_{phys}^3$ is the density. Thus, for a given number of particles the size of the box, $L_{phys} = b_s L$, should be considered a fixed physical scale, and the lattice spacing, or number of lattice points $L$, will control spatial discretization errors. To reach the continuum limit we require $L \gg 1/(b_s k_F)$.

Finite volume errors of the form $a/L$ may be eliminated by tuning the scattering length according to the L\"uscher formalism, which relates the energy levels of two particles in a finite box to the desired scattering phase shift. Note that the L\"uscher formalism requires $L \gg r_0/b_s$. As discussed below, a lattice theory will intrinsically induce $r_0 \sim b_s$, placing some constraint on the volumes which may be considered. However, corrections to the L\"uscher method due to finite range are exponentially suppressed, while the non-zero effective range also induces errors which obey power-law behavior to the energies of unitary fermions, so that in general these corrections may be ignored. Finally, we will not discuss the issue of finite size effects in the approach to the thermodynamic limit as these are not particular to the discussion of lattice methods. 

Discretization errors arise from two sources. The first is due to the discretization of the kinetic operator. As discussed in Sec.~\ref{Sec:Generalities}, one may improve this discretization by attempting to reproduce the correct single-particle dispersion relation, either by adding higher-order derivative terms and matching the energies order-by-order, or by calculating the kinetic operator directly in momentum space, allowing one to reproduce the exact continuum dispersion relation to all orders up to a momentum cutoff. 

The second source of discretization errors comes from the finite range of the interaction induced by the finite lattice spacing. These types of errors may be reduced by improving the two-particle sector. One method for doing so is to introduce momentum-dependent interactions tuned according to L\"uscher's formula, as discussed at the end of Sec.~\ref{Sec:Generalities}. By tuning higher-order energy levels in the two-body spectrum one also effectively tunes higher-order terms in the effective range expansion of $p \cot \delta$ to zero. This can be verified by plugging the resulting lattice energy eigenvalues back into L\"uscher's formula to recover the effective $p \cot \delta$ seen by the two-particle system. This is shown in Fig.~\ref{Fig:pcotdGI}, where for higher numbers of tuned coefficients $p \cot \delta$ becomes progressively closer to zero for all momenta.

%
%
%
%

%
%

\begin{figure}[h]
\includegraphics[width=0.65\columnwidth]{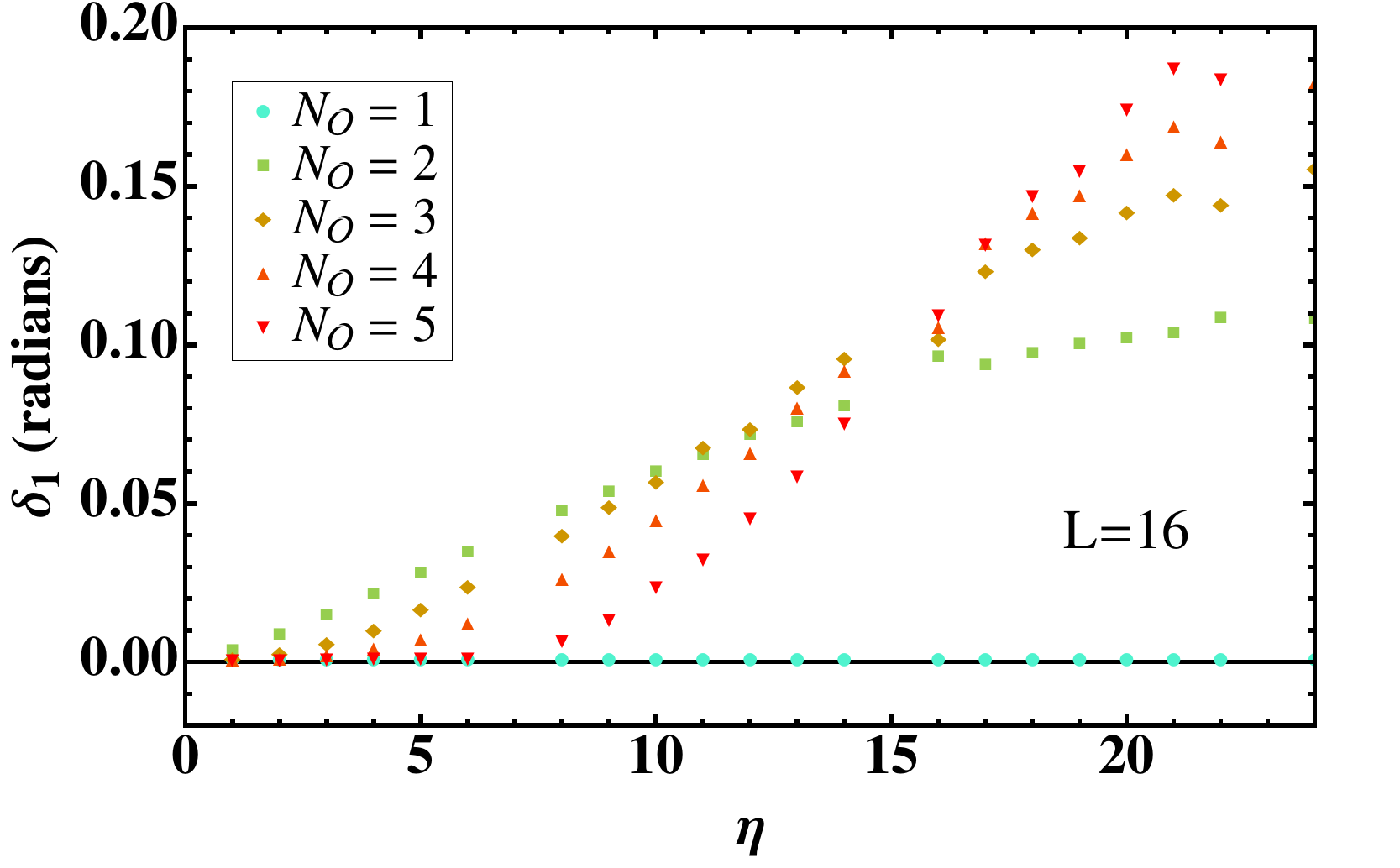}
\caption{\label{fig:deltas}%
(Color online) $\delta_1$ as a function of the dimensionless parameter $\eta$ for unitary fermions of mass
$M = 5$ on an $L = 16$ lattice obtained from Eq.~\ref{eq:luschers_formula_pwave}. Figure from \cite{EKLN4}.
}
\end{figure}

While reducing the effects of unwanted $s$-wave scattering contributions, the introduction of momentum-dependent couplings also 
induces interactions corresponding to higher partial waves. The size of the effects of interactions with angular momenta $\ell>0$ may be deduced by studying the energies of the two-body lattice eigenstates classified by their irreducible representation (irrep) under the octahedral group $O$ (see Appendix I). 
%
%
Using the generalization of L\"uscher's formula for $s$-wave scattering, we may also determine the scattering phase shifts for 
the higher partial waves. For $p$-wave scattering, if one assumes $\tan\delta_4 \ll \tan\delta_1$, one finds (see, for example, 
\cite{Luu:2011ep}):
\begin{eqnarray}
\label{eq:luschers_formula_pwave}
p^3 \cot\delta_1(p) = \left(\frac{2\pi}{L}\right)^3 \frac{1}{2\pi^2} \, \eta\, S(\eta)\ , 
\end{eqnarray}
where $\eta$ and  $S(\eta)$ are defined in Eq.~\ref{Luscherformula}.
Plugging the lattice eigenvalues obtained for the $T_1^-$ irrep 
into the right-hand side of Eq.~\ref{eq:luschers_formula_pwave}, we obtain a lattice prediction for $p^3 \cot\delta_1$. This phase shift is plotted in Fig.~\ref{fig:deltas}, as a function of $\eta$ for $L=16$, and for $N_\mathcal O = 1,2,3,4$ and $5$.

%


%
\begin{figure}[h!]
\includegraphics[width=0.65\columnwidth]{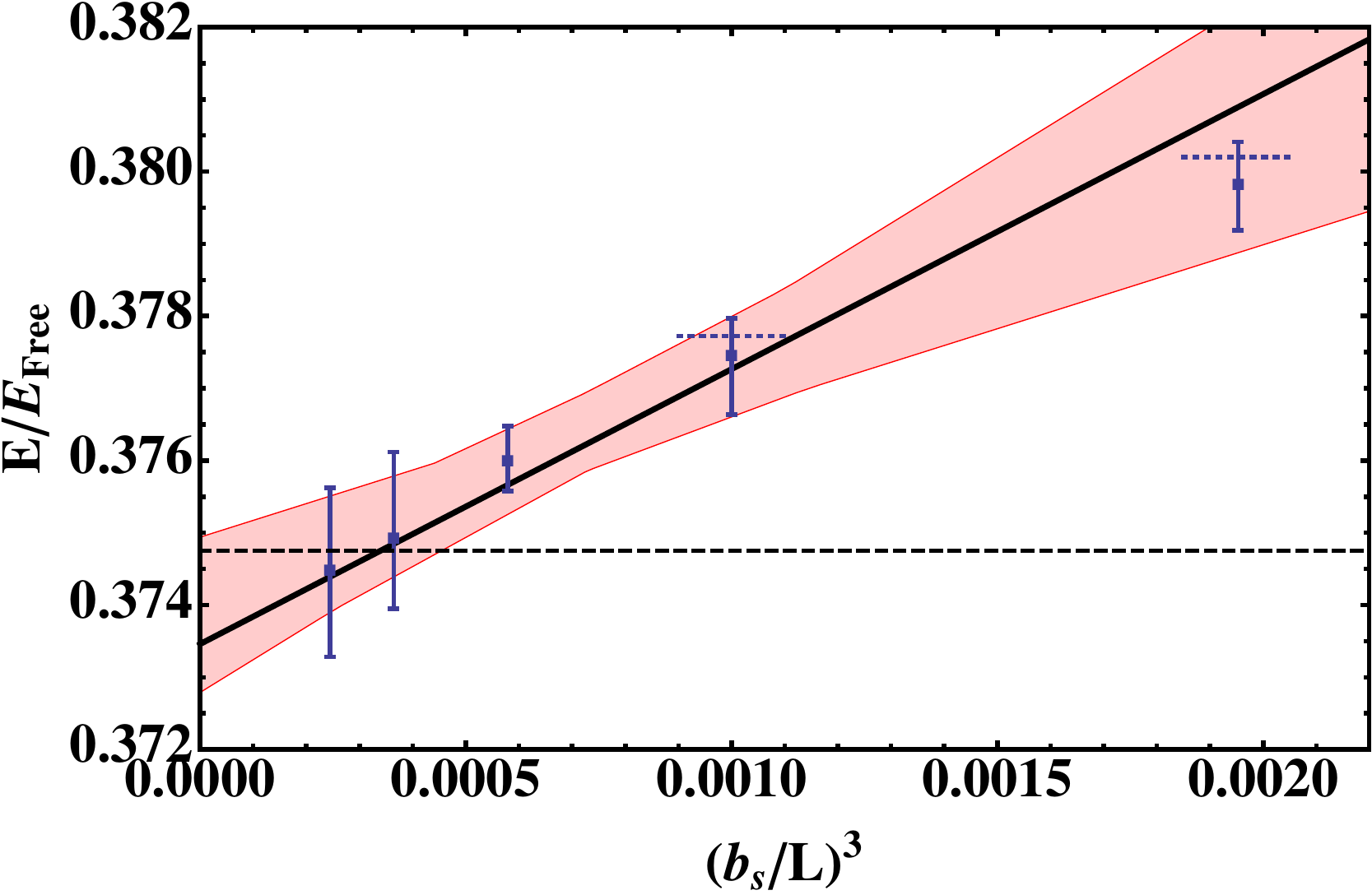}
\caption{
\label{fig:N3_untrapped}%
Energy of $N=2+1$ unitary fermions in a zero total momentum eigenstate in units of $E_{\mathrm{free}} = 4 \pi^2/(ML^2)$ as a function of $1/L^3$.
Blue data points and associated error bars were obtained from numerical simulation, short blue dotted lines at $L=8$ and $L=10$ indicate results from exact diagonalization of the three fermion transfer matrix.
Red error band indicates the infinite volume extrapolation result reported in \cite{EKLN1,EKLN4} using simulation data.
Black dashed line indicates the infinite volume result of Pricoupenko and Castin reported in \cite{2007JPhA...4012863P}. Figure from \cite{EKLN1}.
}
\end{figure}

In addition to introducing a finite range to the two-body interaction, a non-zero lattice spacing also induces $N$-body interactions. Tuning away the effects from these interactions is in general non-trivial, thus, knowledge of the sizes of these errors and how they scale with $L$ are useful, particularly if an extrapolation to the continuum is desired. One tool for doing so is to construct the Symanzik effective 
action~\cite{Symanzik1,Symanzik2,Symanzik3,Symanzik4,EKLN4}. This is done by considering all possible operators 
formed using the fields and temporal or spatial derivative operators, multiplied by unknown, dimensionless coefficients 
and appropriate powers of the lattice spacings to restore the proper dimension. The first operator to appear, assuming 
the leading two-body $s$-wave contributions have been tuned away, will be the two-body $p$-wave 
interaction discussed above, scaling as $L^{-3}$. That this scaling behavior dominates in the three-body system was verified 
in \cite{EKLN1} (see Fig.~\ref{fig:N3_untrapped}). In principle, one may tune away such $p$-wave contributions in the same way as the $s$-wave 
contributions, by introducing $p$-wave operators and tuning the coefficients according to the L\"uscher formalism.

The next class of operators to appear corresponds to $N$-body interactions, the lowest being the zero-derivative three-body 
operator. Such an operator will not obey a naive scaling law due to the conformal nature of the system, but may be shown to 
scale as $L^{-3.55}$ and $L^{-4.33}$ for $\ell=1,0$, respectively, using an operator-state 
correspondence~\cite{2004cond.mat.12764T,NishidaSonConformal}. Additional operators corresponding to the four-derivative 
$p$-wave and $d$-wave two-body operators come in at order $L^{-5}$. For as few as four particles it was shown that effects from the $p$-wave three body operator are non-negligible \cite{EKLN4}, and inclusion of a term scaling as $L^{-3.55}$ was required to successfully extrapolate to the continuum.


\subsection{External potentials}

The introduction of an external potential, such as a harmonic trap, may also add new scales to the problem. Furthermore, the way the potential is included in the transfer matrix affects the size of temporal discretization errors. As a simple example, one may use Trotter's product formula in the absence of interactions to eliminate the leading order $b_{\tau}$ contribution,
\beq
\mathcal{T}_{pot} = e^{-b^{}_{\tau} K/2}e^{-b^{}_{\tau} U}  e^{-b^{}_{\tau} K/2}\, .
\eeq

When interactions are turned on, the order at which temporal discretization errors appear depends on how well the interaction has been tuned to remove discretization errors. If we consider the transfer matrix for unitary fermions without an external potential,
\beq
\mathcal{T}_{unitary} = e^{- b^{}_{\tau} K/2} e^{- b^{}_{\tau} \mathcal{V}_{int}} e^{-b^{}_{\tau} K/2} \equiv e^{-b^{}_{\tau} \mathcal{H} } \, ,
\eeq
then we may write the full transfer matrix as 
\beq
\mathcal{T} = e^{-b^{}_{\tau} K/2 } e^{-b^{}_{\tau} U/2 } e^{-b^{}_{\tau} \mathcal{V}_{int}} e^{-b^{}_{\tau} U/2 } e^{-b^{}_{\tau} K/2 } \, .
\eeq
If discretization errors in $\mathcal{H}$ have been completely eliminated by improvement, one may show that temporal discretization errors again come in at $\mathcal{O}(b_{\tau}^2)$ \cite{EKLN1}. The remaining error from temporal discretization may be controlled by tuning the energy scale introduced by the potential of interest; for the case of a harmonic potential, this scale is the trap frequency, $\omega$.

\begin{figure}[h]
\includegraphics[width=0.65\columnwidth]{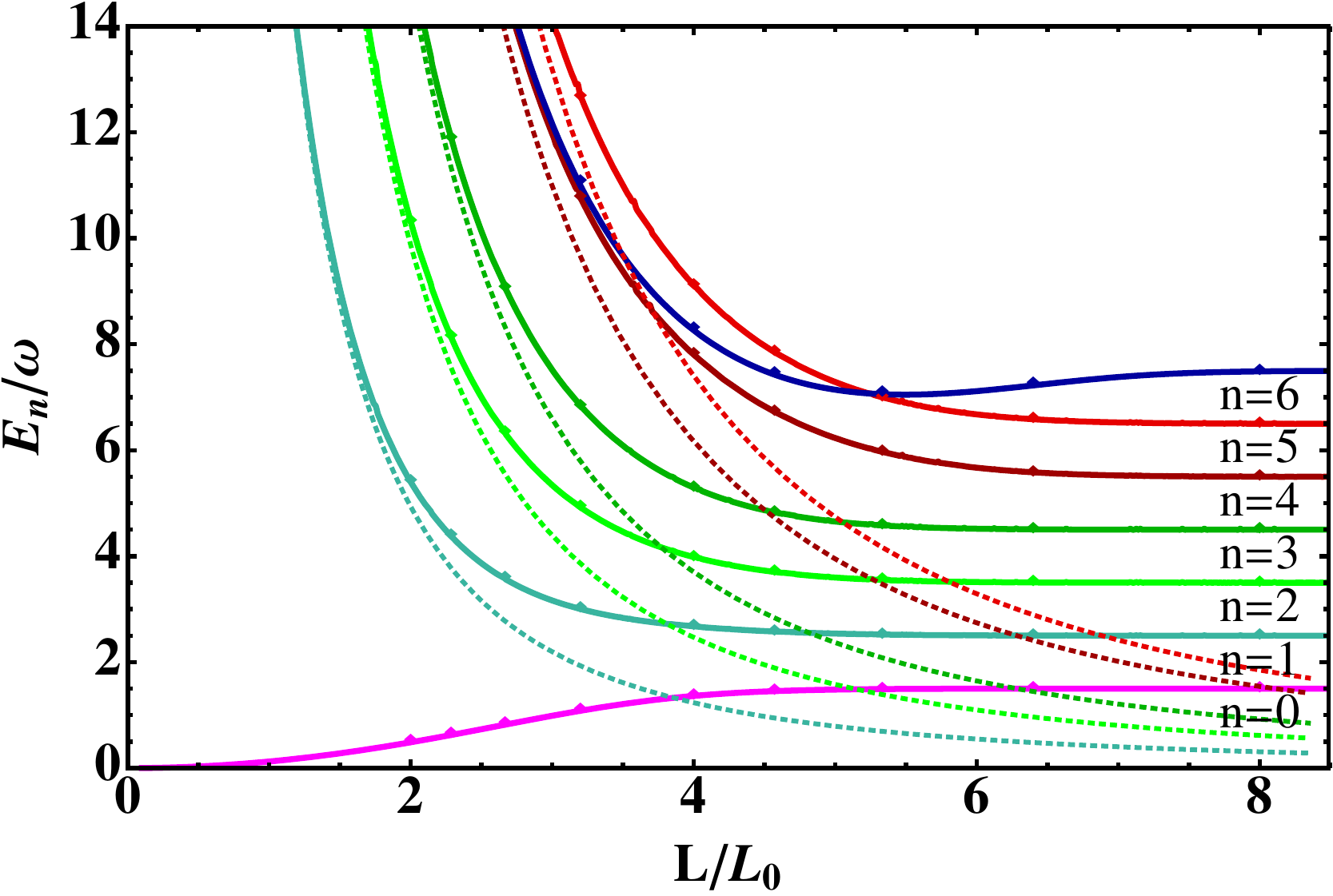}
\caption{
\label{fig:sho_finite_vol}%
$L/L_0$ dependence of the SHO energies $E_n$ ($n = \sum_j n_j$) corresponding to the single fermion states $\mathbf n = (0,0,0), (1,0,0), (1,1,0), (1,1,1), (2,1,0), (3,1,1)$, and $(4,2,0)$, where $L_0 = (M \omega)^{1/4}$.
Solid lines indicate an exact continuum limit calculation, whereas the data-points indicate simulation results for $\omega=0.005$ and $L_0\ge2$. Dotted lines correspond to free fermions in a finite box (small $L/L_0$ limit). Figure from \cite{EKLN1}.
}
\end{figure}

For the simple harmonic oscillator (SHO), a new length scale is introduced, $L_0 = (M \omega)^{1/4}$. Spatial discretization errors are now controlled by the dimensionless ratio $L_0/b_s$, while finite volume errors are controlled by the independent combination $L/L_0$. In the continuum limit, finite volume errors for the noninteracting system may be computed analytically, since the SHO is separable. A plot of the energy dependence of the SHO on $L/L_0$ is shown in Fig.~\ref{fig:sho_finite_vol} for several low energy single fermion states; at large $L/L_0$, the energies in units of $\omega$ are just an integer plus the zero point energy ($3/2$ for a three-dimensional SHO). For very small volumes, the harmonic potential plays no role and the system is effectively a free particle in a finite box, with energies increasing proportional to $ \frac{1}{2} \left(\frac{2\pi }{L/L_0}\right)^2 $  with decreasing $L/L_0$. The dashed lines in Fig.~\ref{fig:sho_finite_vol} indicate this limiting behavior for several SHO states.

When interactions are turned on, in general the errors as a function of $L,L_0$ must be explored numerically. Benchmark results, which will be presented in the next subsection, are useful in determining values for $b_s/L_0$ and $L/L_0$ for which finite volume and discretization errors are minimal. 

%
%

%
%

\subsection{Comparison with benchmark results}

From testing new code to estimating the size of systematic errors for a given parameter set, benchmark results, whether calculated using lattice or other methods such as GFMC, are indispensable when developing new methods. Due to the reduced computational cost and complexity, few-body systems often serve in this role. Such results exist for unitary fermions in both the homogeneous and trapped cases. Verification of these results as the variety of methods grows serves to both strengthen and constrain the results and give validity to new methods.

\begin{figure}[h]
\includegraphics[width=0.70\columnwidth]{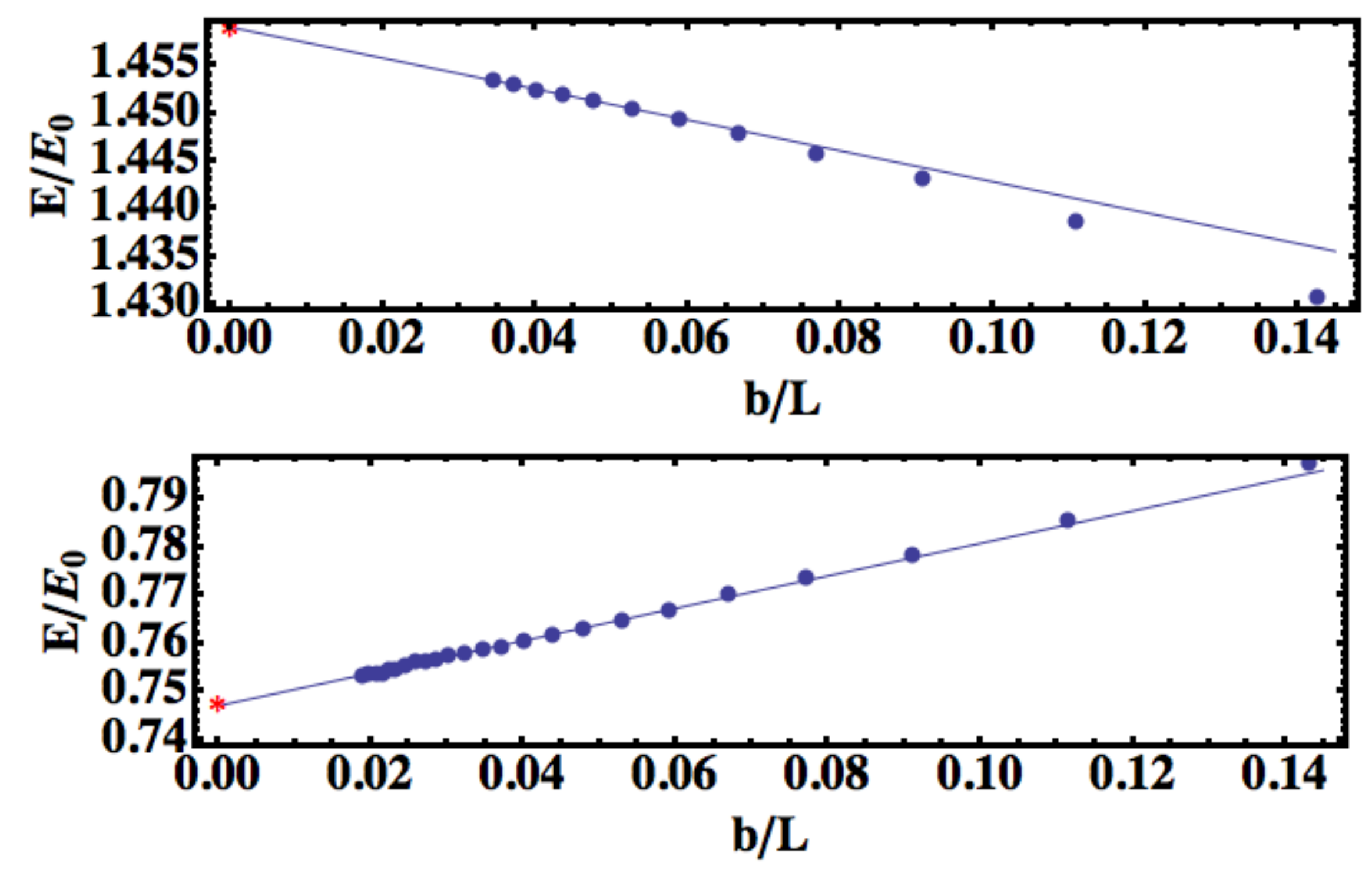}
\caption{
\label{Fig:PCResult}Ground (bottom) and first excited (top) state energies for $N=2+1$ unitary fermions in a box with zero total angular momentum 
as a function of $b^{}_s/L$ as calculated in \cite{2007JPhA...4012863P}. Energies are in units of $E_0 = 4 \pi^2/(2ML^2)$. Lines represent a linear 
fit to the data for $b^{}_s/L \leq 1/15$, and stars show the energies predicted by the Bethe-Peierls model \cite{1935RSPSA.148..146B}. 
}
\end{figure}

The simplest system to investigate is that for three unitary fermions in a box. The energies for the first few zero-momentum states were calculated to high accuracy on very large lattices using an unimproved four-fermion interaction in \cite{2007JPhA...4012863P}, along with a linear extrapolation to the continuum limit (Fig.~\ref{Fig:PCResult}). This measurement was confirmed in \cite{EKLN1,EKLN4} using an improved interaction with four tuned coefficients. An extrapolation based on the $L^{-3}$ scaling of the untuned p-wave operators was performed to eliminate the leading contribution to the remaining systematic error (Fig.~\ref{fig:N3_untrapped}).

\begin{figure}
\includegraphics[width=\columnwidth]{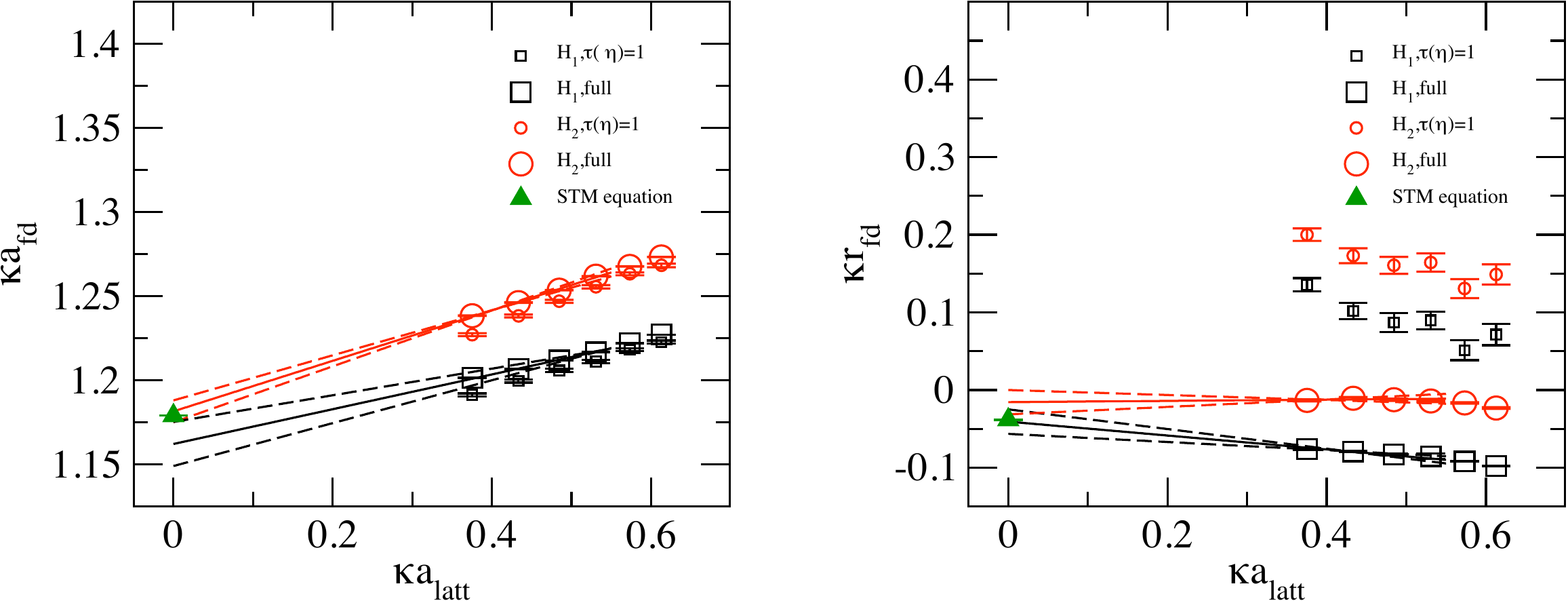}
\caption{
\label{fig:DLScatt}%
Lattice results and continuum extrapolation with statistical error bands for the fermion- dimer scattering length (left) and effective range (right). All units are given by the binding momentum of the dimer, $\kappa$. The full calculation accounts for the topological volume factor, $\tau(\eta) \neq 1$. For comparison the continuum result obtain via the Skorniakov-Ter-Martirosian equation is also presented. Figure from \cite{Bour:2012hn}.
}
\end{figure}

Additional benchmarks for three-body observables, the fermion-dimer scattering length and effective range, have been calculated using lattice methods in combination with the L\"uscher formalism (see Sec.~\ref{Sec:Observables}), modified to account for topological corrections to the dimer binding energy  \cite{Bour:2011ef,Bour:2012hn}. The results found using two different lattice Hamiltonians with a linear extrapolation to the continuum limit are shown in Fig.~\ref{fig:DLScatt}. Also shown is the data computed without correcting for the topological volume factor, $\tau(\eta)$; it is clear that the topological factor has a sizeable impact, particularly on the result for the effective range. Because the scattering amplitude can be reduced to a two-body problem, it can be solved nearly exactly through an integral equation formulation, known as the Skorniakov-Ter-Martirosian (STM) integral equation \cite{STM:1956,PhysRevA.67.010703,Petrov:2005zz}. Results from the STM equation are included in Fig.~\ref{fig:DLScatt}, showing excellent agreement with the lattice results.

\begin{figure}[h]
\includegraphics[width=0.65\columnwidth]{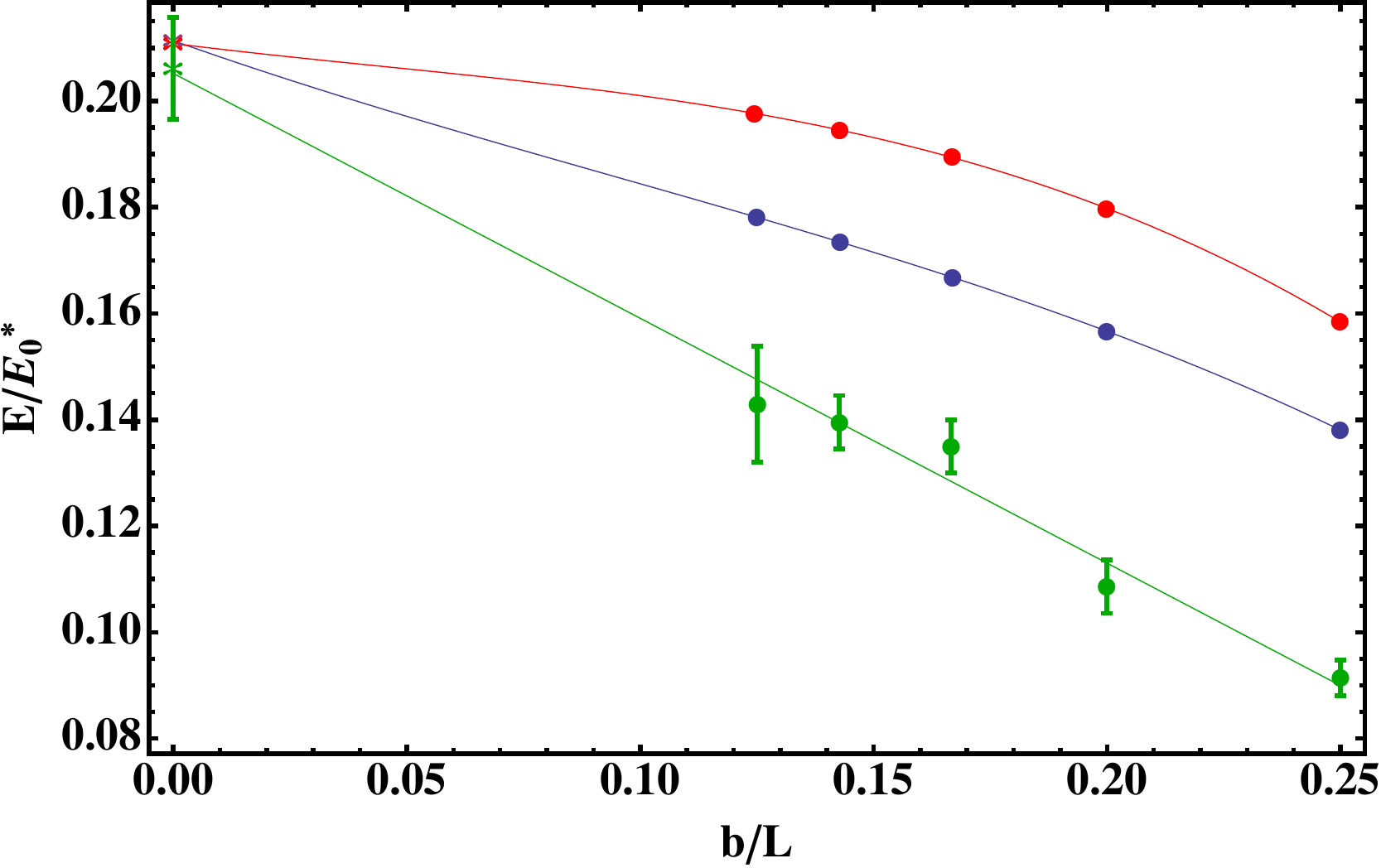}
\caption{
\label{Fig:DL4fermions}
(Color online)Ground state energy for $N=2+2$ unitary fermions in a box as a function of $b/L$ as calculated in \cite{Bour:2011xt} in units of $E_0^{*}  = 4 \pi^2/(ML^2)$. The results are extracted using three methods: an exact Hamiltonian lattice method with two different Hamiltonians (blue, red), and a Euclidean lattice method using auxiliary field Monte Carlo (green). Lines represent extrapolations to the infinite volume limit using a third-order polynomial fit (blue, red) and a linear fit (green), with the extrapolated results shown as stars.}
\end{figure}

The next system to consider is that for four unpolarized unitary fermions in a box. The authors of \cite{Bour:2011xt} performed an extensive analysis of this system using four different techniques: hamiltonian lattice formalism using iterated eigenvector methods, Euclidean lattice formalism with auxiliary-field projection Monte Carlo, and continuum diffusion Monte Carlo with fixed and released nodes. The results of polynomial extrapolations to the continuum for the lattice methods are shown in Fig.~\ref{Fig:DL4fermions}. The agreement between four different methods gives a very robust benchmark. Additionally, this result was verified in \cite{EKLN4} using both an unimproved interaction in conjunction with a linear extrapolation to the continuum limit and an improved interaction with an extrapolation based on the untuned terms in the Symanzik action discussed above.

%

%

Finally, there exist exact results for three unitary fermions in a trap \cite{2004cond.mat.12764T,2006PhRvL..97o0401W}, as well as high-precision solutions to the Schr\"odinger equation for trapped systems of unitary fermions for up to $N=6$ \cite{2007PhRvL..99w3201B,2011CRPhy..12...86B}. These results have been used to determine the parameters $L,L_0$ for which systematic errors are minimized for the lattice theory described in \cite{EKLN1}. 
The final results from the lattice theory agree with the benchmark results to within the approximately $1\%$ error bars.

%
%


\section{Selected Results \label{Sec:Results}}

In this section we present a few selected lattice results for few- and many-fermion systems in the unitary limit.
This is only a sample of what the methods can accomplish and is by no means fully representative of all the efforts in the field.

\subsection{Unitary fermions in a trap}

As a first demonstration of the techniques outlined in this article to many-body systems we present results for the ground-state energies of up to $N=70$ unpolarized unitary fermions in a harmonic trap from \cite{EKLN1}. Agreement with benchmark results for this system using an improved action with four tuned couplings were presented in Sec.~\ref{Sec:Systematics}.

\begin{figure}[h]
\includegraphics[width=0.65\columnwidth]{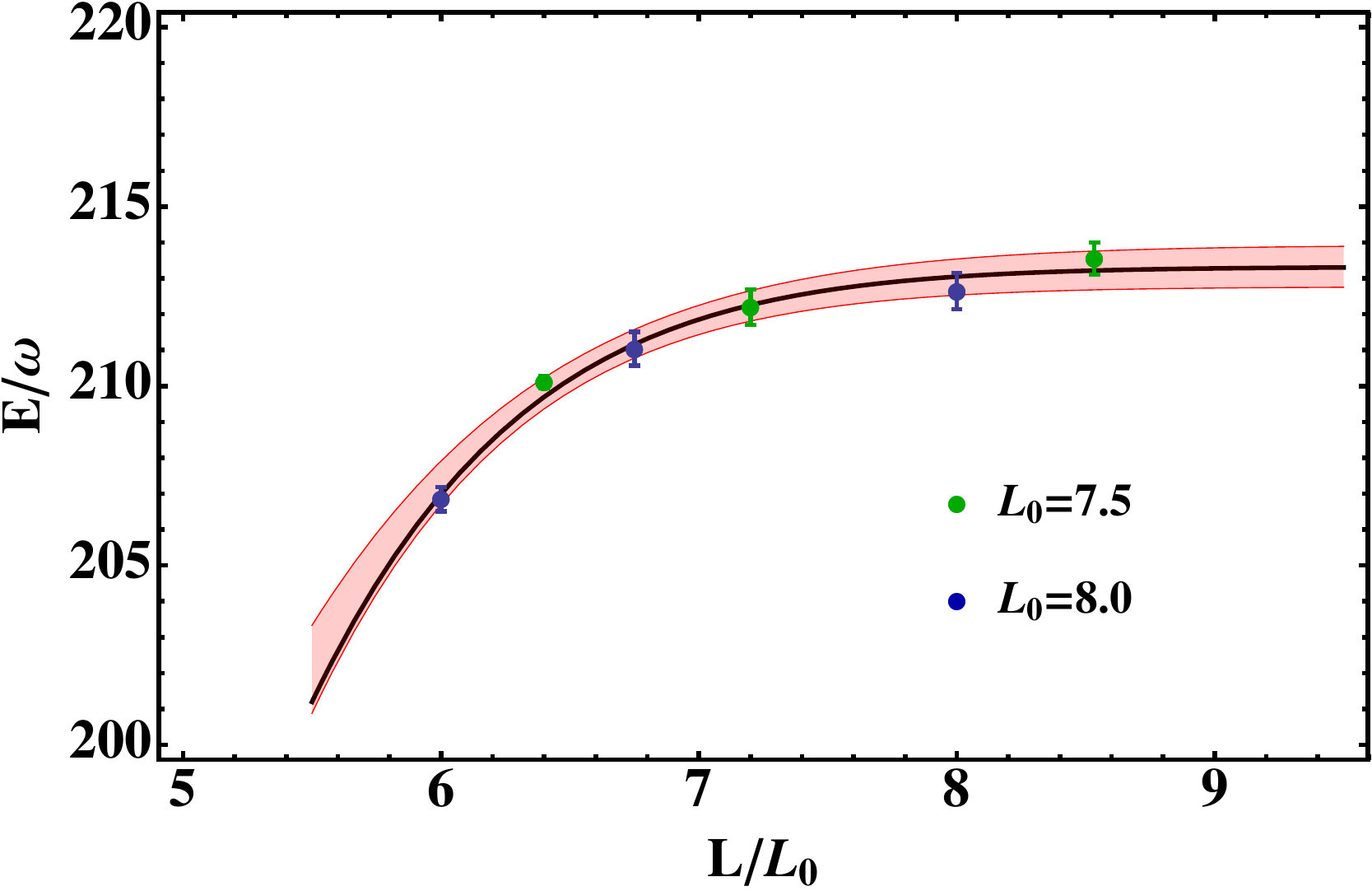}
\caption{%
\label{fig:ExtrapN70}
Volume dependence of the ground state energies (in units of $\omega$) for $N = 70$.
The data points indicate the individual results for six values of $L/L_0$, corresponding to two values of $L_0=7.5,8.0$. The infinite volume extrapolation is shown as a solid line, while the band represents the combined statistical and fitting systematic errors from extrapolations performed for the two values of $L_0$ both separately and combined using Eq.~(\ref{Eq:VolExtrapTrap}). Figure from \cite{EKLN1}.
}
\end{figure}

In addition to the increased computational time associated with larger systems, many of the issues discussed in previous sections become amplified as the number of particles is increased. For example, as more particles are added to the trapped system, the wavefunction spreads out in coordinate space, increasing sensitivity to the periodic boundary at $x=L$. Therefore, finite volume errors increase and in general one needs to perform an infinite volume extrapolation. To aid in this extrapolation, one may use the knowledge that asymptotically the wavefunction for unitary fermions in a trap has the same behavior as that for free particles in a trap. Thus, finite volume errors will fall off as a Gaussian function, and one may perform a fit of the energies using the following form:
\beq
\label{Eq:VolExtrapTrap}
f(L/L_0) = E_{\infty}\left(1-Ae^{-B(L/L_0)^2}\right) \, ,
\eeq
where $E_{\infty}$ is the energy at infinite volume, $L_0$ is the characteristic size of the trap, and $A,B$ are fitting parameters. An example of such an extrapolation is shown in Fig.~\ref{fig:ExtrapN70}.

\begin{figure}[h]
\includegraphics[width=0.65\columnwidth]{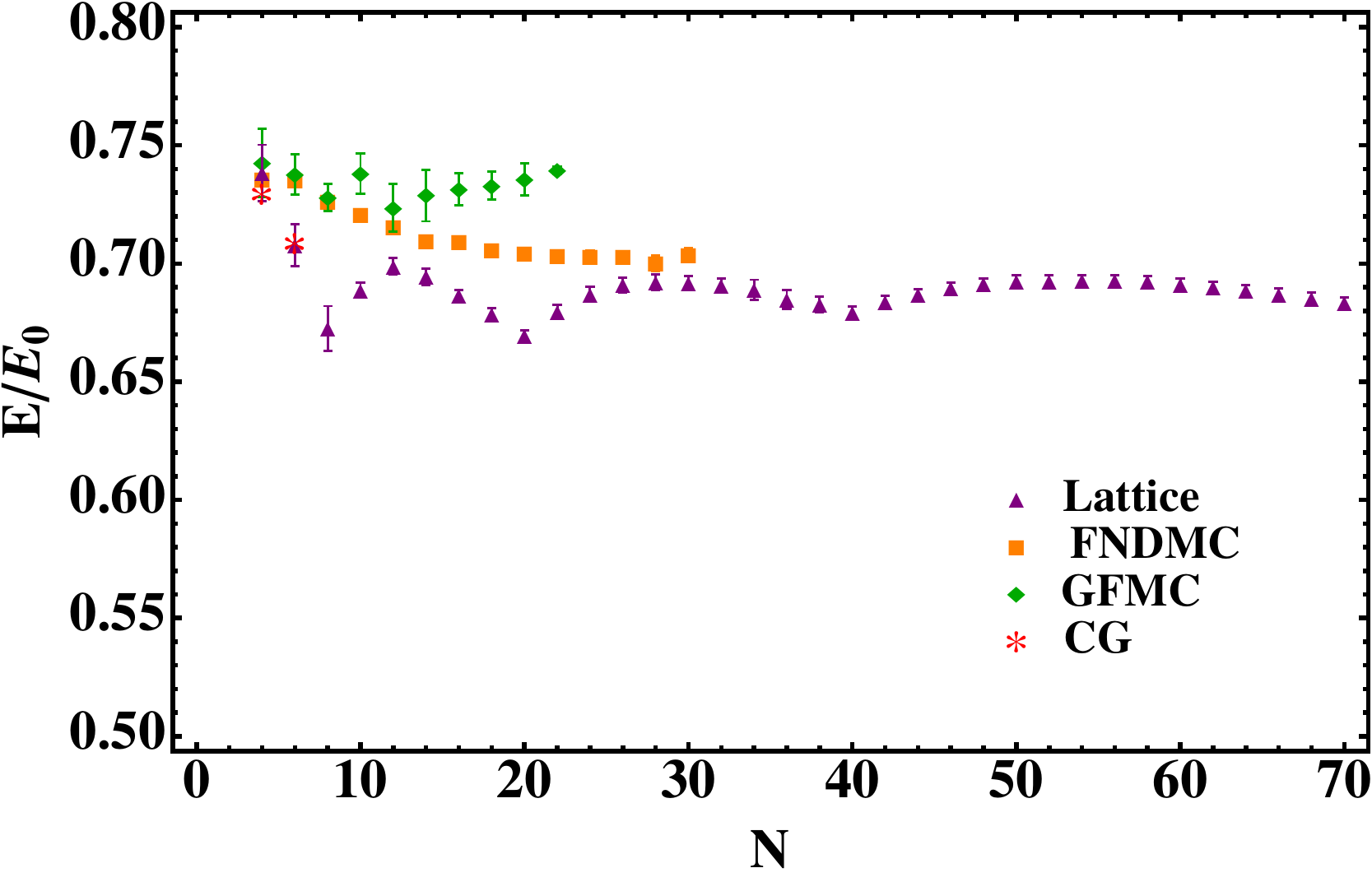}
\caption{%
\label{fig:bertschtrap}
Ground state energies of $N$ trapped unitary fermions in units of the corresponding energies of $N$ trapped noninteracting fermions as a function of $N$. For comparison, we show results from GFMC~\cite{Chang:2007zzd}, FN-DMC~\cite{2007PhRvL..99w3201B}, and CG~\cite{2011CRPhy..12...86B} methods. Figure from \cite{EKLN1}.
}
\end{figure}

The final results for the extrapolated energies as a function of $N$ are shown in Fig.~\ref{fig:bertschtrap} in units of the corresponding energies for $N$ noninteracting fermions in a trap. These results were generated with the same improved action as that for the few-body calculations discussed above using the heat-bath algorithm and utilizing the cumulant expansion Eq.~(\ref{CumulantExp}) to control the large overlap problem. The combined statistical, fitting systematic, discretization, and finite volume errors are well under the $1\%$ level for large $N$. 

A comparison with two fixed-node calculations, Diffusion Monte Carlo (FN-DMC) and Green's Function Monte Carlo (GFMC), are also shown, along with the benchmark results for $N=4,6$. In addition to having generally lower results than the two variational approaches, the lattice data shows marked shell structure, potentially indicating that $N=70$ is not sufficient to reach the thermodynamic limit for the trapped system.

%

\subsection{Unitary fermions in a box: energy}

\begin{figure}[h]
\includegraphics[width=0.95\columnwidth]{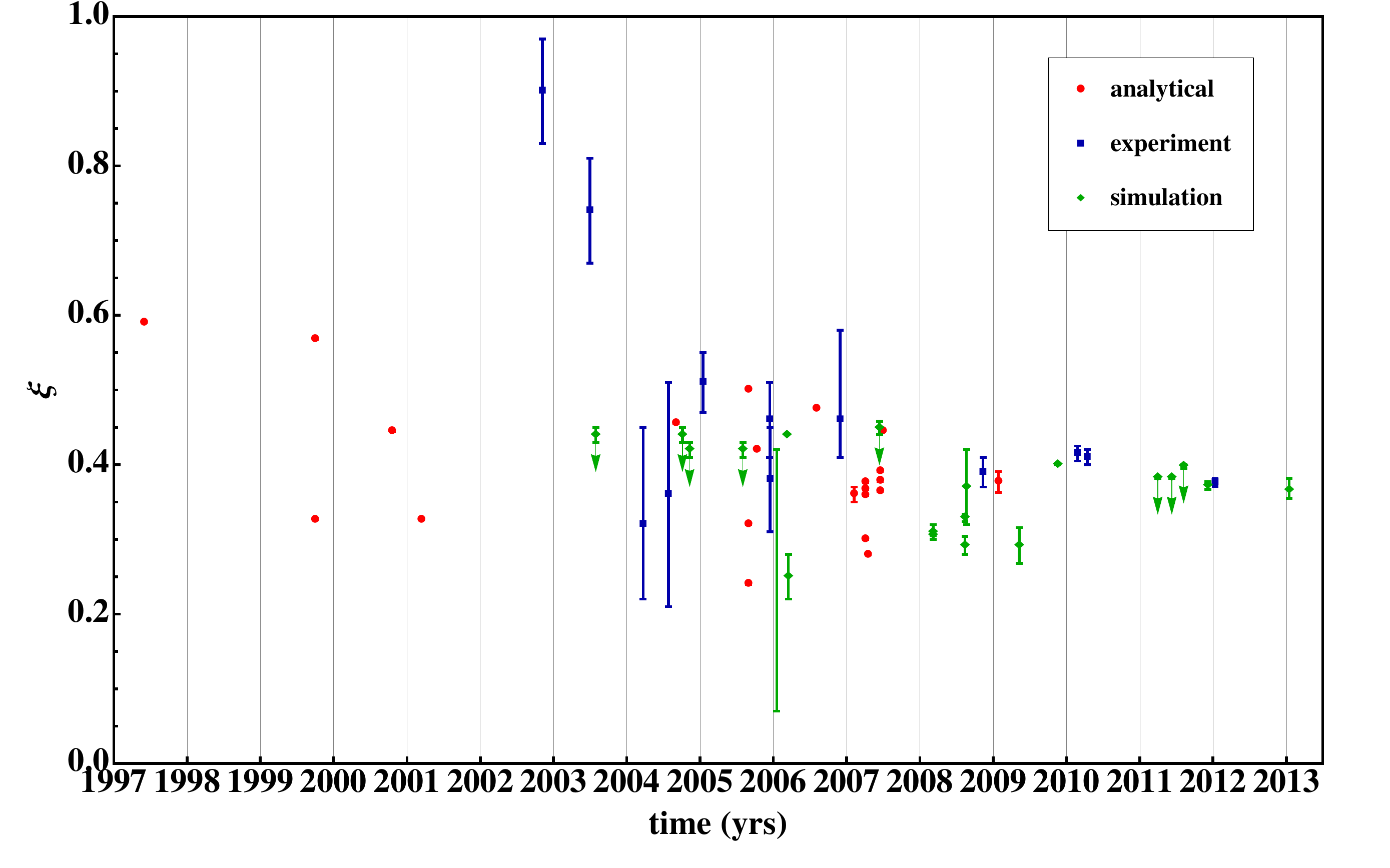}
\caption{
\label{fig:bertsch_history}%
Historical results for the Bertsch parameter determined experimentally, by analytic calculation, and by numerical simulation.
Numerical values and citations are tabulated in Table~\ref{tab:bertsch_history}. Figure from \cite{EKLN4}.
}
\end{figure}

\begin{figure}[h]
\includegraphics[width=0.65\columnwidth]{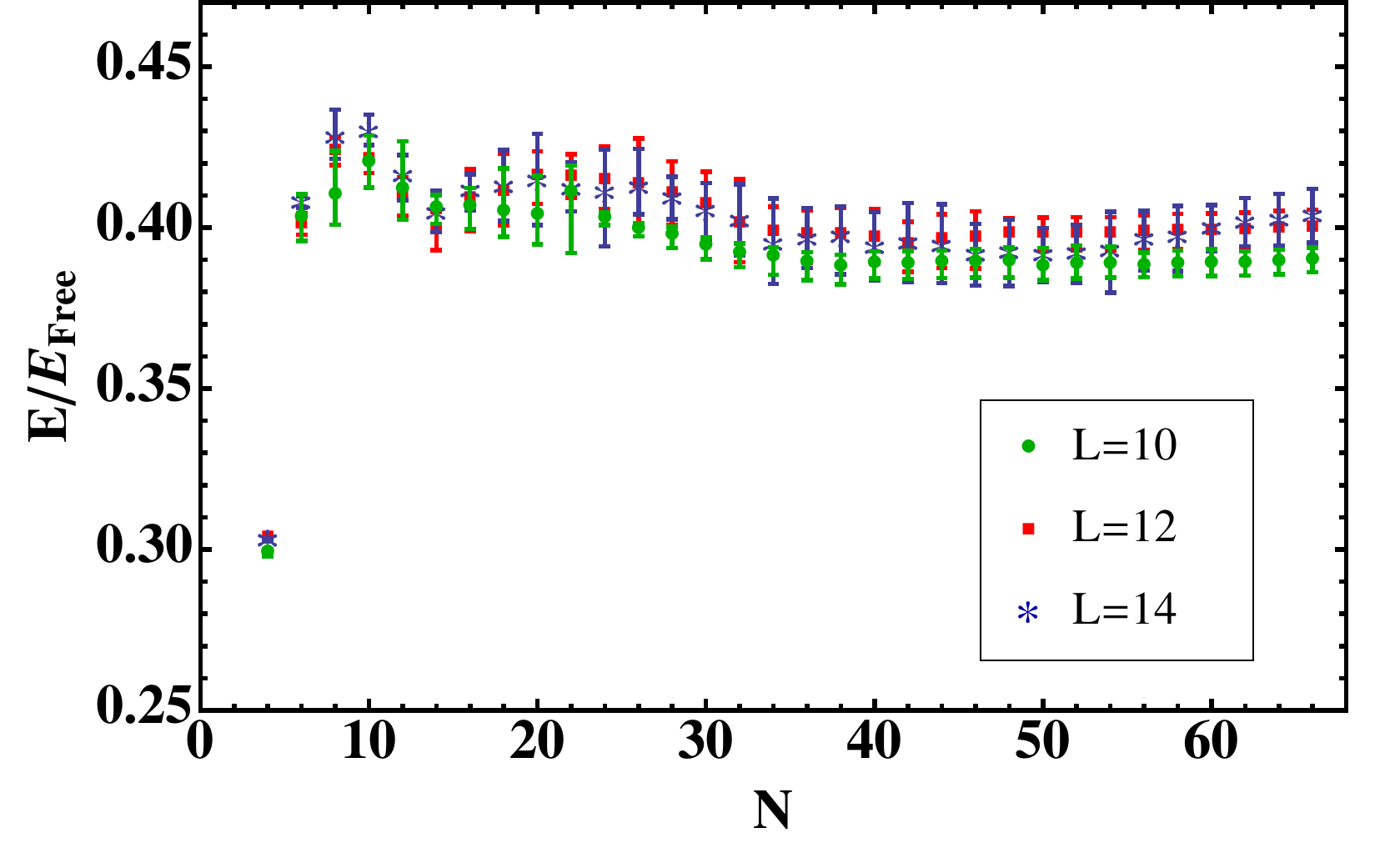}
\caption{
\label{fig:untrapped_bertsch}%
Ground state energy of untrapped unitary fermions in units of $E_{Free}(\rho) = 3N E_{F}(\rho)/5$, where $E_{F}(\rho) = (3\pi^2\rho)^{2/3}/2M$, as a function of N for $L=10, 12$ and $L=14$. Data from \cite{EKLN4}.
}
\end{figure}

Lattice results from \cite{EKLN4} for the ground-state energies of up to $N=66$ unpolarized unitary fermions in a periodic box in units of the corresponding noninteracting energies are presented in Fig. \ref{fig:untrapped_bertsch}. As discussed in the Introduction, this ratio of energies is an important universal quantity known as the Bertsch parameter. There have been many calculations of this quantity, both numerical and theoretical, as well as extractions from experiment. A plot of the values reported by various groups is shown in Fig.~\ref{fig:bertsch_history}, with values and references for the points in Table~\ref{tab:bertsch_history}.

\begingroup
\begin{table}
\caption{%
\label{tab:bertsch_history}%
Historical results for the Bertsch parameter $\xi$ determined experimentally (exp.), by numerical simulation (sim.) and by analytic calculation (anal.), along with publication (pub.) date.
Values obtained variationally are upper bounds, and are indicated with an asterisk; simulation results without a quoted error bar should be regarded as approximate.
}
\begin{tabular}{ccl|ccl|ccc}
pub. date & $\xi$ (exp.) & ref. & pub. date &  $\xi$ (sim.) & ref. & pub. date &  $\xi$ (anal.) & ref. \\ 
\hline
2002-11-07 & 0.90(7)            & \cite{exp_Duke}                         &  2003-07-31 & $0.44(1)^*$    & \cite{Carlson}                   & 1997-06-01 & 0.59      & \cite{PhysRevB.55.15153}      \\
2003-07-02 & 0.74(7)            & \cite{exp_Duke2}                     &  2004-10-05 & $0.44(1)^*$    & \cite{Chang}                      & 1999-10-01 & 0.326     & \cite{PhysRevC.60.054311}     \\
2004-07-27 & 0.36(15)           & \cite{PhysRevLett.93.050401}                  &  2004-11-10 & $0.42(1)^*$    & \cite{PhysRevLett.93.200404}                   & 1999-10-01 & 0.568     & \cite{PhysRevC.60.054311}     \\
2004-03-23 & $0.32^{+13}_{-10}$ & \cite{exp_Innsbruck}                  &  2005-08-02 & $0.42(1)^*$    & \cite{PhysRevLett.95.060401}                   & 2000-10-19 & 4/9       & \cite{2000nucl.th..10066S}    \\
2005-01-16 & 0.51(4)            & \cite{exp_Duke3}                         &  2006-01-18 & 0.07-0.42      & \cite{PhysRevC.73.015202}                      & 2001-03-14 & 0.326     & \cite{Heiselberg}     \\
2005-12-14 & 0.46(5)            & \cite{exp_Rice}                      &  2006-03-10 & 0.44           & \cite{PhysRevLett.96.090404}                   & 2004-09-03 & 0.455     & \cite{PhysRevLett.93.100404}  \\
2005-12-16 & 0.38(7)            & \cite{PhysRevLett.95.250404}                  &  2006-03-17 & 0.25(3)        & \cite{PhysRevB.73.115112}                      & 2005-08-30 & 0.32      & \cite{2005NuPhA.762...82S}    \\
2006-11-30 & $0.46^{+12}_{-5}$  & \cite{PhysRevLett.97.220406}                  &  2007-06-14 & $0.449(9)^*$   & \cite{1367-2630-9-6-163}                       & 2005-08-30 & 0.24      & \cite{2005NuPhA.762...82S}    \\
2008-11-11 & 0.39(2)            & \cite{springerlink:10.1007/s10909-008-9850-2} &  2008-03-07 & 0.31(1)        & \cite{springerlink:10.1140/epja/i2008-10537-2} & 2005-08-30 & 0.5       & \cite{2005NuPhA.762...82S}    \\
2010-04-15 & 0.41(1)            & \cite{Navon07052010}                          &  2008-03-07 & 0.306(1)       & \cite{springerlink:10.1140/epja/i2008-10537-2} & 2005-10-11 & 0.42      & \cite{PhysRevA.72.041603}     \\
2010-02-25 & 0.415(10)          & \cite{2010Natur.463.1057N}                    &  2008-08-13 & 0.292(12)      & \cite{Lee1}                      & 2006-08-04 & 0.475     & \cite{PhysRevLett.97.050403}  \\
2012-01-12 & 0.376(4)           & \cite{Ku12012012}                             &  2008-08-13 & 0.329(5)       & \cite{Lee1}                      & 2007-02-08 & 0.36(1)   & \cite{PhysRevA.75.023610}     \\
       &                &                                        &  2008-08-21 & 0.37 (5)       & \cite{BDM_4}                      & 2007-04-18 & 0.279     & \cite{PhysRevA.75.043614}     \\
           &                    &                                               &  2009-05-11 & 0.292(24)      & \cite{PhysRevC.79.054003}                      & 2007-04-05 & 0.300     & \cite{ArnoldDrutSon}     \\
           &                    &                                               &  2009-11-19 & 0.4            & \cite{PhysRevLett.103.210403}                  & 2007-04-05 & 0.367     & \cite{ArnoldDrutSon}     \\
           &                    &                                               &  2011-04-01 & $0.383(1)^*$   & \cite{Stefano}                      & 2007-04-05 & 0.359     & \cite{ArnoldDrutSon}     \\
           &                    &                                               &  2011-06-10 & $0.383(1)^*$   & \cite{PhysRevLett.106.235303}                  & 2007-04-05 & 0.376     & \cite{ArnoldDrutSon}     \\ 
           &                    &                                               &  2011-08-08 & $0.398(3)^*$   & \cite{PhysRevA.84.023615}                      & 2007-06-18 & 0.391     & \cite{PhysRevA.75.063617}     \\
           &                    &                                               &  2011-12-07 & 0.372(5)       & \cite{PhysRevA.84.061602}                      & 2007-06-18 & 0.364     & \cite{PhysRevA.75.063617}     \\
           &                    &                                               &       2013-02-15       &    0.366(16)  &           \cite{EKLN4}              & 2007-06-18 & 0.378     & \cite{PhysRevA.75.063617}     \\
           &                    &                                               &          &               &                                              & 2007-07-01 & 4/9       & \cite{0256-307X-24-7-011}     \\
           &                    &                                               &             &                &                                                & 2009-01-27 & 0.377(14) & \cite{PhysRevA.79.013627}     \\
\end{tabular}
\end{table}
\endgroup

The results from this work were generated using an improved action with five tuned couplings on lattices ranging from $L = 10-16$. The calculation was performed using the heat-bath algorithm in combination with the cumulant expansion to control the overlap problem. 

Note that the use of a highly improved action gives results with little variation in the energies with the volume. The remaining finite volume effects were removed through an infinite volume extrapolation using the Symanzik effective action to guide the fit function, as described in Section 6.1. The result of a fit to the extrapolated energies for the region $n\geq 40$ (representing the thermodynamic limit) is shown as the last point in Fig.~\ref{fig:bertsch_history}. 

%
	
\subsection{Unitary fermions in a box: contact}

To illustrate the effect of improved operators in realistic lattice calculations, this subsection
presents the results of ground-state MC calculations of the contact, for an unpolarized 
system at unitarity.

Results are shown for 80 particles (40 per spin) in a volume of $10^3$ lattice points, for 
various improvement levels $N^{}_\mathcal{O} = 1 - 4$.
In each case, calculations were performed for time directions of extent 
$\beta \epsilon_F^{} = 2.0 - 8.0$ (corresponding to $N^{}_{\tau}$ roughly between 40 and 200), 
subsequently extrapolating to the $\beta \to \infty$ limit. 
We have taken $b_\tau = 0.05$, and a Slater determinant of plane 
waves as the starting guess for the ground-state wavefunction. For each value of $\beta$, we obtained 
approximately $400$ samples of the auxiliary field $\phi$, and the statistics was enhanced by 
a factor of $\sim 20$ by inserting operators at every other time slice.

\begin{figure}[h]
\includegraphics[width=0.65\columnwidth]{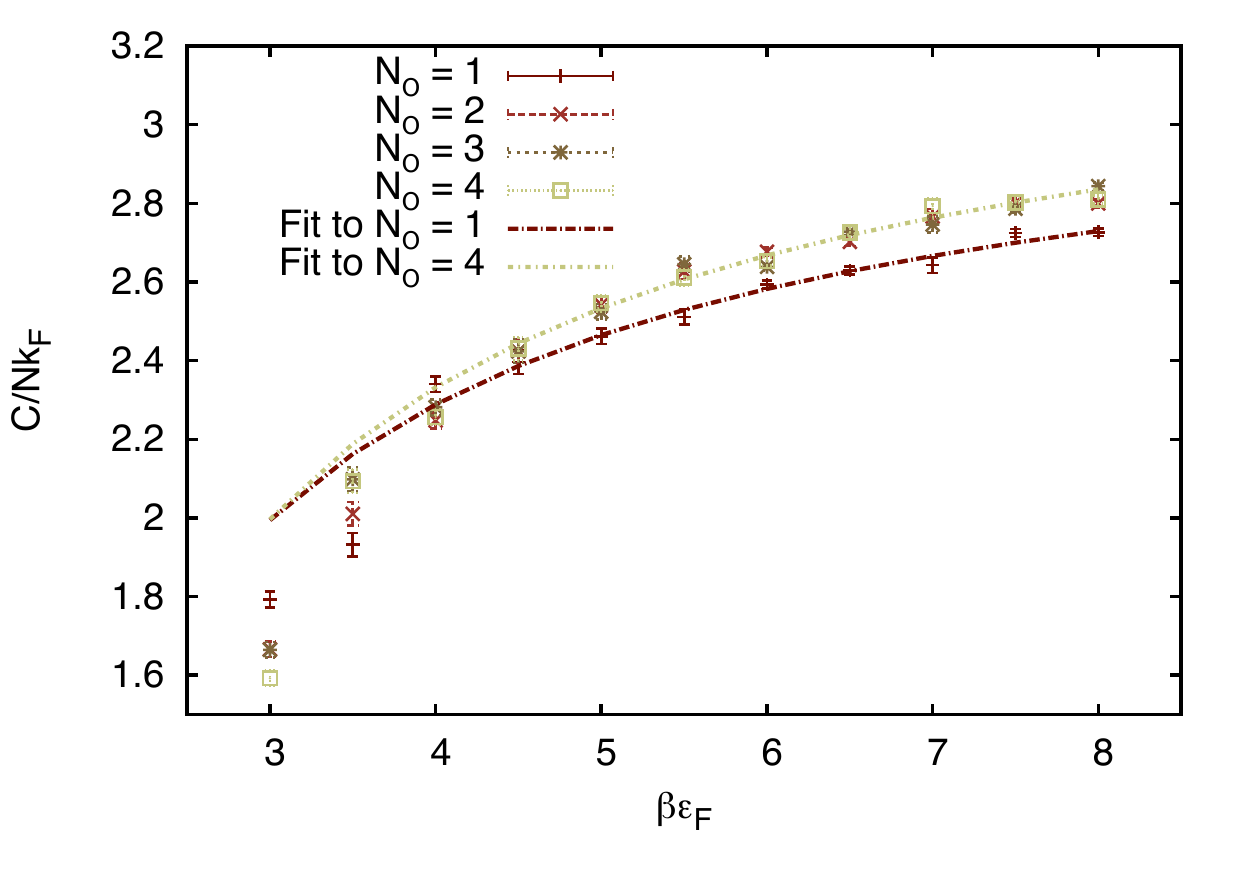}
\caption{
\label{Fig:Contact_40}
(Color online) Contact, in units of $N k_F^{}$, where $k_F^{} = (3 \pi^2 N/V)^{1/3}$ is the Fermi momentum, 
as a function of the extent of the time direction $\beta \epsilon_F^{} = \tau N_\tau \epsilon_F^{}$,
for 80 particles and levels of improvement $N^{}_\mathcal{O} = 1 - 4$. Also shown are fits to the $N^{}_\mathcal{O}\!=\!1$ and
$N^{}_\mathcal{O}\!=\!4$ datasets using the first two terms in the asymptotic form of the previous section (see text for details).}
\end{figure}

Using extrapolation formulae at leading plus next-to leading order, 
we obtain $C/(N k_F^{}) = 3.17(3)$ in the unimproved case, and $C/(N k_F^{}) = 3.31(4), 3.34(4)$ and $3.30(4)$
for $N_{\mathcal O} = 2,3,4$, respectively (discarding three data points at the lowest values of $\beta \epsilon_F^{}$ 
to capture the asymptotic behavior at large $\beta$; the fits are stable against further removal of points at low 
$\beta \epsilon_F^{}$). 
While the latter extrapolations clearly overlap when taking the uncertainty into account, they do not overlap with the unimproved case,
which is clearly consistent with what we see in Fig.~\ref{Fig:Contact_40} at large $\beta$. 
The remaining systematic errors, likely due to volume effects for the most part, appear to be only as large as 
$3\%$, if we consider the most recent estimates of the contact in the ground state ($\approx 3.39$, see Ref.~\cite{Stefano}). 
Remarkably such a small volume as $10^3$ already yields a result that is close to the best current estimate.


\section{Summary \label{Sec:Summary}}

In this short review article we have covered some of the most recent progress on the 
technical aspects of calculating the ground state properties of strongly interacting 
Fermi gases. We have focused on non-relativistic Hamiltonians with two-body interactions, as 
relevant for ultracold atom experiments and dilute neutron matter among other systems.

Once the problem is placed on a space-time lattice, the partition function is written as 
a path integral over the HS auxiliary field. In order to compute expectation values of 
operators in a non-perturbative fashion, one resorts to stochastic methods to evaluate
path integrals, assuming that a useful positive semi-definite probability can be defined. 
We have discussed various sampling algorithms to accomplish this, from the simplest approaches 
based on local updates of the auxiliary field, to more sophisticated HMC methods, and a very 
recent technique based on heat-bath-like ideas.

To reduce lattice-spacing effects one must take the low-density limit and/or implement 
improved actions, along the lines of similar efforts in Lattice QCD. We have explained how 
this can be accomplished by designing improved transfer matrices. Furthermore,
we have shown that that approach can be extended to the construction of improved 
operators.

We have also presented a small set of illustrative results.

Finally, the Appendix contains an abridged discussion of some elementary aspects of 
point group theory, as relevant for the discussion of angular momentum (or its remnants) 
on the lattice.

\ack
We gratefully acknowledge discussions with J. Braun, A. Bulgac, J. Carlson, D. Lee, D. B. Kaplan, and M. W. Paris.
Part of this work was supported by the U.S. Department of Energy under grants DE-FG02-93ER-40762 and
DE-FC02-07ER41457 (UNEDF SciDAC).

\newpage
\section{Appendix I: Group theory on the lattice \label{Sec:GroupTheory}}
\subsection{Introduction and basic definitions}

When placing a field theory on the lattice, i.e. when replacing the spacetime continuum with a mesh, one 
obviously loses a fair amount of symmetry. This is the price to pay for going from a problem with infinitely 
many degrees of freedom to one with a finite number. While we have argued that lower densities and 
improved actions take us closer to the continuum limit, we have left for this Appendix the more technical 
discussion of the isometry group associated with the hypercubic lattice.

In order to understand the degree to which the lattice breaks (or preserves) the symmetries of the original theory,
it is central to discuss the point group which remains as the unbroken group of rotations, namely the hypercubic 
group $O_h$, and its irreducible representations (irreps). Indeed, a subspace of states that transform under a 
given irrep should remain invariant under time evolution if the original (continuum) Hamiltonian was rotationally 
invariant. This allows one to classify the scattering states according to the discrete equivalent of angular 
momentum. As we shall see, there is only a finite number of irreps of $O_h$, such that the continuum irreps of
the rotation group $O(3)$, characterized by an integer $l$, are decomposed into direct sums of the irreps of $O_h$.

Following closely the book by Tinkham~\cite{TinkhamBook}, we shall first review a series of important definitions and theorems 
(without proof) in the theory of group representations. We will bypass the most elementary definitions concerning 
the groups themselves.

By definition, a representation of a group is a group consisting of concrete mathematical entities, such as numbers or 
matrices. In fact, since we are dealing with quantum mechanics, the representations we will discuss are either operators
that act on an infinite-dimensional Hilbert space (when speaking about the continuum limit), or actual matrices (when 
discussing problems on the lattice). We will have in mind the latter case, restricted to square matrices in particular. 
More specifically, a representation associates a matrix ${\bf \Gamma}(A)$ to a given element $A$ of the group in question,
such that, given any other element $B$ of the group,
\beq
{\bf \Gamma}(A){\bf \Gamma}(B) = {\bf \Gamma}(AB),
\eeq
i.e. the matrices in the representation inherit the multiplication rules from the group. It is easy to convince oneself that,
in particular, the identity element $E$ must be mapped onto the identity matrix, i.e. ${\bf \Gamma}(E) = {\bf E}$.
It is also easy to see that from a given representation $\bf \Gamma$ one can obtain another one $\bf \Gamma'$ by 
a similarity transformation on all the matrices: ${\bf \Gamma}' = S{\bf \Gamma}S^{-1}$. Those two representations are 
therefore said to be {\it equivalent}.

In general, representations will be {\it reducible}, by which we mean that it is possible to bring them to a block-diagonal
form by a unitary transformation (one and the same transformation for the whole original representation), or otherwise
{\it irreducible} if this is not possible.

In this context, the following lemma is often useful: {\it Any representation by matrices with non-vanishing determinant
is equivalent through a similarity transformation to a representation in terms of unitary matrices.}

This lemma is useful in particular when proving the famous Schur's lemma: {\it Any matrix which commutes with all the
matrices of an irreducible representation must be a multiple of the identity matrix}. If such a non-constant matrix exists,
the representation must be reducible; if it does not exist, then the representation is irreducible.

\subsection{Orthogonality theorems, characters and conjugacy classes}

The above are very important, but somewhat formal statements. They allow, however, to prove an identity 
concerning the orthogonality property of the representations that is extremely useful in practice:
\beq
\sum_{R} {\bf \Gamma}^{(i)}(R)^{*}_{\mu \nu} {\bf \Gamma}^{(j)}(R)^{}_{\alpha \beta} = 
\frac{h}{l_i} \delta_{ij} \delta_{\mu\alpha} \delta_{\nu\beta},
\eeq
where the sum is over all the group elements, $i,j$ denote particular representations, $l_i$ is the dimension of the 
$i$-th representation, and $h$ is the order of the group (i.e. the number of elements in it). This equation is valid when
$i,j$ denote {\it irreducible}, {\it inequivalent}, {\it unitary} representations of the group in question.

As we shall see, this statement is quite powerful. It can be interpreted geometrically if we set $i\!=\!j$ and regard $R$ as 
denoting a coordinate in a vector space of dimension $h$. The above is then a statement of orthogonality between as 
many vectors as matrix elements in the $i$-th representation, of which there are $l_i^2$. This is valid for all representations,
so the above indicates that there are $\sum_i l_i^2$ orthogonal vectors. But if the dimension of space is $h$, then we must have
$\sum_i l_i^2 \leq h$. As it turns out, the equality is valid, so we have
\beq
\sum_i l_i^2 = h,
\eeq
which dramatically constrains the possible dimensions of the collection of irreducible, inequivalent, unitary representations of the group.

We may also obtain a statement about the traces $\chi^{(i)}(R) = \Tr  {\bf \Gamma}^{(i)}(R)$ of the matrices in the representation. 
These are usually referred to as {\it characters} and they have a special name because they are invariant under similarity 
transformations, such that they are a specific signature of each representation. Summing over $\mu = \nu$ and $\alpha = \beta$, we see 
that characters are orthogonal:
\beq
\sum_{R} {\chi}^{(i)}(R)^{*} {\chi}^{(j)}(R) = 
{h} \delta_{ij}.
\eeq

Within a group it is generally possible to identify subsets of elements which, when taken together, form a partition 
(i.e. the subsets are such that each element belongs to one and only one subset, and the union of all the subsets yields 
the full group). Each subset is defined by the following (equivalence) relation: 
Two elements $A$ and $B$ in the group are said to be {\it conjugate} to each other if an element $X$ exists such that
\beq
A = X B X^{-1}.
\eeq
The subsets in the partition are therefore referred to as {\it conjugacy classes}, and contain elements connected to each other by the 
above relation. Clearly, the identity element $E$ is in a class by itself, as it cannot be conjugate to any other element. In the context 
of groups of symmetry transformations, which is the application we have in mind here, conjugacy classes roughly correspond to different kinds 
of operations (e.g. rotations about a particular axis).

With the above definition, it is clear that the characters of two elements that are in the same class will be the same. We may therefore
consider the characters of the classes ${\mathcal C}_k$ within a given representation $i$ instead of those of each of the elements $R$:
\beq
\sum_{R} {\chi}^{(i)}(R)^{*} {\chi}^{(j)}(R) = 
\sum_{k} {\chi}^{(i)}({\mathcal C}_k)^{*} {\chi}^{(j)}({\mathcal C}_k)N^{}_k =
{h} \delta_{ij},
\eeq
where $N^{}_k$ is the number of elements in the $k$-th class. Once again, a geometric interpretation helps: If we regard the
characters as vectors with entries in the space of conjugacy classes, then the above indicates that there exist as many orthogonal
vectors in this space as irreducible representations. But clearly this number cannot exceed the dimensionality of the space. Therefore,
the number of inequivalent, unitary irreps $N_{irreps}$ is less than or equal to the number of classes $N^{}_{\mathcal C}$. As it turns out, 
they are equal:
\beq
N_{irreps} = N^{}_{\mathcal C}.
\eeq

Using the above identities one can also show another orthogonality relation between characters:
\beq
\sum_{i} {\chi}^{(i)}({\mathcal C}_k)^{*} {\chi}^{(i)}({\mathcal C}_l) =
\frac{h}{N^{}_k} \delta_{kl}.
\eeq
%

\subsection{Decomposition of reducible representations}

If you were wondering when all of the above would become useful, this is it.

The character of a reducible representation is obviously the sum of the characters of all the representations it
contains, i.e.
\beq
{\chi}(R) = \sum_{j} a_j {\chi}^{(j)}(R)
\eeq
where $a^{}_j$ is the multiplicity of the $j$-th irrep, i.e. the number of times it appears. 
The above orthogonality relations can therefore help us identify the irrep content of a given representation:
\beq
a_j  = h^{-1} \sum_{R} {\chi}^{(j)}(R)^{*} {\chi}(R) = 
h^{-1} \sum_{k} N^{}_k {\chi}^{(j)}({\mathcal C}_k)^{*} {\chi}({\mathcal C}_k).
\eeq
%

\subsection{Projection operators for irreducible representations}

Let the basis of the $j$-th irrep be denoted by $\varphi_\alpha^{(j)}$, where $\alpha = 1,...,l^{}_j$.
Given a particular basis element $\varphi_\kappa^{(j)}$, the other ones are usually referred to as {\it partners}.
By definition, the action of an element of the group on $\varphi_\kappa^{(j)}$ can be expanded as a linear combination
in the basis of that function and its partners:
\beq
P^{}_R \varphi_\kappa^{(j)} = \sum^{l^{}_j}_{\lambda=1}\varphi_\lambda^{(j)}{\bf \Gamma}^{(j)}(R)_{\lambda,\kappa}.
\eeq
Multiplying by ${\bf \Gamma}^{(i)}(R)^*_{\lambda',\kappa'}$, summing over $R$, and using the orthogonality theorems, we obtain
\beq
\sum_R {\bf \Gamma}^{(i)}(R)^*_{\lambda',\kappa'} P_R^{} \varphi_\kappa^{(j)} = 
\frac{h}{l^{}_j} \delta^{}_{ij} \delta^{}_{\kappa,\kappa'} \varphi_{\lambda'}^{(j)}.
\eeq
Clearly then, applying to a function the operator
\beq
\label{Eq:ProjectionOp}
\mathscr{P}_{\lambda \kappa}^{(j)} = \frac{l^{}_j}{h} \sum_R {\bf \Gamma}^{(j)}(R)^*_{\lambda,\kappa} P_R^{} 
\eeq
yields zero unless the function has a projection on the $\kappa$-th basis element of the $j$-th representation, 
in which case the result is transferred onto the $\lambda$-th direction. For this reason these operators are sometimes 
referred to as {\it transfer} operators. The special case $\lambda = \kappa$ yields a {\it projection} operator, as in that case 
we have
\beq
\mathscr{P}_{\kappa \kappa}^{(j)} \varphi_{\kappa}^{(j)} = \varphi_{\kappa}^{(j)}.
\eeq

While the above identities are useful when we already know the explicit form of the $\Gamma$ matrices, one can still obtain
practical results by knowing just the characters. Indeed, setting $\lambda = \kappa$ in Eq.~\ref{Eq:ProjectionOp} and summing over
$\kappa$ we obtain a new projection operator
\beq
\mathscr{P}_{}^{(j)} = \sum_{\kappa} \mathscr{P}_{\kappa \kappa}^{(j)} = 
\frac{l^{}_j}{h} \sum_R {\chi}^{(j)}(R)^*P_R^{} 
\eeq
which projects onto the $j$-th irrep.

\subsection{The octahedral group}


By far the most common choice in lattice field theory calculations is to discretize spacetime as a hypercubic mesh.
Therefore the point group one is concerned with is the so-called {\it octahedral} group $O$. This group contains the proper 
(no parity transformations) rotations which take a cube or an octahedron into itself. In this case $h=24$, i.e. it has 24 elements, 
which include, apart from the identity ($E$), eight rotations about cube body diagonals ($8 C^{}_3$), nine rotations around the $X$,$Y$,$Z$ 
axes ($3 C^{}_2$ plus $6 C^{}_4$), and six rotations about axes parallel to face diagonals ($6 C^{}_2$). Should we choose to include 
the inversion operation as well, which form the group $i$, we will then obtain the group $O^{}_h$, where $O^{}_h = O \times i$. This is 
referred to as the {\it full octahedral} group, and with 48 elements it is the largest of all the point groups.

\begin{table}[h!]
\begin{center}
\caption{\label{Table:CharactersO} Table of characters of the octahedral group $O$.}
\begin{tabularx}{\columnwidth}{@{\extracolsep{\fill}}c c c c c c}
       \hline
       $$ & $E$ & $8 C^{}_3$ & $3 C^{}_2$ & $6 C^{}_2$ & $6 C^{}_4$\\
       \hline
       \hline
$A^{}_1$   &  1    &  1   &  1    &  1   &  1  \\
$A^{}_2$   &  1    & 1     & 1    &  -1  & -1     \\
$E$           &  2    & -1    & 2    & 0    & 0     \\
$T^{}_1$   &  3     & 0   & -1    & -1     & 1     \\
$T^{}_2$   &  3     & 0   &  -1     & 1    & -1    \\
\hline
\end{tabularx}
\end{center}
\end{table}

The table of characters of $O$ is shown in Table~\ref{Table:CharactersO}. The classes appear in the top row, and the
irreps along the first column. The dimensionality of each irrep can be read from the second column.

While we have only 5 irreps for $O$, in the continuum we have $SO(3)$, and therefore an infinite number of irreps, parametrized
by an integer $\ell$, with dimensions given by $2\ell + 1$ and spanned by the spherical harmonics $Y_\ell^{m}$. It is therefore natural to 
ask: within each irrep of $SO(3)$, how is the dimensionality split among the finite number of irreps of the octahedral group when we
put the problem on the lattice?

In order to answer this question, we may simply use the character orthogonality relations on the $\ell$-th representation of $SO(3)$.
We need only know the character of the elements of $O$, now interpreted as elements of $SO(3)$, in that representation, which are 
given by
\beq
\chi^{\ell}(\alpha) = \frac{\sin(\ell + \frac{1}{2})\alpha}{\sin(\alpha/2)}.
\eeq
%

Thus, 
\bea
\chi(C^{}_2) &=& \chi(\pi) = (-1)^\ell_{} \nonumber \\
\chi(C^{}_3) &=& \chi(2\pi/3) = 
\left\{ 
   \begin{array}{l l}
1  \quad \mathrm{\ell=0,3,\dots} &\\
0  \quad \mathrm{\ell=1,4,\dots} &\\
1  \quad \mathrm{\ell=2,5,\dots} &
    \end{array} 
\right. \nonumber \\
\chi(C^{}_4) &=& \chi(\pi/2) = 
\left\{ 
   \begin{array}{l l}
1  \quad \mathrm{\ell=0,1,4,5,\dots} &\\
-1  \quad \mathrm{\ell=2,3,6,7,\dots} &
    \end{array} 
\right. \nonumber .
\eea

Using these results one may then decompose the $s$-wave, $p$-wave, etc., representations of $SO(3)$ in terms of the irreps of the
octahedral group. The (partial) result of such a decomposition is shown in Table~\ref{Table:SO3decomposition}. A considerably more 
thorough treatment of this problem appears in Ref.~\cite{Luu:2011ep}. From the table we see that only the $A^{}_1$ representation
contributes to $s$-wave scattering, but it also appears in higher waves. For $\ell>4$ the various irreps of $O$ will in general appear
more than once.

\begin{table}[h!]
\begin{center}
\caption{\label{Table:SO3decomposition} Characters and decomposition of irreps of $SO(3)$ into irreps of the octahedral group $O$.}
\begin{tabularx}{\columnwidth}{@{\extracolsep{\fill}}c c c c c c c}
       \hline
       $\ell$ & $E$ & $8 C^{}_3$ & $3 C^{}_2$ & $6 C^{}_2$ & $6 C^{}_4$ & Decomposition \\
       \hline
       \hline
$0$   &  1    &  1   &  1    &  1   &  1 & $A^{}_1$ \\
$1$   &  3    & 0     & -1    &  -1  & 1  & $T^{}_1$  \\
$2$   &  5    & -1    & 1    & 1    & -1   & $E + T^{}_2$ \\
$3$   &  7     & 1   & -1    & -1     & -1  & $A_2 + T^{}_1 + T^{}_2$  \\
$4$   &  9     & 0   &  1     & 1    & 1  & $A_1 + E + T^{}_1 + T^{}_2$ \\
\hline
\end{tabularx}
\end{center}
\end{table}


\newpage
\section*{References}

\bibliographystyle{unsrt.bst}
\bibliography{Latticemethods}


\end{document}